\documentclass[sigconf,9pt]{acmart}

\newcommand*{\ARCHIVE}{}

\ifdefined\ARCHIVE
    \renewcommand\footnotetextcopyrightpermission[1]{}
    \setcopyright{none}
\else
    \copyrightyear{2020}
    \acmYear{2020}
    \setcopyright{acmcopyright}
    \acmConference[SIGCOMM '20]{Annual conference of the ACM Special Interest Group on Data Communication on the applications, technologies, architectures, and protocols for computer communication}{August 10--14, 2020}{Virtual Event, NY, USA}
    \acmBooktitle{Annual conference of the ACM Special Interest Group on Data Communication on the applications, technologies, architectures, and protocols for computer communication (SIGCOMM '20), August 10--14, 2020, Virtual Event, NY, USA}
    \acmPrice{15.00}
    \acmDOI{10.1145/3387514.3405886}
    \acmISBN{978-1-4503-7955-7/20/08}
\fi

\usepackage[english]{babel}

\usepackage{amsmath}
\usepackage{bm}
\usepackage{amsfonts}

\usepackage{booktabs}
\usepackage{multirow}

\usepackage{graphicx}
\usepackage{subcaption}
\graphicspath{ {figures/} }

\usepackage{setspace}

\usepackage[T1]{fontenc}
\usepackage[utf8]{inputenc}
\usepackage{pgfplots}
\pgfplotsset{compat=newest}
\usetikzlibrary{plotmarks}
\usetikzlibrary{arrows.meta}
\usetikzlibrary{patterns}
\usepgfplotslibrary{patchplots}

\makeatletter
\def\input@path{ {charts/}{figures/} }
\makeatother

\usepackage{nidanfloat} 

\usepackage{amsmath}
\usepackage{mathtools}
\usepackage{bm}
\usepackage{amsfonts}

\usepackage[linesnumbered,ruled,vlined]{algorithm2e}
\SetKw{Break}{break}
\let\oldnl\nl
\newcommand{\nonl}{\renewcommand{\nl}{\let\nl\oldnl}}

\usepackage{comment}
\usepackage{xspace}
\usepackage{xstring}

\usepackage[normalem]{ulem}

\renewcommand{\paragraph}[1]{\vspace{3pt}\noindent\textbf{#1.}}
\newcommand*{\RELEASE}{} 
\newcommand{\ouralgorithm}[0]{NuevoMatch\xspace}
\newcommand{\MSNN}[0]{RQ-RMI\xspace}
\newcommand{\MSNNs}[0]{RQ-RMI\xspace}
\newcommand{\iset}[0]{iSet\xspace}
\newcommand{\isets}[0]{iSets\xspace}
\newcommand{\capital}[0]{\expandafter\MakeUppercase}
\newcommand{\CS}[0]{{\textsc{cs}}\xspace}
\newcommand{\NC}[0]{{\textsc{nc}}\xspace}
\newcommand{\TM}[0]{{\textsc{tm}}\xspace}
\newcommand{\NM}[0]{{\textsc{nm}}\xspace}

\newcommand{\ms}[1]{{\color{red}{\sc MARK: #1}}}

\newcommand{\changes}[1]{#1}
\newcommand{\removed}[1]{}

\renewenvironment{itemize}{
   \begin{list}{\labelitemi}{
     \setlength{\topsep}{0.5ex}
     \setlength{\parsep}{0pt}
     \setlength{\partopsep}{0pt}
     \setlength{\itemsep}{-0pt}
     \setlength{\itemindent}{0pt}
     \setlength{\leftmargin}{\labelwidth}
     \addtolength{\leftmargin}{-8pt}}
}{\end{list}}

\ifdefined\ARCHIVE
\else
\fi

\begin{document}

\title{A Computational Approach to Packet Classification} 

\author{Alon Rashelbach}
\affiliation{\institution{Technion}}
\email{alonrs@campus.technion.ac.il}

\author{Ori Rottenstreich}
\affiliation{\institution{Technion}}
\email{or@technion.ac.il}

\author{Mark Silberstein}
\affiliation{\institution{Technion}}
\email{mark@ee.technion.ac.il}

\begin{abstract}
Multi-field packet classification is a crucial component in modern software-defined data center networks. To achieve high throughput and low latency, state-of-the-art algorithms strive to fit the rule lookup data structures into on-die caches; however, they do not scale well with the number of rules.

We present a novel approach, \emph{\ouralgorithm{}}, which   improves the memory scaling of existing methods. A new data structure, \emph{Range Query Recursive Model Index} (\MSNN), is the key component that enables \ouralgorithm{} to replace most of the accesses to main memory with model inference computations. We describe an efficient training algorithm that guarantees the correctness of the \MSNN-based classification.
The use of \MSNN allows the rules to be compressed into model weights that fit into the hardware cache. Further, it takes advantage of the growing support for fast neural network processing in modern CPUs, such as wide vector instructions, achieving a rate of tens of nanoseconds per lookup.

Our evaluation using 500K multi-field rules from the standard ClassBench benchmark shows a geometric mean compression factor of 4.9$\times$, 8$\times$, and 82$\times$, and average performance improvement of 2.4$\times$, 2.6$\times$, and 1.6$\times$ in throughput compared to CutSplit, NeuroCuts, and TupleMerge, all state-of-the-art algorithms\footnote{This work does not raise any ethical issues.}.

\end{abstract}

\ifdefined\ARCHIVE
\else

\fi

\sloppypar
\maketitle

\section{Introduction}
\emph{Packet classification} is a cornerstone of packet-switched networks. Network functions such as switches use a set of \emph{rules} that determine which action
they should take for each incoming packet. The rules originate in higher-level domains, such as routing, Quality of Service, or security policies. They match the packets' metadata, e.g., the destination IP-address and/or the transport protocol. If multiple rules match, the rule with the highest \emph{priority} is used.

Packet classification algorithms have been studied extensively. There are two main classes: those that rely on Ternary Content Addressable Memory (TCAM) hardware~\cite{Taylor2005, Lakshminarayanan2005, Narodytska2019, Katta2016, Liu2008},
and those that are implemented in software ~\cite{Srinivasan1999, Gupta2000, Singh2003, VamananBalajee2011, Wenjun2018, YingSoDa2018, Daly2019, Liang2019}. In this work, we focus on software-only algorithms that can be deployed in virtual network functions, such as forwarders or ACL firewalls, running on commodity X86 servers.

Software algorithms fall into two major categories:  decision-tree based~\cite{Gupta2000, Singh2003, VamananBalajee2011, YingSoDa2018, Wenjun2018, Liang2019} and hash-based~\cite{Srinivasan1999, Daly2019}. The former use decision trees for indexing and matching the rules, whereas the latter perform lookup via hash-tables by hashing the rule's prefixes. 
%
Other methods for packet classification~\cite{Gupta1999, DCFLTaylor2005} are less common as they either require too much memory or are too slow.

A key to achieving high classification performance in modern CPUs is to ensure that the classifier fits into the CPU on-die cache. When the classifier is too large, the lookup involves high-latency memory accesses, which stall the CPU, as the data-dependent access pattern during the lookup impedes hardware prefetching. Unfortunately, as the number of rules grows,  it becomes difficult to maintain the classifier in the cache.
In particular, in decision-tree methods, rules are often replicated among multiple leaves of the decision tree, inflating its memory footprint and affecting scalability. Consequently, recent approaches, notably CutSplit~\cite{Wenjun2018} and NeuroCuts~\cite{Liang2019}, seek to reduce rule replication to achieve better scaling. However, they still \emph{fail to scale} to large rule-sets, which in modern data centers may reach hundreds of thousands of rules~\cite{Firestone2018}.
Hash-based techniques also suffer from poor scaling, as adding rules increases the number of hash-tables and their size.

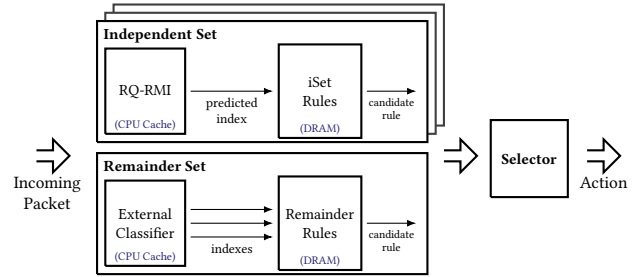
\begin{figure}[t]
	\centering
\ifdefined\RELEASE
\else
\documentclass[conference,compsoc]{IEEEtran}

\usepackage{graphicx}
\usepackage{subfig}

\usepackage[T1]{fontenc}
\usepackage[utf8]{inputenc}
\usepackage{pgfplots}
\usepackage{grffile}
\pgfplotsset{compat=newest}
\usetikzlibrary{plotmarks}
\usetikzlibrary{arrows.meta}
\usepgfplotslibrary{patchplots}

\usepackage{amsmath}
\usepackage{xspace}

\newcommand{\ouralgorithm}[0]{NeuMatch}
\newcommand{\iset}[0]{iSet\xspace}
\newcommand{\MSNN}[0]{RQ-RMI\xspace}
\newcommand{\isets}[0]{iSets\xspace}
\newcommand{\capital}[0]{\expandafter\MakeUppercase}
\begin{document}
\fi

\definecolor{bluish}{HTML}{24246b}
	
\newcommand{\DrawRectangle}[8]{ 
	\draw[-,#5] (#2,#1) -- (#2+#3,#1) -- (#2+#3,#1+#4) -- (#2,#1+#4) -- (#2,#1);
	\draw[] (#2+#3/2,#1+#4/2) node[align=center, font={\small}] {#6}; 
	\draw[] (#2,#1+#4) node[align=left, anchor=north west, font={\small}] {#7};
	\pgfmathsetmacro{\nomargin}{0.05};
	\draw[] (#2+#3/2,#1-\nomargin) node[color=bluish, align=right, anchor=south, font={\scriptsize}] {#8};
}

\newcommand{\DrawDupRectangle}[5]{ 
	\foreach \i in {1,2} {
		\pgfmathsetmacro{\startX}{(#2+\i*#5};
		\pgfmathsetmacro{\endX}{\startX+#3};
		\pgfmathsetmacro{\middleX}{\endX-#5};
		\pgfmathsetmacro{\startY}{#1+#4+1*(\i-1)*#5};
		\pgfmathsetmacro{\middleY}{\startY+#5};
		\pgfmathsetmacro{\endY}{\middleY-#4};
		\draw[-, draw=darkgray] (\startX,\startY) -- (\startX,\middleY) -- (\endX,\middleY) -- (\endX,\endY) -- (\middleX,\endY);
	}
}

\newcommand{\DrawArrow}[7]{ 
	\draw [-Latex, line width=#5] (#1,#2) -- (#3,#4);
	\draw (#1,#2) node[anchor=east] {#6};
	\draw (#3,#4) node[anchor=west] {#7};
}

\newcommand{\BigArrow}[5] { 
	\draw[-] (#1,#2+#4) -- (#3-#5*#4,#2+#4) --
			(#3-#5*#4,#2+#5*#4) -- (#3,#2) -- (#3-#5*#4,#2-#5*#4) --
			(#3-#5*#4,#2-#4) -- (#1,#2-#4);
}

\newcommand{\scale}{0.732}

\newcommand{\arrow}{{Latex[scale=\scale]}}

\begin{tikzpicture}[thick,scale=\scale, every node/.style={scale=\scale}]
	
	
	\BigArrow{-1.1}{0.25}{-0.5}{0.1}{3};
	\draw (-0.9,0.00) node[align=center, anchor=north] {Incoming\\Packet};
	
	\pgfmathsetmacro{\msnnTop}{0.65};
	\pgfmathsetmacro{\msnnLeft}{0.15};
	\pgfmathsetmacro{\msnnWidth}{1.45};
	\pgfmathsetmacro{\msnnHeight}{1.55};
	
	\pgfmathsetmacro{\arrowMargin}{0.1};
	\pgfmathsetmacro{\arrowTop}{\msnnTop+\msnnHeight/2};
	
	\pgfmathsetmacro{\hintLeft}{\msnnLeft+\msnnWidth+\arrowMargin};
	\pgfmathsetmacro{\hintWidth}{1.5};
	
	\pgfmathsetmacro{\rulesHeight}{1.7};
	\pgfmathsetmacro{\rulesTop}{\arrowTop-\rulesHeight/2};
	\pgfmathsetmacro{\rulesLeft}{\hintLeft+\hintWidth+\arrowMargin};
	\pgfmathsetmacro{\rulesWidth}{1.5};
	
	\pgfmathsetmacro{\searchHeight}{0.7};

	\pgfmathsetmacro{\candidateLeft}{\rulesLeft+\rulesWidth+\arrowMargin};
	\pgfmathsetmacro{\candidateWidth}{0.9};

	\pgfmathsetmacro{\externalTop}{-1.75};
	\pgfmathsetmacro{\remainderArrowTop}{\externalTop+\msnnHeight/2};
	\pgfmathsetmacro{\remainderArrowGap}{0.25};
	\pgfmathsetmacro{\remainderArrowWidth}{\rulesLeft-\hintLeft-\arrowMargin};
	
	\pgfmathsetmacro{\remainderRulesHeight}{\rulesHeight};
	\pgfmathsetmacro{\remainderRulesTop}{\remainderArrowTop-\remainderRulesHeight/2};
	\pgfmathsetmacro{\remainderRulesLeft}{\rulesLeft};
	\pgfmathsetmacro{\remainderRulesWidth}{\rulesWidth};
	
	\DrawDupRectangle{0.5}{0}{6.0}{2.2}{0.15};
	\DrawRectangle{0.5}{0}{6.0}{2.2}{}{}{\bf{Independent Set}}{};
	\DrawRectangle{\msnnTop}{\msnnLeft}{\msnnWidth}{\msnnHeight}{}{\MSNN}{}{(CPU Cache)};
	\DrawRectangle{\rulesTop}{\rulesLeft}{\rulesWidth}{\rulesHeight}{}{\iset\\Rules}{}{(DRAM)};

	\draw[-\arrow, thin] (\hintLeft,\arrowTop) -- (\hintLeft+\hintWidth,\arrowTop);
	\draw (\hintLeft+\hintWidth/2, \arrowTop) node [anchor=north, align=center, font={\footnotesize}] {predicted\\index};

	\draw[-\arrow, thin] (\candidateLeft,\arrowTop) -- (\candidateLeft+\candidateWidth,\arrowTop);
	\draw (\candidateLeft+\candidateWidth/2, \arrowTop) node [anchor=north, font={\scriptsize}, align=center] {candidate\\rule};

	\DrawRectangle{-1.9}{0}{6.0}{2.2}{}{}{\bf{Remainder Set}}{};
	\DrawRectangle{\externalTop}{\msnnLeft}{\msnnWidth}{\msnnHeight}{}{External\\Classifier}{}{(CPU Cache)};
	\DrawRectangle{\remainderRulesTop}{\remainderRulesLeft}{\remainderRulesWidth}{\remainderRulesHeight}{}{Remainder\\Rules}{}{(DRAM)};
	
	\draw[-\arrow, thin] (\hintLeft,\remainderArrowTop-\remainderArrowGap) -- (\hintLeft+\remainderArrowWidth,\remainderArrowTop-\remainderArrowGap);
	\draw[-\arrow, thin] (\hintLeft,\remainderArrowTop) -- (\hintLeft+\remainderArrowWidth,\remainderArrowTop);
	\draw[-\arrow, thin] (\hintLeft,\remainderArrowTop+\remainderArrowGap) -- (\hintLeft+\remainderArrowWidth,\remainderArrowTop+\remainderArrowGap);
	\draw (\hintLeft+\remainderArrowWidth/2, \remainderArrowTop-\remainderArrowGap) node [anchor=north, font={\footnotesize}] {indexes};
	
	\draw[-\arrow, thin] (\candidateLeft,\remainderArrowTop) -- (\candidateLeft+\candidateWidth,\remainderArrowTop);
	\draw (\candidateLeft+\candidateWidth/2, \remainderArrowTop) node [anchor=north, font={\scriptsize}, align=center] {candidate\\rule};
	
	\BigArrow{6.3}{0.25}{6.9}{0.1}{3};
	
	\DrawRectangle{-0.5}{7.15}{1.4}{1.4}{}{\bf{Selector}}{}{};
	
	\BigArrow{8.9}{0.25}{9.5}{0.1}{3};
	\draw (9.2,0.00) node[align=center, anchor=north] {Action};
	
\end{tikzpicture}
\ifdefined\RELEASE
\else
\end{document}
\fi
	\caption{\ouralgorithm{} overview. The rules are divided into Independent Sets indexed by \MSNN{}s and the Remainder Set indexed by any classifier.  One \MSNN{} predicts the storage index of the matching rule. The Selector chooses the highest-priority matching rule.}
	\label{fig:lookup_procedure}
\end{figure}

We propose a novel approach to packet classification, \emph{\ouralgorithm{}}, which 
\emph{compresses} the rule-set index dramatically to fit it entirely into the upper levels of the CPU cache (L1/L2) even for large 500K rule-sets. We introduce  a novel \emph{Range Query Recursive Model Index } (\MSNNs) model, and train it to \emph{learn} the rules' matching sets,  turning rule matching into \emph{neural network inference}.
We show that \MSNN achieves out-of-L1-cache execution by reducing the memory footprint on average by
4.9$\times$, 8$\times$, and 82$\times$
compared to recent CutSplit~\cite{Wenjun2018}, NeuroCuts~\cite{Liang2019}, and TupleMerge~\cite{Daly2019} on the standard ClassBench~\cite{TaylorTurner2007} benchmarks, and up to 29$\times$ for real forwarding rule-sets.

To the best of our knowledge, \ouralgorithm{} is the first to perform packet classification using trained neural network models. NeuroCuts also uses neural nets, but it applies them for optimizing the decision tree parameters during the \emph{offline} tree construction phase; their rule matching still uses traditional (optimized) decision trees. In contrast, \ouralgorithm{} performs classification  via \MSNN{}s, which are  more space-efficient than decision trees or hash-tables,  improving scalability by an order of magnitude.

\ouralgorithm{} 
transforms the packet classification task from memory- to compute-bound. This design is appealing because it is likely to scale well in the future, with 
rapid advances in hardware acceleration of neural network inference~\cite{Habana, IntelNervana, NvidiaDLInference}. On the other hand, the performance of both decision trees and hash-tables is inherently limited because of the poor scaling of DRAM access latency and CPU on-die cache sizes (e.g., $1.5\times$  over five years for L1 in Intel's CPUs).

\ouralgorithm{} builds on the recent work on \emph{learned indexes}~\cite{Kraska2018}, which applies a Recursive Model Index (RMI) model to indexing key-value pairs.
The values are stored in an array, and the RMI is \emph{trained to learn} the mapping function between the keys and the indexes of their values in the array. The model is used to \emph{predict} the index given the key.
When applied to databases~\cite{Kraska2018}, RMI boosts performance by compressing the indexes to fit in CPU caches.

Unfortunately,  RMI is not directly applicable for packet classification.
%
%
First, a key (packet field) may not have an exact matching value, but match a \emph{rule range}, whereas RMI can learn only exact key-index pairs.  This is a fundamental property of RMI: it guarantees correctness only for the keys used during training, but provides no such guarantees for non-existing keys (\cite{Kraska2018}, Section 3.4). Thus, for range matching it requires enumeration of all possible keys in the range, making it too slow.
Second, the match is evaluated over multiple packet fields, requiring lookup in a multi-dimensional space. Unfortunately, multi-dimensional RMI~\cite{Kraska2019} requires that the input be flattened into one dimension, which in the presence of wildcards results in an exponential blowup of the input domain,  making it too large to learn for compact models. Finally, a key may match multiple rules, with the highest priority one used as output, whereas RMI retrieves only a single index for each key.

\ouralgorithm{} successfully solves these challenges.

{\noindent \bf RQ-RMI}. We design a novel model which can match keys to ranges, with an efficient training algorithm that does not require exhaustive key enumeration to learn the ranges. The training strives to minimize the prediction error of the index, while maintaining a small model size. We show that the models can store indices of 500K ClassBench rules in 35~KB (\S\ref{sec:eval_compression}). We prove that our algorithm  \emph{guarantees range lookup correctness} (\S\ref{sec:introducing_msnn}). 

{\noindent \bf Multi-field packet classification}.
To enable multi-field matching with overlapping ranges, the rule-set is split into independent sets with non-overlapping ranges, called \emph{\isets}, each associated with a single field and indexed with its own \MSNN model. The \iset partitioning  (\S\ref{sec:generating_isets}) strives to cover the rule-set with  as few \isets as possible, discarding those that are too small. The \emph{remainder set} of the rules not covered by large \isets is indexed via existing classification techniques. In practice, the rules in the remainder constitute a small fraction in representative  rule-sets, so the remainder index fits into a fast cache together with the \MSNN{}s.

 Figure \ref{fig:lookup_procedure} summarizes the complete classification flow. The query of the \MSNN models produces the hints for the secondary search that selects one matching rule per \iset.  The validation stage selects the candidates with a positive match across all the fields, and a selector chooses the highest priority matching rule.  


Conceptually, \ouralgorithm{} can be seen as an \emph{accelerator} for existing packet classification techniques  and thus complements them.  In particular, the \MSNN model is best used for indexing rules with high value diversity that can be partitioned into fewer \isets. We show that the \iset construction algorithm is effective for selecting the rules that can be indexed via \MSNN, leaving the rest in the remainder (\S\ref{sec:eval_rule_mixture}).  The performance benefits of \ouralgorithm{} become evident when it indexes more than 25\% of the rules. Since the remainder is only a fraction of the original rule-set, it can be indexed efficiently with smaller decision-trees/hash-tables or will fit smaller TCAMs.

Our experiments\footnote{\changes{The source code of \ouralgorithm{} is available in \cite{sourcecode}}.
} show
that \ouralgorithm{} outperforms all the state-of-the-art algorithms on synthetic and real-life rule-sets. For example, 
 it is faster than CutSplit, NeuroCuts, and TupleMerge, by
 2.7$\times$, 4.4$\times$ and 2.6$\times$ in latency and 2.4$\times$, 2.6$\times$, and 1.6$\times$ in throughput   respectively, averaged over 12 rule-sets of 500K ClassBench-generated rules, and by $7.5\times$ in latency and $3.5\times$ in throughput vs. TupleMerge for the real-world Stanford backbone forwarding rule-set. 

 \ouralgorithm{} supports rule updates by removing the updated rules from the \MSNNs and adding them to the remainder set indexed by another algorithm that supports fast updates, e.g., TupleMerge. This approach requires periodic retraining to maintain a small remainder set; hence it does not yet support more than a few thousands of updates (\S\ref{sec:updates}).  
The algorithmic solutions to directly update \MSNNs are deferred for future work.

In summary, our contributions are as follows.
\begin{itemize}
	\item
	We present an novel \MSNN{} model and a training technique for learning packet classification rules.

	\item
    We demonstrate the application of \MSNNs to multi-field packet classification.

	\item
	\ouralgorithm{} outperforms existing techniques in terms of memory footprint, latency, and throughput on challenging rule-sets with up to 500K rules, compressing them to fit into small caches of modern processors.

\end{itemize}

\section{Background}

This section describes the packet classification problem and surveys existing solutions.

\subsection{Classification algorithms}

Packet classification is the process of locating a single rule that is satisfied by an input packet among a set of rules. A rule contains a few fields in the packet's metadata. Wildcards define \emph{ranges}, i.e., they match multiple values. Ranges may overlap with each other, i.e.,  a packet may match several rules, but only the one having the highest priority is selected. Figure \ref{fig:packet_classification} illustrates a classifier with two fields and five overlapping matching rules. An incoming packet matches two rules ($R^3, R^4$), but $R^3$ is used as it has a higher priority.





Packet classification performance becomes difficult to scale  as the number of rules and the number of matching fields grow. Therefore, it has received renewed interest with increased complexity of software-defined data center networks, featuring hundreds of thousands of rules per virtual network function~\cite{Firestone2017} and tens of matching fields (up to 41 in OpenFlow 1.4~\cite{McKeown2008}).

\paragraph{Decision Tree Algorithms} \label{sec:decision_tree_algorithms}
The rules are viewed as hyper-cubes and packets as points in a multi-dimensional space. The axes of the \emph{rule space} represent different fields and hold non-negative integers. 
A recursive partitioning technique divides the rule space into subsets with at most \emph{binth} rules. Thus, to match a rule,  
a tree traversal finds the smallest subset for a given packet, while a secondary search scans over the subset's rules  to select the best match.

Unfortunately, a \emph{rule replication} problem may hinder performance in larger rule-sets by dramatically increasing the tree's memory footprint when a rule spans several subspaces. Early works, such as HiCuts~\cite{Gupta2000} and HyperCuts~\cite{Singh2003} both suffer from this issue.
More recent EffiCuts~\cite{VamananBalajee2011} and CutSplit~\cite{Wenjun2018},
suggest that the rule set should be split into groups of rules that share similar properties and generate a separate decision-tree for each. NeuroCuts \cite{Liang2019}, the most recent work in this domain, uses reinforcement learning for optimizing decision tree parameters to reduce its memory footprint,  or the number of memory accesses during traversal, by efficiently exploring a large tree configuration space.

\paragraph{Hash-Based Algorithms}
 Tuple Space Search \cite{Srinivasan1999} and recent TupleMerge \cite{Daly2019} partition the rule-set into subsets according to the number of prefix bits in each field. As all rules of a subset have the same number of prefix bits, they can act as keys in a hash table. The classification is performed by extracting the prefix bits, in all fields, of an incoming packet, and checking all hash-tables for matching candidates. A secondary search eliminates false-positive results and selects the rule with the highest priority.

Hash-based techniques are effective in an \emph{online classification} problem with frequent rule updates, whereas decision trees are not. However, decision trees have been  traditionally considered faster in classification. Nevertheless, the recent TupleMerge hash-based algorithm closes the gap and achieves high classification throughput while supporting high performance updates.

\begin{figure}[t]
	\centering
	\ifdefined\RELEASE
\else
\documentclass[conference,compsoc]{IEEEtran}

\usepackage{graphicx}
\usepackage{subfig}

\usepackage[T1]{fontenc}
\usepackage[utf8]{inputenc}
\usepackage{pgfplots}
\usepackage{grffile}
\pgfplotsset{compat=newest}
\usetikzlibrary{plotmarks}
\usetikzlibrary{arrows.meta}
\usepgfplotslibrary{patchplots}
\usepackage{amsmath}

\newcommand{\ouralgorithm}[0]{Plasmus}
\begin{document}
\fi

\newcommand{\DrawRectangle}[7]{ 
	\draw[-,#5] (#1,#2) -- (#1+#3,#2) -- (#1+#3,#2-#4) -- (#1,#2-#4) -- (#1,#2);
	\draw[] (#1+#3/2,#2-#4/2) node[align=center, font={#6}] {#7}; 
}

\newcommand{\BigArrowDown}[2] { 
	
	\pgfmathsetmacro{\height}{0.4};
	\pgfmathsetmacro{\gap}{0.07};
	\pgfmathsetmacro{\biggap}{0.2};
	
	\draw[-] (#1-\gap,#2) --
	(#1-\gap,#2-\height+\biggap) -- (#1-\biggap,#2-\height+\biggap) --
	(#1, #2-\height) --
	(#1+\biggap,#2-\height+\biggap) -- (#1+\gap,#2-\height+\biggap) --
	(#1+\gap,#2);
}

\newcommand{\BigArrowRight}[2] { 
	\pgfmathsetmacro{\width}{2.5};
	\pgfmathsetmacro{\gap}{0.07};
	\pgfmathsetmacro{\biggap}{0.2};
	\pgfmathsetmacro{\tox}{#1+\width-\gap};
	\draw[-] (#1,#2+\gap) -- (\tox-\biggap,#2+\gap) --
	(\tox-\biggap,#2+\biggap) -- (\tox,#2) -- (\tox-\biggap,#2-\biggap) --
	(\tox-\biggap,#2-\gap) -- (#1,#2-\gap);
}

\newcommand{\Router}[3] { 
	
	\pgfmathsetmacro{\arrowSize}{#3/3};
	\pgfmathsetmacro{\arrowBase}{#3/7};
	\pgfmathsetmacro{\arrowInner}{(\arrowBase/2)};
	\pgfmathsetmacro{\arrowOuter}{(\arrowBase+\arrowSize/2)};

	\definecolor{bgColor}{rgb}{0.92,0.92,0.92};
	\definecolor{arrowColor}{rgb}{0.8,0.8,0.8};
	
	\draw [fill=bgColor] (#1, #2) circle (#3);
	
	\draw [-, fill=arrowColor] 	
				(#1-\arrowBase, #2+\arrowBase) --
				(#1-\arrowInner, #2+\arrowBase+\arrowSize) -- (#1-\arrowOuter, #2+\arrowBase+\arrowSize) --
				(#1, #2+\arrowBase+2*\arrowSize) --
				(#1+\arrowOuter, #2+\arrowBase+\arrowSize) --  (#1+\arrowInner, #2+\arrowBase+\arrowSize) --
				(#1+\arrowBase, #2+\arrowBase) --
				(#1+\arrowBase+\arrowSize, #2+\arrowInner) -- (#1+\arrowBase+\arrowSize, #2+\arrowOuter) --
				(#1+\arrowBase+2*\arrowSize, #2) --
				(#1+\arrowBase+\arrowSize, #2-\arrowOuter) -- (#1+\arrowBase+\arrowSize, #2-\arrowInner) --				
				(#1+\arrowBase, #2-\arrowBase) --
				(#1+\arrowInner, #2-\arrowBase-\arrowSize) -- (#1+\arrowOuter, #2-\arrowBase-\arrowSize) --
				(#1, #2-\arrowBase-2*\arrowSize) --
				(#1-\arrowOuter, #2-\arrowBase-\arrowSize) --  (#1-\arrowInner, #2-\arrowBase-\arrowSize) --
				(#1-\arrowBase, #2-\arrowBase) --
				(#1-\arrowBase-\arrowSize, #2-\arrowInner) -- (#1-\arrowBase-\arrowSize, #2-\arrowOuter) --
				(#1-\arrowBase-2*\arrowSize, #2) --
				(#1-\arrowBase-\arrowSize, #2+\arrowOuter) -- (#1-\arrowBase-\arrowSize, #2+\arrowInner) --	
				(#1-\arrowBase, #2+\arrowBase);
}

\newcommand{\scale}{1}

\newcommand{\arrow}{{Latex[scale=\scale]}}

\begin{tikzpicture}[scale=\scale, every node/.style={scale=\scale}]

	\definecolor{mycolor1}{rgb}{0.8,0.8,0.8}
	\pgfmathsetmacro{\w}{1};
	\pgfmathsetmacro{\h}{0.4};
	
	
	\draw (0, 0) 		node[anchor=south, align=center, font={\footnotesize}]{
	IPv4 Address};
	\draw (2*\w, 0) 	node[anchor=south, align=center, font={\footnotesize}]{
	Port};
	\draw (4*\w, 0) 	node[anchor=south, align=center, font={\footnotesize}]{Priority};
	\draw (5.5*\w, 0) 	node[anchor=south, align=center, font={\footnotesize}]{Action};
		
	\draw (-1*\w, +0*\h-0.5*\h) node [anchor=east, align=center, font={\footnotesize}] {$R^0$};
	\DrawRectangle{-1*\w}{+0*\h}{2*\w}{\h}{}{\footnotesize}{10.10.*.*};
	\DrawRectangle{+1*\w}{+0*\h}{2*\w}{\h}{}{\footnotesize}{10-18};
	\DrawRectangle{+3*\w}{+0*\h}{2*\w}{\h}{}{\footnotesize}{1 (highest)};
	\DrawRectangle{+5*\w}{+0*\h}{1*\w}{\h}{}{\footnotesize}{$a_1$};
	
	\draw (-1*\w, -1*\h-0.5*\h) node [anchor=east, align=center, font={\footnotesize}] {$R^1$};
	\DrawRectangle{-1*\w}{-1*\h}{2*\w}{\h}{}{\footnotesize}{10.10.1.*};
	\DrawRectangle{+1*\w}{-1*\h}{2*\w}{\h}{}{\footnotesize}{15-25};	
	\DrawRectangle{+3*\w}{-1*\h}{2*\w}{\h}{}{\footnotesize}{2};
	\DrawRectangle{+5*\w}{-1*\h}{1*\w}{\h}{}{\footnotesize}{$a_2$};
	
	\draw (-1*\w, -2*\h-0.5*\h) node [anchor=east, align=center, font={\footnotesize}] {$R^2$};
	\DrawRectangle{-1*\w}{-2*\h}{2*\w}{\h}{}{\footnotesize}{10.*.*.*};
	\DrawRectangle{+1*\w}{-2*\h}{2*\w}{\h}{}{\footnotesize}{5-8};
	\DrawRectangle{+3*\w}{-2*\h}{2*\w}{\h}{}{\footnotesize}{3};
	\DrawRectangle{+5*\w}{-2*\h}{1*\w}{\h}{}{\footnotesize}{$a_3$};
	
	\draw (-1*\w, -3*\h-0.5*\h) node [anchor=east, align=center, font={\footnotesize}] {$R^3$};
	\DrawRectangle{-1*\w}{-3*\h}{2*\w}{\h}{}{\footnotesize}{10.10.3.*};
	\DrawRectangle{+1*\w}{-3*\h}{2*\w}{\h}{}{\footnotesize}{7-20};
	\DrawRectangle{+3*\w}{-3*\h}{2*\w}{\h}{}{\footnotesize}{4};
	\DrawRectangle{+5*\w}{-3*\h}{1*\w}{\h}{}{\footnotesize}{$a_4$};
	
	\draw (-1*\w, -4*\h-0.5*\h) node [anchor=east, align=center, font={\footnotesize}] {$R^4$};
	\DrawRectangle{-1*\w}{-4*\h}{2*\w}{\h}{}{\footnotesize}{10.10.3.100};
	\DrawRectangle{+1*\w}{-4*\h}{2*\w}{\h}{}{\footnotesize}{19};
	\DrawRectangle{+3*\w}{-4*\h}{2*\w}{\h}{}{\footnotesize}{5 (lowest)};
	\DrawRectangle{+5*\w}{-4*\h}{1*\w}{\h}{}{\footnotesize}{$a_5$};
	
	\pgfmathsetmacro{\b}{-1.9*\h};
	
	
	\draw (-0.1*\w, \b-1.35) node[anchor=north, align=center, font={\footnotesize}]{Incoming packet};
	\draw (-0.1*\w, \b-1.75) node[anchor=north, align=center, font={\footnotesize}]{10.10.3.100:19};
		
	\BigArrowRight{1.5*\w}{\b-1.9};
	
	\draw (5.1*\w, \b-1.35) node[anchor=north, align=center, font={\footnotesize}]{Action to take \\$a_4$};
	
\end{tikzpicture}

\ifdefined\RELEASE
\else
\end{document}
\fi
	\caption{Packet classification with two fields: IP address and port.}
	\label{fig:packet_classification}
\end{figure}

\subsection{Poor performance with large rule-sets}
The packet classification performance of all the existing techniques does not scale well with the number of rules.  This happens because their indexing structures spill out of the fast L1/L2 CPU caches into L3 or DRAM. Indeed, as we show in our experiments (\S\ref{sec:results}), TupleMerge and NeuroCuts exceed the 1MB L2 cache with 100K rules and CutSplit with 500K rules. However, keeping the \emph{entire} indexing structure in fast caches is critical for performance. The inherent lack of access locality in hash and tree data structures, combined with the data-dependent nature of the accesses, make hardware prefetchers ineffective for hiding memory access latency. Thus, the performance of all lookups drops dramatically.

The performance drop is significant even when the data structures fit in the L3 cache. This cache is shared among all the cores, whereas L1 and L2 caches are
per-core. Thus, L3 is not only slower (up to 90 cycles in recent X86 CPUs), but
also suffers from cache contention, e.g., when another core runs a
cache-demanding workload and causes cache trashing. We observe the effects of
L3 contention in~\S\ref{sec:eval_compression}.

\ouralgorithm{} aims to provide more space efficient representation of the rule
index to scale to large rule-sets.

\section{\ouralgorithm{} construction} \label{sec:algorithm}

We first explain the RMI model for learned indexes which we use as the basis, explain its limitations, and then show our solution that overcomes them.

\subsection{Recursive Model Index} \label{background_ml_for_indexing}

Kraska et al.~\cite{Kraska2018} suggest using machine-learning models for storing key-value pairs instead of conventional data structures such as B-trees or hash tables. The values are stored in a \emph{value array}, and a \emph{Recursive Model Index} (RMI) is used to retrieve the value given a key. Specifically, RMI \emph{predicts the index} of the corresponding value in the value array using a model that \emph{learned} the underlying key-index mapping function.  

The main insight is that any index structure can be expressed as a continuous monotonically increasing function $y=h(x):[0,1]\mapsto[0,1]$, where $x$ is a  key scaled down uniformly into $[0,1]$, and $y$ is the  index of the respective value in the value array scaled down uniformly into $[0,1]$. RMI is trained to learn $h(x)$. The resulting \emph{learned index model} $\widehat{h}(x)$ performs lookups in two phases:  first it computes the \emph{predicted} index $\widehat{y}=\widehat{h}(key)$, and then performs a \emph{secondary search} in the array, in the vicinity $\epsilon$ of the predicted index, where $\epsilon$ is the \emph{maximum index prediction error} of the model, namely $|\widehat{h}(key)-h(key)| \leq \epsilon$.

\begin{figure}[t]
	\centering
	\ifdefined\RELEASE
\else
\documentclass[conference,compsoc]{IEEEtran}

\usepackage{graphicx}
\usepackage{subfig}

\usepackage{amsfonts}

\usepackage[T1]{fontenc}
\usepackage[utf8]{inputenc}
\usepackage{pgfplots}
\usepackage{grffile}
\pgfplotsset{compat=newest}
\usetikzlibrary{plotmarks}
\usetikzlibrary{arrows.meta}
\usepgfplotslibrary{patchplots}
\usepackage{amsmath}

\newcommand{\ouralgorithm}[0]{Plasmus}
\begin{document}
\fi
	
	\newcommand{\submodel}[4] { 
		\draw [-] (#1,#2) rectangle (#1+\width,#2-\height);
		\draw (#1+\width/2,#2-\height/2) node [align=center,font={\footnotesize}] {$m_{#3,#4}(x)$};
	}

	\newcommand{\connect}[6] { 
	    \pgfmathsetmacro{\fromX}{#1+0.5*(\width)};
	    \pgfmathsetmacro{\fromY}{#2-(\height)-\arrowGap};
	    \pgfmathsetmacro{\toX}{#3+0.5*(\width)};
	    \pgfmathsetmacro{\toY}{#4+\arrowGap};
    	\pgfmathsetmacro{\labelX}{(\fromX*0.5+\toX*0.5};
		\pgfmathsetmacro{\labelY}{(\fromY*0.5+\toY*0.5};
		\draw [-latex] (\fromX,\fromY) -- (\toX, \toY);
		\draw ( \labelX, \labelY ) node [anchor=#5, font={\scriptsize}] {#6};
	}

	\newcommand{\connectToUnknown}[4] { 
	    \pgfmathsetmacro{\fromX}{#1+0.5*(\width)};
	    \pgfmathsetmacro{\fromY}{#2-(\height)-\arrowGap};
	    \pgfmathsetmacro{\toX}{#3+0.5*(\width)};
	    \pgfmathsetmacro{\toY}{#4+\arrowGap};
	    
        \draw [-] (\fromX,\fromY) -- (\toX, \toY);
	    
	}

	\newcommand{\dotdotdot}[2] { 
		\draw [fill=black] (#1,#2-\height/2) circle (\radius);
		\draw [fill=black] (#1-\dotGap,#2-\height/2) circle (\radius);
		\draw [fill=black] (#1+\dotGap,#2-\height/2) circle (\radius);
	}

	\newcommand{\dotdotdotvertical}[2] { 
		\draw [fill=black] (#1,#2-\height/2) circle (\radius);
		\draw [fill=black] (#1,#2-1*\dotGap-\height/2) circle (\radius);
		\draw [fill=black] (#1,#2-2*\dotGap-\height/2) circle (\radius);
	}
	
	\newcommand{\pairs}[3] { 
		\foreach \i in {0,...,#3} {
			\draw (#1+\pairWidth*\i,#2) rectangle (#1+\pairWidth*\i+\pairWidth, #2-\height);
		}
		\foreach \i in {0,...,2} {
			\draw (#1+\pairWidth*\i+\pairWidth*0.5,#2-\height/2) node [align=center, font={\footnotesize}] {$v_\i$};
		}
		\draw (#1+\pairWidth*#3+\pairWidth*0.5,#2-\height/2) node [align=center, font={\footnotesize}] {$v_{|I|-1}$};
	}
	
\newcommand{\scale}{1.1}
\begin{tikzpicture}[scale=\scale, every node/.style={scale=\scale}]
	
	\pgfmathsetmacro{\stageHeight}{-.6};
	\pgfmathsetmacro{\stageLabelX}{-1};
	\pgfmathsetmacro{\submodelXGap}{0.1};
	
	\pgfmathsetmacro{\width}{1.7};
	\pgfmathsetmacro{\height}{0.4};
	\pgfmathsetmacro{\arrowGap}{0.00};
	
	\pgfmathsetmacro{\radius}{0.02};
	\pgfmathsetmacro{\dotGap}{0.135};
	
	\pgfmathsetmacro{\pairWidth}{0.83};
	
	\submodel{\width*1.2}{-.1*\stageHeight}{0}{0};
	\draw (\stageLabelX, 0*\stageHeight-0.5*\height) node [font={}] {$s_0$};
	
	\submodel{0.0*\width+0.0*\submodelXGap}{1*\stageHeight}{1}{0};
	\submodel{1.0*\width+1.0*\submodelXGap}{1*\stageHeight}{1}{1};
	\dotdotdot{2.0*\width+4*\submodelXGap}{1*\stageHeight};
	\submodel{2.0*\width+3.0*\submodelXGap+3*\dotGap}{1*\stageHeight}{1}{W_1-1};
	\draw (\stageLabelX, 1*\stageHeight-0.5*\height) node [font={}] {$s_1$};
	
	\connect{\width*0.8}{-.1*\stageHeight}{0.0*\width+0.0*\submodelXGap}{1*\stageHeight}{south east}{};
	\connect{\width*1.1}{-.1*\stageHeight}{1.0*\width+1.0*\submodelXGap}{1*\stageHeight}{west}{};
	\connect{\width*1.6}{-.1*\stageHeight}{2.0*\width+3.0*\submodelXGap+3*\dotGap}{1*\stageHeight}{south west}{};
	
	\dotdotdotvertical{\stageLabelX}{1.7*\stageHeight};
	\draw [dashed] (0*\width, 1.9*\stageHeight) rectangle (3.4*\width,1.5*\stageHeight-\height*1.7);
	\draw (1.7*\width, 1.7*\stageHeight-\height*0.85) node [font={\footnotesize}] {More Stages};
	
	\connectToUnknown{0.0*\width+0.0*\submodelXGap}{1*\stageHeight}{0.1*\width+0.0*\submodelXGap}{1.9*\stageHeight};
	\connectToUnknown{1.0*\width+1.0*\submodelXGap}{1*\stageHeight}{0.9*\width+1.0*\submodelXGap}{1.9*\stageHeight};
	\connectToUnknown{2.0*\width+3.0*\submodelXGap+3*\dotGap}{1*\stageHeight}{1.8*\width+3.0*\submodelXGap+3*\dotGap}{1.9*\stageHeight};

	\submodel{0.2*\width+0.0*\submodelXGap}{3*\stageHeight}{n-1}{0};
	\dotdotdot{1.2*\width+1.0*\submodelXGap+3*\dotGap}{3*\stageHeight};
	\pgfmathsetmacro{\largeRectX}{1.2*\width+1.0*\submodelXGap+6*\dotGap};
	\pgfmathsetmacro{\oldWidth}{\width};
	\pgfmathsetmacro{\width}{2.6};
	\submodel{\largeRectX}{3*\stageHeight}{n-1}{W_{n-1}-1};
	\draw (\stageLabelX, 3*\stageHeight-0.5*\height) node [font={}] {$s_{n-1}$};
	\pgfmathsetmacro{\width}{\oldWidth};
	
	\connect{0.7*\width+0.0*\submodelXGap}{2*\stageHeight}{0.2*\width+0.0*\submodelXGap}{3*\stageHeight}{east}{};
	\connect{1.2*\width+0.0*\submodelXGap}{2*\stageHeight}{1.3*\width+0.0*\submodelXGap}{3*\stageHeight}{west}{};
	
	\pairs{0}{4*\stageHeight}{6};
	\draw (\stageLabelX,4*\stageHeight) node [anchor=north, align=center, font={\scriptsize}] {Values};

	\connect{0.3*\width}{3*\stageHeight}{0.1*\width}{4*\stageHeight}{west}{};
	
	\draw (\stageLabelX,0*\stageHeight+0.2*\height-0.2) node [align=center, anchor=south, font={\footnotesize}] {Stage};
	
	\end{tikzpicture}
	
	\ifdefined\RELEASE
	\else
\end{document}
\fi
	\caption{RMI model structure and inference~\cite{Kraska2018}.}
	\label{fig:rmi_evaluation_process}
\end{figure}
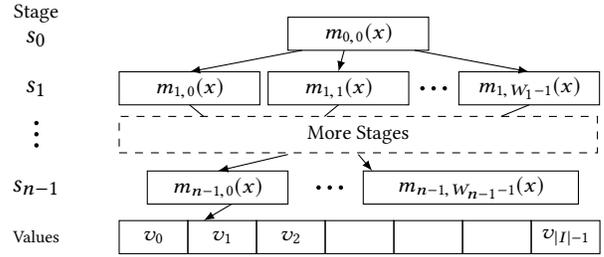

\paragraph{Model structure} RMI is a hierarchical model made of several ($n$) stages (Figure~\ref{fig:rmi_evaluation_process}). Each stage $i$ includes $W_i$ \emph{submodels} $m_{i,j}, \; j<W_i$, where $W_i$ is the \emph{stage width}. The first stage has a single submodel. Each successive stage has a larger width. The submodels in each stage are trained on a progressively smaller subset of the input keys,  \emph{refining} the index prediction toward the submodels in the leaves. Thus, each key-index pair is learned by one submodel at each stage, with the leaf submodel producing the index prediction.

RMI is a generic structure; a variety of machine learning models or data structures can be used as submodels, such as \changes{regression models or B-trees.} 
\removed{linear regression or neural networks.}
The type of the submodels, the number of stages and the width of each stage are configured prior to training. 

\paragraph{Training} Training is performed stage by stage. 

\vspace{2pt}\emph{First stage.} The submodel in stage $m_{0,0}$ is trained on the \emph{whole} data set. Then, the input key-index pairs are split into $W_1$ disjoint subsets. The input partitioning is performed as follows. \underline{For each input key-index pair} $\{key:idx\}$  we compute the submodel prediction $\widehat{j}=m_{0,0}(key)$, satisfying $\widehat{j} \in [0,1)$. The output $\widehat{j}$ is used to obtain $j=\lfloor\,\widehat{j}\cdot{W_1}\rfloor$ which is the index of the submodel in stage 1, $m_{1,j}$, to be used for learning $\{key:idx\}$.  We call the subset of the input to be learned by model $m_{i,j}$ as model \emph{input responsibility domain} $R_{i,j}$, or responsibility for short. $R_{0,0}$ is the whole input.
    
\vspace{2pt}\emph{Internal stages.} The submodels in stage $i$,  $m_{i,j}$, are trained on the keys in $R_{i,j}\; (j<W_i)$. After training, the responsibilities of the submodels in  stage $i+1$ are computed, and the process continues until the last stage.
    
\vspace{2pt}\emph{Last stage.} The submodels of the last stage must predict the actual index of the matching value in the value array. However, a submodel may have a prediction error. Therefore, RMI uses the model prediction as a \emph{hint}. The matching value is found by searching in the value array in the vicinity of the predicted index, as defined by the maximum error bound $\epsilon$ of the model. Note that $\epsilon$  should be valid for \emph{all} input key-index pairs. To compute $\epsilon$, RMI exhaustively computes the submodel prediction  \underline{for each input key} in its responsibility. Submodels with a high error bound are retrained \changes{or converted to B-trees}. 

\paragraph{Inference} Given a key, we iteratively evaluate each submodel stage after stage, from $m_{0,0}$. We use the prediction in stage $i-1$ to select a submodel in stage $i$, until we reach the last stage. The last selected submodel predicts the index in the value array. This index $\widehat{i}$  determines the range for the secondary search in the value array that spans [$\widehat{i}-\epsilon,\widehat{i}+\epsilon$].

\subsection{RMI limitations}

Direct application of RMI to indexing packet classification rules is not possible for the following reasons:

\paragraph{No support for range matching\footnote{The RMI paper also uses the term \emph{range index} while applying RMI to range index data structures (i.e., B-trees) that can quickly retrieve all stored keys in a requested numerical range. Our work is fundamentally different: given a key it retrieves \changes{\emph{the index of its matching range}.} \removed{\emph{the indexes of matching ranges}.}}}  RMI allows only an exact match for a given key, whereas packet classification requires retrieving rules with matching \emph{ranges} as defined by wildcards. This problem is fundamental: RMI exhaustively enumerates \emph{all} the keys in all the ranges to calculate the submodel responsibility and the maximum model prediction error (see the underlined parts of the training algorithm). In other words, all the values in the range must be materialized into key-index pairs for RMI to learn them, since RMI \emph{does not guarantee correct lookup for keys not used in training}~\cite{Kraska2018}. The original paper sketches a few possible solutions, however, they either rely on model monotonicity (while we do not) or use smarter yet still expensive enumeration techniques.

\paragraph{Slow multi-dimensional indexing}
RMI is ineffective for multi-dimensional indexes because the proposed solution \cite{Kraska2019} leads to generating an exponential number of rules in the presence of wildcards. For example, a single rule with wildcards in destination IP (0.0.0.*), port (10-100), and protocol (TCP/UDP) results in 46,592 distinct key-index pairs. Since the input domain becomes too large, it requires a large model that exceeds the CPU cache.

In the following we outline the solutions to these challenges
\changes{We first discuss Range-Query RMI (\MSNN), which extends RMI to perform \emph{range-value queries} in a one-dimensional index where ranges do not overlap (\S\ref{sec:introducing_msnn}-\S\ref{sec:full_msnn_training}). We then show how to apply \MSNN in multi-dimensional index space with overlaps (\S\ref{sec:partitioning}-\S\ref{sec:remainder}).}

\subsection{One-dimensional \MSNN} \label{sec:introducing_msnn}

We first seek to find a way to perform range matching over a set of non-overlapping ranges in one dimension. 


There are two basic ideas:

\paragraph{Sampling} Each submodel $m_{i,j}$ is trained  by generating a uniform sample of  key-index pairs from input ranges in its responsibility. The samples are generated on-the-fly for each submodel (\S\ref{sec:sampling}).

\paragraph{Analytical error bound estimation for ranges} We eliminate the RMI's requirement for exhaustive key-value enumeration during training by making the following observation: \emph{if a submodel is a piece-wise linear function, the worst-case error bound $\epsilon$ can be computed analytically}, thereby enabling efficient  learning of ranges.

%

The intuition behind this observation is illustrated in Figure~\ref{fig:transition_inputs}.  
It shows the graph of some piece-wise linear function which represents a submodel $M$ whose outputs are quantized into integers  in $[0,4)$, i.e., $M$ predicts the index in an array of size 4. We call the inputs for which this function changes its quantized output \emph{transition inputs} $t_i\in{T}$. 
In turn, transition inputs determine the region of inputs with the \emph{same} quantized output. Therefore, given an input range in the model's responsibility, to compute the model's maximum prediction error for any key in that range, it suffices to evaluate the prediction error in the transition inputs that fall in the range.  We describe the training algorithm that relies on these observations in Section~\ref{sec:training}. We now provide a more formal description, but leave most of the proofs in the Appendix.

\begin{figure}[t]
	\centering
	

\ifdefined\RELEASE
\else
\documentclass[conference,compsoc]{IEEEtran}

\usepackage{graphicx}
\usepackage{subfig}

\usepackage[T1]{fontenc}
\usepackage[utf8]{inputenc}
\usepackage{pgfplots}
\usepackage{grffile}
\pgfplotsset{compat=newest}
\usetikzlibrary{plotmarks}
\usetikzlibrary{arrows.meta}
\usepgfplotslibrary{patchplots}
\usetikzlibrary{decorations.markings}
\usepackage{amsmath}

\newcommand{\ouralgorithm}[0]{Plasmus}
\begin{document}
\fi

\newcommand{\scale}{1}

\begin{tikzpicture}[scale=\scale, every node/.style={scale=\scale}]

	\tikzstyle{densely dashed}= [dash pattern=on 3pt off 1.3pt]
	
	
	\pgfmathsetmacro{\base}{6.5};
	\pgfmathsetmacro{\width}{6};
	
	\pgfmathsetmacro{\stateHeight}{0.5};
	
	\draw[draw=none, fill=gray, opacity=0.7](\base,0.0) rectangle (\base+\width, 1*\stateHeight);
	\draw[draw=none, fill=gray, opacity=0.5](\base,1*\stateHeight) rectangle (\base+\width, 2*\stateHeight);
	\draw[draw=none, fill=gray, opacity=0.3](\base,2*\stateHeight) rectangle (\base+\width, 3*\stateHeight);
	\draw[draw=none, fill=gray, opacity=0.1](\base,3*\stateHeight) rectangle (\base+\width, 4*\stateHeight);
	
	\draw[-latex, line width=0.8] (\base,0) -- (\base+\width+0.3,0) node [anchor=west, align=left] {$x$};
	\draw[-latex, line width=0.8] (\base,0) -- (\base,4.5*\stateHeight) node [anchor=west, align=left] {$M(x), \left\lfloor M(x) \cdot 4\right\rfloor$};
	
	\draw[-, line width=0.5] (\base,4*\stateHeight) -- (\base-0.07, 4*\stateHeight) node [anchor=east, font={\normalsize}] {1, 4};
	\draw[-, line width=0.5] (\base,3*\stateHeight) -- (\base-0.07, 3*\stateHeight) node [anchor=east, font={\normalsize}] {0.75, 3};
	\draw[-, line width=0.5] (\base,2*\stateHeight) -- (\base-0.07, 2*\stateHeight) node [anchor=east, font={\normalsize}] {0.5, 2};
	\draw[-, line width=0.5] (\base,1*\stateHeight) -- (\base-0.07, 1*\stateHeight) node [anchor=east, font={\normalsize}] {0.25, 1};
	\draw[-, line width=0.5] (\base,0.0) -- (\base-0.07, 0.0) node [anchor=east, font={\normalsize}] {0, 0};

	\pgfmathsetmacro{\a}{\base+0.2};
	\pgfmathsetmacro{\b}{\base+0.9};
	\pgfmathsetmacro{\c}{\base+2.81};
	\pgfmathsetmacro{\d}{\base+3.3};
	\pgfmathsetmacro{\e}{\base+4.9};
	\pgfmathsetmacro{\f}{\base+5.5};
	
	\coordinate (A) at (\a, 0.3);
	\coordinate (B) at (\b, 0.3);
	\coordinate (C) at (\c, 0.9);
	\coordinate (D) at (\d, 0.2);
	\coordinate (E) at (\e, 1.6);
	\coordinate (F) at (\f, 1.7);
		
	\draw[-] (A) -- (B) -- (C) -- (D) -- (E) -- (F);
	
	\pgfmathsetmacro{\a}{\base+1.55};
	\pgfmathsetmacro{\b}{\base+3.1};
	\pgfmathsetmacro{\c}{\base+3.65};
	\pgfmathsetmacro{\d}{\base+4.2};
	\pgfmathsetmacro{\e}{\base+4.8};
	
	\draw[densely dashed] (\a,1*\stateHeight) -- (\a,0) node [anchor=north, font={\normalsize}] {$t_0$};
	\draw[densely dashed] (\b,1*\stateHeight) -- (\b,0) node [anchor=north, font={\normalsize}] {$t_1$};
	\draw[densely dashed] (\c,1*\stateHeight) -- (\c,0) node [anchor=north, font={\normalsize}] {$t_2$};
	\draw[densely dashed] (\d,2*\stateHeight) -- (\d,0) node [anchor=north, font={\normalsize}] {$t_3$};
	\draw[densely dashed] (\e,3*\stateHeight) -- (\e,0) node [anchor=north, font={\normalsize}] {$t_4$};
	
	
\end{tikzpicture}

\ifdefined\RELEASE
\else
\end{document}
\fi
	\caption{ 
	Transition inputs ($t_0,...,t_4$) for a piece-wise linear function  with the output domain=4.}
    \label{fig:transition_inputs}
\end{figure}
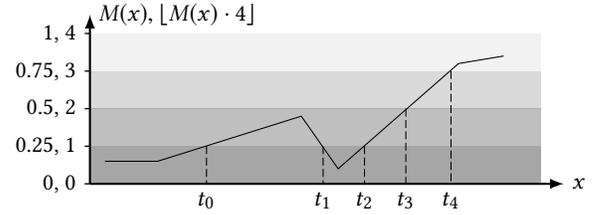

\subsection{Using a neural network as a submodel} 
\label{sec:training}

We choose to use a 3-layer fully-connected neural network (NN) with a single hidden layer and ReLU activation $A$.  Such NNs have been suggested in the original RMI paper~\cite{Kraska2018}; however, they did not leverage their properties for accelerating error bound computations. 

We denote a submodel as $m_{i,j}$, and define it as follows.

\begin{definition} [\MSNN submodel]

    Denote the output of a 3-layer fully-connected neural network as:
    \begin{equation} \nonumber
    N_{i,j}(x)= A\big( x \cdot \bm{w_1} + \bm{b_1}  \big) \times \bm{w_2} + b_2
    \end{equation}

	\noindent
	where $x$ is a scalar input, $\bm{w_1}, \bm{b_1}$ are the weight and bias row-vectors for layer $1$ (hidden layer), and $\bm{w_2}, b_2$ are the weight column-vector and bias scalar for layer $2$. Note that $N_{i,j}(x)$ is a scalar. The ReLU function $A$ applies a function $a$ on each element of an input vector:
	\begin{equation} \nonumber
	a(x)=
	\begin{cases} 
	x & x\geq0 \\
	0 & x<0.
	\end{cases} 
	\end{equation}
	
    \noindent
	The submodel output, denoted $M_{i,j}(x)$,  is defined as follows:
	\begin{equation} \nonumber
	M_{i,j}(x)= H\big( N_{i,j}(x)\big)
	\end{equation}
	
	\noindent
	where $H(x)$ trims the output domain to be in $[0,1)$. 
\end{definition}

\begin{corollary} \label{corollary:piecewise_linear}
	$M_{i,j}(x)$  is a piece-wise linear function.
\end{corollary}

\subsection{\MSNN training} \label{sec:full_msnn_training}

We use Corollary~\ref{corollary:piecewise_linear} to compute the transition inputs and the responsibility of the submodels. We provide a simplified description; see Appendix for the precise explanation.

\changes{
\subsubsection{Overview}
 \MSNN training is similar to RMI's. It is performed stage by stage. 
 Figure \ref{fig:submodel_training_last_stage} illustrates the training process for one stage.
 We start by training the single submodel in the first stage using the entire input domain. Next, we calculate its transition inputs (\S\ref{computing_transition_inputs}) and use them to find the responsibilities of the submodels in the following stage (\S\ref{computing_responsibilities_in_next_stage}). We proceed by training submodels in the subsequent stage using designated datasets we generate based on the submodels' responsibilities (\S\ref{sec:sampling}). We repeat this process until all submodels in all internal stages are trained.
 For the submodels in the leaves (last stage), there is an additional phase (dashed lines in Figure \ref{fig:submodel_training_last_stage}). After training, we calculate their error bounds and retrain the submodels that do not satisfy a predefined error threshold (\S\ref{computing_error_bounds}).
}

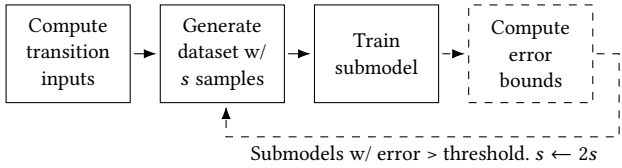
\begin{figure}[t]
	\centering
	\ifdefined\RELEASE
\else
\documentclass[conference,compsoc]{IEEEtran}
\usepackage{graphicx}
\usepackage{subfig}
\usepackage{amsfonts}
\usepackage[T1]{fontenc}
\usepackage[utf8]{inputenc}
\usepackage{pgfplots}
\usepackage{grffile}
\usepackage{xstring}
\pgfplotsset{compat=newest}
\usetikzlibrary{plotmarks}
\usetikzlibrary{arrows.meta}
\usepgfplotslibrary{patchplots}
\usepackage{amsmath}
\begin{document}
	\fi
	
	\newcommand{\TopDownSideArrow}[7]{
		
		\pgfmathsetmacro{\srcX}{#1};
		\pgfmathsetmacro{\srcY}{#2};
		\pgfmathsetmacro{\dstX}{#3};
		\pgfmathsetmacro{\dstY}{#4};
		\pgfmathsetmacro{\thick}{#5};
		\pgfmathsetmacro{\tail}{#6};
		
		\pgfmathsetmacro{\tailSign}{abs(\tail)/\tail};
		\pgfmathsetmacro{\shortTail}{\tailSign*(abs(\tail)-abs(\thick))};
		\pgfmathsetmacro{\arrowHeadWidth}{0.35};

		\pgfmathsetmacro{\srcXTail}{\srcX+\tail};
		\pgfmathsetmacro{\dstXShortTail}{\dstX+\shortTail};
		\pgfmathsetmacro{\arrowHeadSrcX}{#1};
		
		\IfEqCase{#7}{
			{up} {
				\pgfmathsetmacro{\arrowHeadY}{\srcY};
				\pgfmathsetmacro{\srcX}{#1+\arrowHeadWidth*\tailSign-\tailSign*0.1};
			}
			{down} {
				\pgfmathsetmacro{\arrowHeadY}{\dstY};
				\pgfmathsetmacro{\dstX}{#3+\arrowHeadWidth*\tailSign-\tailSign*0.1};
			}
		}
		
		\draw [-, fill=arrowfill]
		(\srcX,       \srcY+0.5*\thick) -- 
		(\srcXTail,   \srcY+0.5*\thick) -- 
		(\srcXTail,   \srcY+0.5*\thick) -- 
		(\srcXTail,   \dstY-0.5*\thick) -- 
		(\dstX,       \dstY-0.5*\thick) -- 
		(\dstX,       \dstY+0.5*\thick) -- 
		(\dstXShortTail, \dstY+0.5*\thick) -- 						
		(\dstXShortTail, \dstY+0.5*\thick) --					
		(\dstXShortTail, \srcY-0.5*\thick) --
		(\srcX,          \srcY-0.5*\thick) --
		(\srcX,          \srcY+0.5*\thick);
		
		\draw [-, fill=arrowfill]
		(\arrowHeadSrcX+\arrowHeadWidth*\tailSign, \arrowHeadY+0.5*\thick) --
		(\arrowHeadSrcX+\arrowHeadWidth*\tailSign, \arrowHeadY+1.25*\thick) --
		(\arrowHeadSrcX,                           \arrowHeadY) --
		(\arrowHeadSrcX+\arrowHeadWidth*\tailSign, \arrowHeadY-1.25*\thick) --
		(\arrowHeadSrcX+\arrowHeadWidth*\tailSign, \arrowHeadY-0.5*\thick);
	}

	\begin{tikzpicture}
	
	\definecolor{arrowfill}{HTML}{cccccc}
	
	\newcommand{\BigArrowRight}[4] { 
		\pgfmathsetmacro{\x}{#1};
		\pgfmathsetmacro{\y}{#2};
		\pgfmathsetmacro{\width}{#3};
		\pgfmathsetmacro{\gap}{0.5/5};
		\pgfmathsetmacro{\height}{0.5/2};
		\pgfmathsetmacro{\taillength}{0.5/1.5};
		\draw[-, fill=arrowfill]
		(\x,\y+\gap) --
		(\x+\width-\taillength,\y+\gap) --
		(\x+\width-\taillength,\y+\height) --
		(\x+\width,\y) --
		(\x+\width-\taillength,\y-\height) --
		(\x+\width-\taillength,#2-\gap) --
		(\x,\y-\gap) --
		(\x,\y+\gap);
		\draw (\x+0.5*\width,\y+2*\gap)
		node [align=center, font={\small}, anchor=south]
		{#4};
	}
	
	\newcommand{\ReturnArrowRight}[6] { 
		
		\pgfmathsetmacro{\x}{#1};
		\pgfmathsetmacro{\y}{#2};
		\pgfmathsetmacro{\width}{#3};
		\pgfmathsetmacro{\leftwidth}{#3/2};
		\pgfmathsetmacro{\gap}{0.5/5};
		\pgfmathsetmacro{\thick}{0.5/2.5};
		\pgfmathsetmacro{\height}{0.5/2};
		\pgfmathsetmacro{\taillength}{0.5/1.5};
		\pgfmathsetmacro{\longpartwidth}{#4};
		\pgfmathsetmacro{\longpartheight}{#5};
		
		\draw[-, fill=arrowfill]
		(\x,\y-\gap) --
		(\x,\y+\gap) --
		(\x+\leftwidth,\y+\gap) --
		(\x+\leftwidth,\y+\longpartheight) --
		(\x-\longpartwidth-\width,\y+\longpartheight) --
		(\x-\longpartwidth-\width,\y+\gap) --
		(\x-\longpartwidth-\taillength,\y+\gap) --
		(\x-\longpartwidth-\taillength,\y+\height) --
		(\x-\longpartwidth,\y) --
		(\x-\longpartwidth-\taillength,\y-\height) --
		(\x-\longpartwidth-\taillength,\y-\gap) --
		(\x-\longpartwidth-\width-\thick,\y-\gap) --
		(\x-\longpartwidth-\width-\thick,\y+\thick+\longpartheight) --
		(\x+\leftwidth+\thick,\y+\thick+\longpartheight) --
		(\x+\leftwidth+\thick,\y-\gap) --
		(\x,\y-\gap);
		
		\node [align=center, font={\small}, anchor=south]
		at (\x-0.5*\longpartwidth,\y+\longpartheight+\thick)
		{#6};
	}

	\pgfmathsetmacro{\arrowWidth}{0.3};
	\pgfmathsetmacro{\boxArrowGap}{0.05};
	\pgfmathsetmacro{\boxWidth}{1.65};
	\pgfmathsetmacro{\boxHeight}{1.3};
	\pgfmathsetmacro{\positionX}{0};
	\pgfmathsetmacro{\positionY}{0};
	
	\draw (\positionX,\positionY) rectangle (\positionX+\boxWidth,\boxHeight)
		node [pos=.5, align=center, font=\small]
		{Compute \\ transition \\ inputs};
	
	\pgfmathsetmacro{\positionX}{\positionX+\boxWidth+\boxArrowGap};
	\draw[-Latex] 
		(\positionX, \boxHeight/2) -- 
		(\positionX+\arrowWidth, \boxHeight/2);
	\pgfmathsetmacro{\positionX}{\positionX+\arrowWidth+\boxArrowGap};
	
	\draw (\positionX,\positionY) rectangle (\positionX+\boxWidth,\boxHeight)
		node [pos=.5, align=center, font=\small]
		{Generate \\ dataset w/ \\ $s$ samples};
	
	\pgfmathsetmacro{\positionX}{\positionX+\boxWidth+\boxArrowGap};
	\draw[-Latex] 
	(\positionX, \boxHeight/2) -- 
	(\positionX+\arrowWidth, \boxHeight/2);
	\pgfmathsetmacro{\positionX}{\positionX+\arrowWidth+\boxArrowGap};
	
	\draw (\positionX,\positionY) rectangle (\positionX+\boxWidth,\boxHeight)
		node [pos=.5, align=center, font=\small]
		{Train \\ submodel};
	\pgfmathsetmacro{\positionX}{\positionX+\boxWidth+\boxArrowGap};
	
	\draw[-Latex,dashed] 
		(\positionX, \boxHeight/2) -- 
		(\positionX+\arrowWidth, \boxHeight/2);
	\pgfmathsetmacro{\positionX}{\positionX+\arrowWidth+\boxArrowGap};
	
	\draw[dashed] (\positionX,\positionY) rectangle (\positionX+\boxWidth,\boxHeight)
		node [pos=.5, align=center, font=\small]
		{Compute \\ error \\ bounds};
		
	\pgfmathsetmacro{\positionX}
	  {\positionX+\boxWidth+\boxArrowGap};
	\pgfmathsetmacro{\lastPositionX}
	  {\positionX-\boxWidth*0.5-\boxWidth*2-\boxArrowGap*4-\arrowWidth*2};
	\pgfmathsetmacro{\arrowTop}
	  {-\boxHeight*0.35};
		  
	\draw[-Latex,dashed] 
		(\positionX, \boxHeight/2) -- 
		(\positionX+\arrowWidth, \boxHeight/2) --
		(\positionX+\arrowWidth, \arrowTop) --
		(\lastPositionX,         \arrowTop) --
		(\lastPositionX,         -\boxArrowGap);
	\pgfmathsetmacro{\positionX}{\positionX+\arrowWidth+\boxArrowGap};

	\node [anchor=north, align=center, font=\small] at
			(\positionX*0.5+\lastPositionX*0.5, \arrowTop)
			{Submodels w/ error > threshold. $s \gets 2s$};

	\end{tikzpicture}
	\ifdefined\RELEASE
	\else
\end{document}
\fi
	\caption{
	  The submodel training process. The additional phase for training submodels in the leaves is depicted with dashed lines.
	}
	\label{fig:submodel_training_last_stage}
\end{figure}

\subsubsection{Computing transition inputs} \label{computing_transition_inputs}
Given a trained submodel $m$ we can analytically find all its linear regions, and respectively the inputs delimiting them, which we call \emph{trigger inputs} $g_l$. For all inputs in the region $[\,g_l,g_{l+1}]$, the model function, denoted as $M(x)$, is linear by construction. On the other hand, the uniform output quantization defines a step-wise function $Q=\lfloor M(x)\cdot W \rfloor/W$, where $W$ is the size of the quantized output domain \changes{(Figure~\ref{fig:transition_inputs})}. Thus, for each input region $[\,g_l,g_{l+1}]$, the set of transition inputs $t_{l} \in T$ are those where $M(x)$ and $Q$ intersect.

\subsubsection{Computing the responsibilities of submodels in the following stage.} \label{computing_responsibilities_in_next_stage}
Given a trained submodel $m_{i,j}$ in an internal stage $i$, we say that it \emph{maps} a $key$ to a submodel $m_{i+1,k}$,\, $k<W_{i+1}$, if $\lfloor M_{i,j}(key)\cdot W_{i+1} \rfloor=k$. As discussed informally earlier, the responsibility $R_{i+1,k}$ of $m_{i+1,k}$ is defined as all the inputs which are mapped by submodels in stage $i$ to $m_{i+1,k}$.  In other words, the trained submodels at stage $i$ define the responsibility of untrained submodels at stage $i+1$.

Knowing the responsibility of a submodel is crucial, as it determines the subset of the inputs  used to train the submodel. RMI exhaustively evaluates all the inputs, which is inefficient. Instead, we compute $R_{i+1,k}$ using the transition inputs of $m_{i,j}$. In the following,  we assume for clarity that $R_{i,j}$ is contiguous, and $m_{i,j}$ is the only submodel at stage $i$. 

We compute $R_{i+1,k}$ by observing that it is composed of all the inputs in the regions  $(t_l,t_l+1)$ that map to submodel $m_{i+1,k}$, where $t_l\in T_{i,j}$ are transition inputs of $m_{i,j}$. By construction, the inputs in the region between two adjacent transition points map to the same output. Then,  it suffices to compute the output of $m_{i,j}$ for its transition points, and choose the respective input ranges that are mapped to $m_{i+1,k}$.

\subsubsection{Training a submodel with ranges using sampling}
\label{sec:sampling}

Up to this point, we used only key-index pairs as model inputs.
Now we focus on training on input \emph{ranges}. A range can be represented as all the keys that fall into the range, all associated with the \emph{same} index of the respective rule. For example, 10.1.1.0-10.1.1.255 includes 256 keys. Our goal is to train a model such that given a key in the range, the model predicts the correct index. Enumerating all the keys in the ranges is inefficient. Instead, we use sampling as follows.

\removed{The training is performed in three steps repeated as needed: sample generation, submodel training, maximum error computation. If the maximum error is too large, the model is retrained.
{\noindent \bf Sample generation.}}

We generate the training key-index pairs by uniformly sampling the submodel's responsibility. We start with a low sampling frequency. A sample is included in the training set if there is an input rule range that matches the sampled key. Thus, the number of samples per input range is proportional to its relative size in the submodel's responsibility. Note that some input ranges (or individual keys) might not be sampled at all. Nevertheless, they will be matched correctly as we explain further. 

\changes{
\subsubsection{Submodel training} \label{sec:submodel_traiaing_supervised}
We train submodels on the generated datasets using supervised learning and Adam optimizer \cite{Kingma2014} with a mean squared error loss function.
}

\removed{
{\noindent \bf Neural network training.}
We train the submodel using the generated samples via standard stochastic gradient descent \removed{training} with mean squared error loss function.
{\noindent \bf Maximum error computation.} Given a trained submodel,
}

\changes{
\subsubsection{Computing error bounds} \label{computing_error_bounds}
Given a trained submodel in the last stage,
}
we compute the prediction error bound for \emph{all inputs in its responsibility} by evaluating the submodel on its transition inputs. The prediction error is computed also for the inputs that were not necessarily sampled, guaranteeing match correctness.
%
%
If the error is too large, we double the number of samples, regenerate the key-index pairs, and retrain the submodel. Training continues until the target error bound is attained \changes{or after a predefined number of attempts}. If training does not converge, the target error bound \removed{is increased} \changes{may be increased by the operator}. The error bound determines the \emph{search distance of the secondary search}; hence a larger bound causes lower system performance. We evaluate this tradeoff later (\S\ref{sec:training_time}).

\removed{
\subsubsection{Training an \MSNN}
The whole training process is similar to RMI's and is performed stage by stage. For each stage, we train each submodel separately according to ~\S\ref{sec:sampling}. We then compute the responsibilities of the submodels in the next stage, and train them, and so on, until the last stage.
}

\subsection{Handling multi-dimensional queries with range overlaps} \label{sec:generating_isets} \label{sec:partitioning}

\ouralgorithm{} supports overlapped ranges and matching over multiple dimensions, i.e., packet fields, by combining two simple ideas:  partitioning the rule-set into disjoint independent sets (\emph{\isets}),  and performing multi-field validation of each rule. In the following, we use the terms dimension and field interchangeably. 

\paragraph{Partitioning} Each \iset contains rules that do not overlap in \emph{one specific dimension}. We refer to the \emph{coverage} of an \iset as the fraction of the rules it holds out of those in the input. One \iset may cover all the rules if they do not overlap in at least one dimension, whereas the same dimension with many overlapping ranges may require multiple \isets.
Figure~\ref{fig:projections} shows the \isets for the rules from Figure~\ref{fig:packet_classification}. 

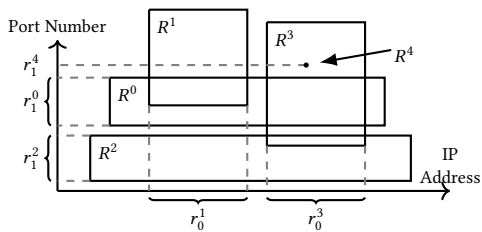
\begin{figure}[t]
	\centering
	\ifdefined\RELEASE
\else
\documentclass[conference,compsoc]{IEEEtran}

\usepackage{graphicx}
\usepackage{subfig}

\usepackage[T1]{fontenc}
\usepackage[utf8]{inputenc}
\usepackage{pgfplots}
\usepackage{grffile}
\pgfplotsset{compat=newest}
\usetikzlibrary{plotmarks}
\usetikzlibrary{arrows.meta}
\usepgfplotslibrary{patchplots}
\usepackage{amsmath}

\newcommand{\ouralgorithm}[0]{Plasmus}
\begin{document}
	\fi

\newcommand{\Rule}[6]{ 
	\draw[-] (#2,#1) -- (#2+#3,#1) -- (#2+#3,#1-#4) -- (#2,#1-#4) -- (#2,#1) node[anchor=north west, font={\small}] {$R^#5$};
	
	\ifnum#6=0 
		\draw[dashed, color=gray] (#2+#3,#1-#4) -- (#2+#3,0);
		\draw[dashed, color=gray] (#2,#1-#4) -- (#2,0);
		\draw[decorate,decoration={brace,amplitude=2pt}] (#2+#3,-0.1) -- (#2,-0.1);
		\draw (#2+#3/2, -0.15) node[align=center, anchor=north, font={\small}] {$r^#5_0$};
	\else 
		\draw[dashed, color=gray] (#2,#1) -- (0,#1);
		\draw[dashed, color=gray] (#2,#1-#4) -- (0,#1-#4);
		\draw[decorate,decoration={brace,amplitude=2pt}] (-0.1, #1-#4) -- (-0.1, #1);
		\draw (-0.15, #1-#4/2) node[align=center, anchor=east, font={\small}] {$r^#5_1$};
	\fi
}

\newcommand{\scale}{0.87}

\begin{tikzpicture}[thick,scale=\scale, every node/.style={scale=\scale}]
	
	\pgfmathsetmacro{\heightFactor}{0.77};
	
	\draw[->] (0,0) -- (6,0) node[anchor=south, font={\small}, align=center] {IP\\Address};
	\draw[->] (0,0) -- (0,3*\heightFactor) node[align=center, anchor=south, font={\small}] {Port Number};
	
	\Rule{2.25*\heightFactor}{0.8}{4.2}{0.95*\heightFactor}{0}{1};
	\Rule{3.6*\heightFactor}{1.4}{1.5}{1.9*\heightFactor}{1}{0};
	\Rule{1.1*\heightFactor}{0.5}{4.9}{0.9*\heightFactor}{2}{1};
	\Rule{3.35*\heightFactor}{3.2}{1.5}{2.45*\heightFactor}{3}{0};
	
	\draw [dashed, color=gray] (3.8, 2.5*\heightFactor) -- (0, 2.5*\heightFactor);
	\draw (-0.15,2.5*\heightFactor) node [anchor=east, color=black, font={\small}] {$r_1^4$};
	
	\draw [-Latex] (5.1, 2.8*\heightFactor) -- (4, 2.55*\heightFactor);
	\draw [fill=black]  (3.8, 2.5*\heightFactor) circle (0.03*\heightFactor);
	\draw (5.3, 2.4*\heightFactor) node[anchor=south, font={\small}] {$R^4$};
	
\end{tikzpicture}

\ifdefined\RELEASE
\else
\end{document}
\fi
	
	\caption{
	    Rules from Figure~\ref{fig:packet_classification} are split into two \isets: $\{R^0,R^2,R^4\}$ (by port), and $\{R^1,R^3\}$ (by IP).}
	\label{fig:projections}
\end{figure}

Each \iset is indexed by one \MSNN.  Thus, to find the match to a query with multiple fields, 
we query all \MSNN{}s (in parallel), each over the field on which it was trained. Then, the highest priority result is selected as the output. 

Each \iset adds to the total memory consumption and computational requirements of \ouralgorithm{}. Therefore, we introduce a heuristic that strives to find the smallest number of \isets that cover the largest part of the  rule-set~(\S\ref{sec:partitioning_algo}).  

\paragraph{Multi-field validation} Since an \MSNN builds an index of the rules over a single field, it might retrieve a rule which does not match against other fields. Hence, each rule returned by an \MSNN is validated across all fields. This enables \ouralgorithm{} to avoid indexing all dimensions, yet obtain correct results.


\subsubsection{\iset partitioning}
\label{sec:partitioning_algo}

%
We introduce a greedy heuristic that repetitively constructs the largest \iset from the input rules, producing a group of \isets.
%
%
%
To find the largest \iset over one dimension, we use a classical \emph{interval scheduling maximization} algorithm~\cite{Kleinberg2006}. The algorithm sorts the ranges by their upper bounds, and repetitively picks the range with the smallest upper bound that does not overlap previously selected ranges. 

We apply the algorithm to find the largest \iset in each field. Then we greedily choose the largest \iset among all the fields and remove its rules from the input set. We continue until exhausting the input. This heuristic is sub-optimal but quite efficient. We plan to improve it in future work.

Having a larger number of fields in a rule-set might help improve coverage. For example, if the rules that overlap in one field do not overlap in another and vice versa, two \isets cover the whole rule-set, requiring more \isets for each field in isolation.


\subsection{Remainder set and external classifiers}
\label{sec:remainder}


Real-world rule-sets may require many \isets for full coverage, with a single rule per \iset in the extreme cases. Using separate \MSNN{}s for such \isets will hinder performance. Therefore, we merge  small \isets into a single \emph{remainder set}. The rules in the remainder set are indexed using an \emph{external} classifier.
Each query is performed on both the \MSNN and the external classifier. 


In essence, \ouralgorithm{} serves as an \emph{accelerator} for the external classifier. Indeed, if rule-sets are covered using a few large \isets, the external classifier needs to index a small remainder set that often fits into faster memory, so it can be very fast. 


Two primary factors determine the end-to-end performance: (1)  the number of
\isets required for high coverage (depends on the rule-set), and; (2) the number of
\isets for achieving high performance (set by an operator). 
%

Our evaluation (\S\ref{sec:eval_rule_mixture}) shows that most of the evaluated rule-sets can be covered with high coverage above 90\% with only 2-3 \isets. This is enough to accelerate the external classifier, as is evident from the performance results.
%
%
On the other hand, the choice of the number of \isets depends on the external classifier properties, in particular, its sensitivity to memory footprint. We analyze this tradeoff in \S\ref{sec:evaluation:analys}.


\paragraph{Worst-case inputs} 
Some rule-sets cannot achieve good coverage with only a few \isets. For example, a rule-set with a single field whose ranges overlap requires too many \isets to be covered. 

To obtain a better intuition about the origins of worst-case inputs, we consider the notion of \emph{rule-set diversity} for rule-sets with exact matches. Rule-set diversity in a field is the number of unique values in it across the rule-set, divided by the total number of rules. \emph{The rule-set diversity is an upper bound on the fraction of rules in the largest \iset of that field}. In other words, low diversity implies that using the field for \iset partitioning would result in poor coverage. 

We can also identify challenging rule-sets with ranges. We define \emph{rule-set centrality} as the maximal number of rules that each pair of them overlap (they all share a point in a multi-dimensional space).
\emph{The rule-set centrality is a lower bound on the number of \isets required for full coverage}. 

The diversity and centrality metrics can indicate the potential of \ouralgorithm{} to accelerate the classification of a rule-set. On the positive side, our \iset partitioning algorithm is effective at segregating the rules that cannot be covered well from the rules that can, thereby accelerating the remainder classifier as much  as possible for a given rule-set. We analyze this property in \S\ref{sec:partitionining_effectiveness}.




\subsection{Putting it all together}
 \label{sec:the_lookup_process}

We briefly summarize all the steps of \ouralgorithm{}.

{\noindent \bf Training}
\begin{enumerate}
    \item Partition the input into \isets and a remainder set
    \item Train one \MSNN on each \iset
    \item Construct an external classifier for the remainder set
\end{enumerate}
 
{\noindent \bf Lookup}
\begin{enumerate}
    \item Query all the \MSNN{}s
    \item Query the external classifier
    \item Collect all the outputs, return the highest-priority rule
\end{enumerate}
 
\subsection{Rule Updates} \label{sec:updates}

%
%
We explain how \ouralgorithm{} can support updates with a limited performance degradation. 

Firstly, an external classifier used for the remainder must support updates. Among the evaluated external classifiers only TupleMerge is designed for fast updates.

Secondly, we distinguish four types of updates:
\emph{(i)} a change in the \emph{rule action}; 
\emph{(ii)} rule  \emph{deletion}
\emph{(iii)} rule  \emph{matching set change};
\emph{(iv)} rule \emph{addition}.

The first two types of updates are supported without performance degradation, and require a lookup followed by an update in the value array. 
However, if an update modifies a rule's matching set or adds a new rule, it might require modifications to the \MSNN model.  We currently do not know an algorithmic way to update \MSNNs without retraining; therefore, an updated rule is always added to the remainder set. 

Unfortunately, this design leads to gradual performance degradation, as updates are likely to increase the remainder set.
%
%
Accordingly, the model is retrained  on the updated rule-set, either periodically or when a large performance degradation is detected. 
Updates occurring while retraining are accommodated in the following batch of updates.   

\paragraph{Estimating sustained update rate}
%
%
Let $r$ and $u$  be the total number of rules  and the number of updates that move a rule to the remainder, respectively; $u$ can be smaller than the real rate of rule updates. We assume that the updates are independent and uniformly distributed among the $r$ rules. 
For each rule update, a rule is modified w.p. (with probability) $\frac{1}{r}$. 
Thus a rule is not modified in any of the updates w.p. $(1 - \frac{1}{r}) ^ u \approx e^{- u / r}$. 
The expected number of unmodified rules is $r \cdot (1 - \frac{1}{r}) ^ u \approx r \cdot e^{- u / r}$. Throughput behaves as a weighted average between that of \ouralgorithm{} and the  remainder implementation, based on the number of rules in each. 

Figure~\ref{fig:throughput_vs_upate} illustrates the throughput over time for different retraining rates given a certain update rate. If retraining is invoked every $\tau$ time units, the slower the training process, the worse the performance degradation. 

\begin{figure}[!t]
	\centering
    \ifdefined\RELEASE
\else
\documentclass[conference,compsoc]{IEEEtran}

\usepackage{graphicx}
\usepackage{subfig}

\usepackage{amsfonts}

\usepackage[T1]{fontenc}
\usepackage[utf8]{inputenc}
\usepackage{pgfplots}
\usepackage{grffile}
\pgfplotsset{compat=newest}
\usetikzlibrary{plotmarks}
\usetikzlibrary{arrows.meta}
\usepgfplotslibrary{patchplots}
\usepackage{amsmath}

\newcommand{\ouralgorithm}[0]{Plasmus}
\begin{document}
\fi

\definecolor{tpt_line}{HTML}{009954}
\definecolor{dark}{HTML}{1E1E1C}
\definecolor{bright}{HTML}{e29a12}

\begin{tikzpicture}

\begin{axis}[
xmax=4,ymax=1.5,
xmin=0,ymin=0,
axis lines=middle,
xlabel = Time,style={font={}},
ylabel = Throughput,
xtick={0,1,2,3,4},
xticklabels={{$0$}, {$\tau$}, {$2\tau$}, {$3\tau$}, {$4\tau$}}, 
yticklabels={}, 
height = 0.22 \textwidth,
width = 0.47  \textwidth,
legend style={
    at={(1,1)}, anchor=north east, font={\footnotesize}, legend columns =-1
}]

\pgfmathsetmacro{\markSize}{1.5};

\addplot[color=dark, mark=diamond*, domain=1/2^6:1,samples=5, mark size=\markSize]  {0.80 * e^(- x / 2) + 0.0};
\addlegendentry{Fast training}; 


\addplot[color=bright, mark=square*, domain=1/2^6:1,samples=5, mark size=\markSize]  {0.60 * e^(- x / 2) + 0.0};
\addlegendentry{Long training}; 

\addplot[color=dark, mark=diamond*, domain=1+1/2^6:2,samples=5, mark size=\markSize]  {0.80 * e^(- 0.5 * (x-1) ) + 0.0};
\addplot[color=bright, mark=square*, domain=1+1/2^6:2,samples=5, mark size=\markSize]  {0.60 * e^(- 0.5 * (x-1) ) + 0.0};

\addplot[color=dark, mark=diamond*, domain=2+1/2^6:3,samples=5, mark size=\markSize]  {0.80 * e^(- 0.5 * (x-2) ) + 0.0};
\addplot[color=bright, mark=square*, domain=2+1/2^6:3,samples=5, mark size=\markSize]  {0.60 * e^(- 0.5 * (x-2) ) + 0.0};

\addplot[color=dark, mark=diamond*, domain=3+1/2^6:4,samples=5, mark size=\markSize]  {0.80 * e^(- 0.5 * (x-3) ) + 0.0};

\addplot[color=bright, mark=square*, domain=3+1/2^6:4,samples=5, mark size=\markSize]  {0.60 * e^(- 0.5 * (x-3) ) + 0.0};


\addplot[color=tpt_line,dashed, domain=1/2^6:4,samples=100, line width=1.4]  {1}; 
\end{axis}
\end{tikzpicture}

\ifdefined\RELEASE
\else
\end{document}
\fi
	\caption{Updates impact on Throughput over time.  An upper bound (in green) is for zero training time.}\label{fig:throughput_vs_upate}
\end{figure}
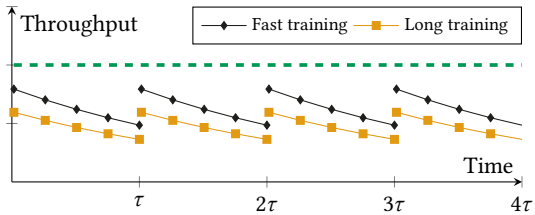

With these update estimates, using the measured speedup as a function of the fraction of the remainder (\S\ref{sec:partitionining_effectiveness}), \ouralgorithm{} can sustain up to 4k updates per second for 500K rule-sets, yielding about half the speedup of the update-free case, assuming a minute-long training.  
These results indicate the need for speeding up training,  but we conjecture there might be a more efficient way to perform updates directly in \MSNN without complete re-training of all submodels. Accelerating updates is left for future work.

\section{Implementation details} \label{sec:implmentation}

\paragraph{\MSNN structure} The number of stages and the width of each stage  depend on the number of rules to index. 
We increase the width of the last stage from 16 for 10K rules to
as much as 512 for 500K. See Table~\ref{table:msnn_structure} in the Appendix.

\paragraph{Submodel structure}
Each submodel is a fully connected 3-layer neural net with 1 input, 1 output, and 8 neurons in the hidden layer with ReLU activation. 
This structure affords an efficient vectorized implementation (see below).

\paragraph{Training} We use TensorFlow~\cite{tensor_flow} to train each submodel
on a CPU.  Training a submodel requires a few seconds, but the whole \MSNN may
take up to a few minutes (see \S\ref{sec:training_time}). We believe, however,
that a much faster  training time could be achieved with more optimizations,
i.e., replacing TensorFlow (known for its poor performance on small models).
We leave it for future work.

\changes{
    \paragraph{\iset partitioning}
    We implement the \iset partitioning algorithm using Python. The partitioning takes at most a few seconds and is negligible compared to \MSNN training time.
}

\paragraph{Inference and secondary search}
We implement \MSNN inference in C++. For each \iset we sort the rules
by the value of the respective field 
to optimize the secondary search. To reduce the number of memory accesses, 
we pack multiple field values from different rules in the same cache line.

\paragraph{Handling long fields}
Both \iset partitioning algorithms and \MSNN{} models map the inputs into single-precision floating-point numbers.
This allows the packing of more scalars in vector operations, resulting in faster inference. 
While enough for 32-bit fields, doing so might cause poor performance for fields
of 64-bits and 128-bits.

We compared two solutions: (1) splitting the fields into 32-bit parts and treating each as a distinct field, and (2) using a single-precision floating-point to express long fields.
The two showed similar results for \iset partitioning with MAC addresses, while
with IPv6, splitting into multiple fields worked better. Note that both the
secondary search and the validation phases are not affected because the rules
are stored with the original fields.

\paragraph{Vectorization}
We accelerate the inference by using wide CPU vector instructions.
Specifically, with 8 neurons in the hidden layer of each submodel,  computing
the prediction involves a handful of vector instructions. Validation is also vectorized.

Table \ref{table:msnn_performace} shows the effectiveness of vectorization. The
use of wider units speeds up inference, highlighting the
potential for scaling \ouralgorithm in future CPUs.

\begin{table}[t]
\small
    \caption{Submodel acceleration via vectorization. Methods are annotated with the number of floats per single instruction.}
    \label{table:msnn_performace}
    \begin{tabular}{c|ccc}    
		\toprule
		Instruction set (width) & Serial(1) & SSE(4) & AVX(8) \\
		\midrule
		Inference Time (ns) & 126 & 62 & 49 \\
		\bottomrule
	\end{tabular}
\end{table}

\paragraph{Parallelization}
\ouralgorithm{} lends itself to parallel execution where \isets and the remainder classifier run in parallel on different CPU cores. The system receives the packets and enqueues each for execution into the worker threads. The threads are statically allocated to run \MSNNs or the external classifier with a balanced load between the cores. 

Note that since \MSNNs are small and fit in L1, running them on a separate core enables L1-cache-resident executions even if the remainder classifier is large. Such an efficient cache utilization could not have been achieved with other classifiers running on two cores.

\changes{
\paragraph{Early termination}
One drawback of the parallel implementation is that the slowest thread determines the execution time. Our experiments show that the remainder classifier is the slowest one. It holds only a small fraction of the rules, so it returns an empty set for most of the queries, which in turn leads to the worst-case lookup time. In TupleMerge, for example, a query which does not find any matching rules results in a search over all tables, whereas in the average case some tables are skipped.

Instead, we query the remainder \emph{after} obtaining the results from the \isets, and terminate the search when we determine that the target rule is not in the remainder.  

To achieve that, we make minor changes to existing classification techniques. Specifically, in decision-tree algorithms, we store in each node the maximum priority of all the sub-tree rules. Whenever we encounter a maximum priority that is lower than that found in the \isets, we terminate the tree-walk. The changes to the hash-based algorithms are similar.

We call this optimization \emph{early termination}. With this optimization, both the \isets and the remainder are queried on the same core. While a parallel implementation is possible, it incurs higher synchronization overheads among threads. 

 }
\section{Evaluation} \label{sec:results}

In the evaluation, we pursued the following goals.
\begin{enumerate}
    \item 
    Comparison of \ouralgorithm{} with the state-of-the-art algorithms TupleMerge~\cite{Daly2019}, CutSplit~\cite{Wenjun2018}, and NeuroCuts~\cite{Liang2019};
    \item Systematic analysis of the performance characteristics, including coverage in challenging data sets, the effect of \MSNN error bound, and training time.
\end{enumerate}


\subsection{Methodology}
We ran the experiments on Intel Xeon Silver 4116 @ 2.1 GHz with 12 cores, 32KB L1, 1024KB L2, and 16MB L3 caches, running Ubuntu 16.04 (Linux kernel 4.4.0). We disable power management for stable measurements.

\removed{
    For evaluating each classifier we generated traces with 700K packets, accessing all matching rules  uniformly to evaluate the worst-case memory access pattern. We processed the trace 6 times, using the first five as warmup and measuring the last. We report the average of 15 measurements.
}

\changes{
\paragraph{Evaluated configurations}
CutSplit
}
(\CS) is set with $\emph{binth}=8$, as suggested in~\cite{Wenjun2018}. 

For NeuroCuts (\NC), we performed a hyperparameter sweep and selected the best classifier per rule-set.
As recommended in~\cite{Liang2019}, we focused on both top-mode partitioning and reward scaling.
We ran the search on three 12-core Intel machines, allocating six hours per configuration to converge.
In total, we ran \NC training for up to 36 hours per rule-set.
In addition, we developed a C++ implementation of \NC for faster evaluation of the generated classifiers,
much faster than the authors' Python-based prototype.

TupleMerge (\TM) is used with the version that supports updates with $\emph{collision-limit}=40$, as suggested in~\cite{Daly2019}.

\ouralgorithm{} (\NM) was trained with a maximum error threshold of 64.
\removed{
    We used two \isets with \CS and \NC as a remainder, and four \isets with \TM. We discarded \isets with coverage below 25\% for comparisons against \CS and \NC, and below 5\% for comparisons against \TM.
}
We present the analysis of the sensitivity to the chosen parameters and training times in \S\ref{sec:evaluation:analysis}.

\changes{\paragraph{Multi-core implementation}} We run a parallel implementation on two cores. \ouralgorithm{} allocates one core for the remainder computations and the second for the \MSNN{}s.
For \CS, \NC, and \TM, we ran two instances of the algorithm in parallel on two cores using two threads (i.e., no duplication of the rules), splitting the input equally between the cores.
\changes{
    We discarded \isets with coverage below 25\% for comparisons against \CS and \NC, and below 5\% for comparisons against \TM.
}
We used batches of 128 packets to amortize the synchronization overheads. Thus, these algorithms achieve almost linear scaling and the highest possible throughput with perfect load-balancing between the cores. 

\changes{
    \paragraph{Single-core implementation}
    We used a single core to measure the performance of \ouralgorithm{} with the early termination optimization. For \NM, we discarded \isets with coverage below 25\%.
 }

    \subsubsection{Packet traces and rule-sets}
    For evaluating each classifier, we generated traces with 700K packets. We processed each trace 6 times, using the first five as warmup and measuring the last. We report the average of 15 measurements.
      
    \paragraph{Uniform traffic}
    We generate traces that access all matching rules uniformly to evaluate the worst-case memory access pattern. 
    

    \paragraph{Skewed traffic}
    For each rule-set we generate traces that follow Zipf distribution with four different skew parameters,  according to the amount of traffic that accounts for the 3\% most frequent flows (e.g., 80\% of the traffic accounts for the 3\% most frequent flows). This is representative of real traffic, as has been shown in previous works~\cite{Katta2016, Sarrar2012}. 
    
    Additionally, we use a real CAIDA trace from the Equinix datacenter in Chicago \cite{CAIDA}. As CAIDA does not publish the rules used to process the packets, we modify the packet headers in the trace to match each evaluated rule-set as follows. For each rule, we generate one matching five-tuple. Then, for each packet in CAIDA, we replace the original five-tuple with a random five-tuple generated from the rule-set, while maintaining a consistent mapping between the original and the generated one. Note that the rule-set access locality of the generated trace is the same or as high as the original trace. 
    

\paragraph{ClassBench rules}
ClassBench~\cite{TaylorTurner2007} is a standard benchmark broadly used for evaluating packet classification algorithms~\cite{Liang2019, Wenjun2018, Daly2019, YingSoDa2018, KoganKirill2014, VamananBalajee2011, Narodytska2019}. It creates rule-sets that correspond to the rule distribution of three different applications: Access Control List (ACL), Firewall (FW), and IP Chain (IPC). We created rule-sets of sizes 500K, 100K, 10K, and 1K, each with 12 distinct applications, all with 5-field rules: source and destination IP, source and destination port, and protocol.

\paragraph{Real-world rules}
We used the Stanford Backbone dataset which contains a large enterprise network configuration~\cite{Zeng2012}. There are four IP forwarding rule-sets with roughly 180K single-field rules each (i.e., destination IP address).

\begin{figure*}[t]
\ifdefined\RELEASE
\else
\documentclass[conference,compsoc]{IEEEtran}

\usepackage{graphicx}
\usepackage{subfig}

\usepackage[T1]{fontenc}
\usepackage[utf8]{inputenc}
\usepackage{pgfplots}
\usepackage{grffile}
\pgfplotsset{compat=newest}
\usetikzlibrary{plotmarks}
\usetikzlibrary{arrows.meta}
\usepgfplotslibrary{patchplots}
\usepackage{amsmath}

\newcommand{\ouralgorithm}[0]{NuevoMatch}
\newcommand{\iset}[0]{iSet}
\newcommand{\isets}[0]{iSets}
\begin{document}
\fi

\definecolor{cutsplit}{HTML}{489ac7}
\definecolor{cutsplit_dark}{HTML}{003242}

\definecolor{neurocuts}{HTML}{ff9c24}
\definecolor{neurocuts_dark}{HTML}{1a0d00}

\definecolor{tuplemerge}{HTML}{f83f3f}
\definecolor{tuplemerge_dark}{HTML}{280606}

\definecolor{graybg}{HTML}{e0e0e0}
	
\pgfmathsetmacro{\barwidth}{0.23}
\pgfmathsetmacro{\barshift}{0.25}
\usetikzlibrary{patterns}

\newcommand{\scale}{1}

\begin{tikzpicture}[scale=\scale, every node/.style={scale=\scale}]

\pgfplotsset{
width=6in,
height=0.7in,
scale only axis,
xmin=0.5, xmax=26.5,
xticklabel style={font={\small}},
xtick style={draw=none},
yticklabel style={font={\small}},
yminorticks=true,
ymajorgrids, yminorgrids,
xtick={1,2,3,4,5,6,7,8,9,10,11,12,13,14,15,16,17,18,19,20,21,22,23,24,25,26},
xticklabels={{1},{2},{3},{4},{5},{6},{7},{8},{9},{10},{11},{12},{GM},{1},{2},{3},{4},{5},{6},{7},{8},{9},{10},{11},{12},{GM}},
ylabel style={font=\small, align=center},
legend style={
at={(0.5,-0.3)}, anchor=north, legend cell align=left, legend columns=-1, font={\small}
},
}

\begin{axis}[
at={(0in,0in)},
ymin=0, ymax=8,
ytick={0,2,4,6,8,10},
ylabel={Latency\\Speedup},
]

\addlegendimage{fill=cutsplit, area legend, postaction={pattern=crosshatch dots, pattern color=cutsplit_dark}};
\addlegendentry{\ouralgorithm{} w/ CutSplit};
\addlegendimage{fill=neurocuts, area legend, postaction={pattern=horizontal lines, pattern color=neurocuts_dark}};
\addlegendentry{\ouralgorithm{} w/ NeuroCuts};
\addlegendimage{fill=tuplemerge, area legend, postaction={pattern=grid, pattern color=tuplemerge_dark}};
\addlegendentry{\ouralgorithm{} w/ TupleMerge};

\addplot[draw=none, opacity=0.4, fill=gray] table {
	25.5 0
	25.5 8
	27 8
	27 0
};
\addplot[draw=none, opacity=0.4, fill=gray] table {
	12.5 0
	12.5 8
	13.5 8
	13.5 0
};
\addplot[color=gray, line width=1] table {
    13.5 0
    13.5 8
};

\addplot[ybar, bar width=\barwidth, fill=cutsplit, draw=none, area legend, postaction={pattern=crosshatch dots, pattern color=cutsplit_dark},bar shift=-\barshift] table {
1 3.55889
2 1.84614
3 1.94042
4 1.58832
5 1.42816
6 2.62557
7 1.67401
8 1.94103
9 1.58683
10 1.82088
11 1.83146
12 2.90981

13 2.0

14 4.97267
15 2.33491
16 2.57074
17 2.24704
18 4.28014
19 2.57983
20 2.14736
21 2.45942
22 2.40644
23 2.66037
24 2.31757
25 2.95184

26 2.7
}; 

\addplot[ybar, bar width=\barwidth, fill=neurocuts, draw=none, area legend, postaction={pattern=horizontal lines, pattern color=neurocuts_dark}] table {
1  2.25287
2  4.4055
3  2.82371
4  1.52101
5  1.44528
6  7.1004
7  5.06835
9  4.82754
10 4.54453
11 6.60652

13 3.6

14 7.71344
15 2.2674
16 5.39338
17 5.17254
18 2.30181
19 8.00961
21 3.61233
23 4.62534
25 4.31366

26 4.4
}; 

\addplot[ybar, bar width=\barwidth, fill=tuplemerge, draw=none, area legend, postaction={pattern=grid, pattern color=tuplemerge_dark},bar shift=+\barshift] table {
1  3.45708
2  2.33877
3  2.00161
4  1.57794
5  2.41239
6  2.29774
7  3.03758
8  3.19196
9  2.86692
10  2.46963
11  2.44173
12  4.19568

13 2.6

14 2.92971
15 2.2986
16 2.36656
17 1.76278
18 3.38572
19 3.62321
20 2.60841
21 2.0853
22 2.78356
23 2.4682
24 3.04271
25 2.92117

26 2.6
}; 

\node [anchor=north west, font={\bf \small}, opacity=0.8] at (axis cs:1.5,8) {100K Classifiers};
\node [anchor=north east, font={\bf \small}, opacity=0.8] at (axis cs:25.5,8) {500K Classifiers};

\end{axis}


\begin{axis}[
at={(0in,0.9in)},
ymin=0, ymax=4,
ytick={0,1,2,3,4,5},
ylabel={Throughput\\Speedup}
]

\addplot[draw=none, opacity=0.4, fill=gray] table {
	25.5 0
	25.5 8
	27 8
	27 0
};
\addplot[draw=none, opacity=0.4, fill=gray] table {
	12.5 0
	12.5 8
	13.5 8
	13.5 0
};
\addplot[color=gray, line width=1] table {
    13.5 0
    13.5 8
};

\addplot[ybar, bar width=\barwidth, fill=cutsplit, draw=none, area legend, postaction={pattern=crosshatch dots, pattern color=cutsplit_dark},bar shift=-\barshift] table {
1 1.71998
2 0.90355
3 0.945425
4 0.777572
5 0.668228
6 1.22068
7 0.81632
8 0.945698
9 0.776304
10 0.880514
11 0.895601
12 1.41662

13 1.0

14 2.45195
15 1.12313
16 1.20399
17 1.11078
18 2.11927
19 1.20132
20 1.07073
21 1.22673
22 1.17777
23 1.28187
24 1.11481
25 1.4208

26 1.3
}; 

\addplot[ybar, bar width=\barwidth, fill=neurocuts, draw=none, area legend, postaction={pattern=horizontal lines, pattern color=neurocuts_dark}] table {
1 1.06851
2 2.18448
3 1.37004
4 0.701008
5 0.679429
6 3.45227
7 2.47763
9 2.37509
10 2.20969
11 3.18089

13 1.7

14 3.83132
15 1.0943
16 2.68699
17 2.51686
18 1.14162
19 3.98158
21 1.69261
23 2.26425
25 2.02377

26 2.2
}; 

\addplot[ybar, bar width=\barwidth, fill=tuplemerge, draw=none, area legend, postaction={pattern=grid, pattern color=tuplemerge_dark},bar shift=+\barshift] table {
1  1.56916
2  1.10351
3  0.953613
4  0.730907
5  1.07815
6  1.081
7  1.42545
8  1.20016
9  1.34696
10  1.17856
11  1.14053
12  1.96741

13 1.2

14 1.37126
15 1.10097
16 1.13047
17 0.846959
18 1.61161
19 1.43343
20 1.2216
21 0.984442
22 1.30366
23 1.15662
24 1.44942
25 1.37313

26 1.2
};

\node [anchor=north west, font={\bf \small}, opacity=0.8] at (axis cs:1.5,4) {100K Classifiers};
\node [anchor=north east, font={\bf \small}, opacity=0.8] at (axis cs:25.5,4) {500K Classifiers};

\end{axis}

\end{tikzpicture}%
\ifdefined\RELEASE
\else
\end{document}
\fi
    \caption{ClassBench: \ouralgorithm{} vs. CutSplit, NeuroCuts, and TupleMerge, \changes{using two CPU cores}. (See rule-set in the Appendix.)}
    \label{fig:performance_speedup_large}
\end{figure*}
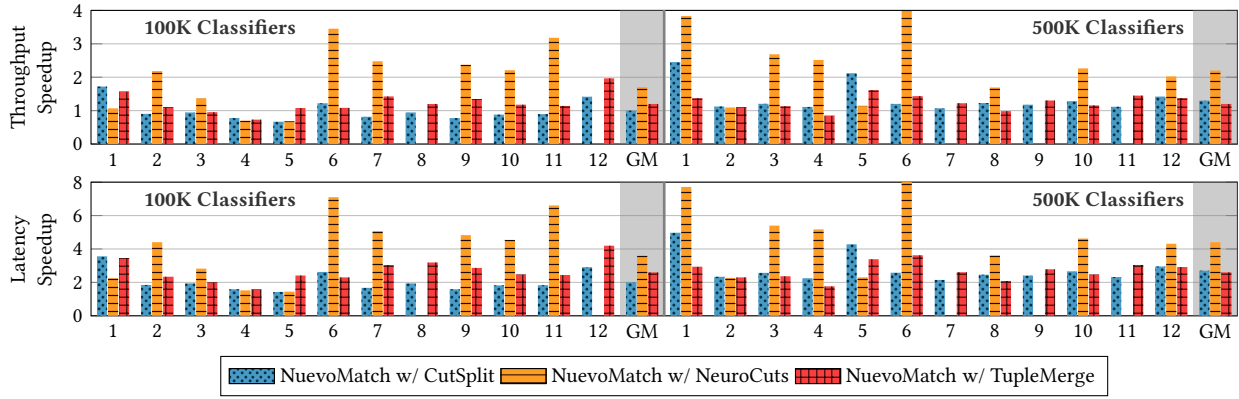

\subsection{End-to-end performance}
\label{sec:performance_eval}
For fair comparison, \ouralgorithm{} used \emph{the same} algorithm for both the remainder classifier and the baseline. For example, we evaluated the speedup produced by \ouralgorithm{} over \CS while also using \CS to index the remainder set.

We present the results for random packet traces, followed by skewed and CAIDA traces.


\paragraph{Large rule-sets: ClassBench: multi-core}
Figure \ref{fig:performance_speedup_large}  shows that, in the largest rule-sets (500K), \changes{the parallel implementation of} \ouralgorithm{} achieves a geometric mean factor of 2.7$\times$, 4.4$\times$, and 2.6$\times$ lower latency and 1.3$\times$, 2.2$\times$, and 1.2$\times$ higher throughput over \CS, \NC, and \TM, respectively.
For the classifiers with 100K rules, the gains are lower but still significant: 2.0$\times$, 3.6$\times$, and 2.6$\times$ lower latency and 1.0$\times$, 1.7$\times$, and 1.2$\times$ higher throughput over \CS, \NC, and \TM, respectively. The performance varies among rule-sets, i.e., some classifiers are up to 1.8$\times$ faster than \CS for 100\.K inputs.

\changes{
\paragraph{Large rule-sets: ClassBench: single core} Figure \ref{fig:performance_early_stopping} shows the throughput speedup of \NM compared to \CS, \NC, and \TM. For 500K rule-sets, \ouralgorithm{} achieves a geometric mean improvement of 2.4$\times$, 2.6$\times$, and 1.6$\times$ in throughput compared to \CS, \NC, and \TM, respectively. For the single core execution the latency and the throughput speedups are the same.
}

\paragraph{Large rule-sets: Stanford backbone: multi-core} Figure~\ref{fig:stanford} shows the speedup of \NM over \TM for the real-world Stanford backbone dataset with 4 rule-sets. \NM achieves $3.5\times$ higher throughput and $7.5\times$ lower latency over \TM on all four rule-sets. 

\paragraph{Small rule-sets: multi-core} 
For rule-sets with 1K and 10K rules, \ouralgorithm{} results in the same or lower throughput, and 2.2$\times$ and 1.9$\times$ on average better latency compared to \CS and \TM. 
The lower speedup is expected, as both \CS and \TM fit into L1 (\S\ref{sec:eval_compression}), so \NM does not benefit from reduced memory footprint, while adding computational overheads. See Appendix for the detailed chart.

The \CS results are averaged over three rule-sets of 1K and six rule-sets for 10K. In the remaining rule-sets, \ouralgorithm{} did not produce large-enough \isets to accelerate the remainder. Note, however, that it promptly identifies the rule-sets expected to be slow and falls back to the original classifier. 

\paragraph{The source of speedups}
The ability to compress the rule-set to fit into faster memory while retaining fast lookup is the key factor underlying the performance benefits of \ouralgorithm{}. To illustrate it, we take a closer look at the performance. We evaluate \TM with and without \NM acceleration as a function of its memory footprint on ClassBench-generated 1K,10K,100K and 500K rule-sets
for one application (ACL). 

Figure~\ref{fig:thpt_tm_nm} shows that the performance of \TM degrades as the number of rules grows, causing the hash tables to spill out of L1 and L2 caches. \NM compresses a large part of the rule-set (see coverage annotations), thereby making the remainder index small enough to fit in the L1 cache, and gaining back the throughput equivalent to \TM's on small rule-sets.

\begin{figure*}[t]
\ifdefined\RELEASE
\else
\documentclass[conference,compsoc]{IEEEtran}

\usepackage{graphicx}
\usepackage{subfig}

\usepackage[T1]{fontenc}
\usepackage[utf8]{inputenc}
\usepackage{pgfplots}
\usepackage{grffile}
\pgfplotsset{compat=newest}
\usetikzlibrary{plotmarks}
\usetikzlibrary{arrows.meta}
\usepgfplotslibrary{patchplots}
\usepackage{amsmath}

\newcommand{\ouralgorithm}[0]{NuevoMatch}
\newcommand{\iset}[0]{iSet}
\newcommand{\isets}[0]{iSets}
\begin{document}
\fi

\definecolor{cutsplit}{HTML}{489ac7}
\definecolor{cutsplit_dark}{HTML}{003242}

\definecolor{neurocuts}{HTML}{ff9c24}
\definecolor{neurocuts_dark}{HTML}{1a0d00}

\definecolor{tuplemerge}{HTML}{f83f3f}
\definecolor{tuplemerge_dark}{HTML}{280606}

\definecolor{graybg}{HTML}{e0e0e0}
	
\pgfmathsetmacro{\barwidth}{0.23}
\pgfmathsetmacro{\barshift}{0.25}
\usetikzlibrary{patterns}

\newcommand{\scale}{1}

\begin{tikzpicture}[scale=\scale, every node/.style={scale=\scale}]

\pgfplotsset{
width=6in,
height=0.8in,
scale only axis,
xmin=0.5, xmax=26.5,
xticklabel style={font={\small}},
xtick style={draw=none},
yticklabel style={font={\small}},
yminorticks=true,
ymajorgrids, yminorgrids,
xtick={1,2,3,4,5,6,7,8,9,10,11,12,13,14,15,16,17,18,19,20,21,22,23,24,25,26},
xticklabels={{1},{2},{3},{4},{5},{6},{7},{8},{9},{10},{11},{12},{GM},{1},{2},{3},{4},{5},{6},{7},{8},{9},{10},{11},{12},{GM}},
ylabel style={font=\small, align=center},
xlabel style={font=\small, align=center},
legend style={
at={(0.5,-0.3)}, anchor=north, legend cell align=left, legend columns=-1, font={\small}
},
}

\begin{axis}[
at={(0in,0in)},
ymin=0, ymax=4.6,
ytick={0,1,2,3,4},
ylabel={Throughput\\Speedup},
]

\addlegendimage{fill=cutsplit, area legend, postaction={pattern=crosshatch dots, pattern color=cutsplit_dark}};
\addlegendentry{\ouralgorithm{} w/ CutSplit};
\addlegendimage{fill=neurocuts, area legend, postaction={pattern=horizontal lines, pattern color=neurocuts_dark}};
\addlegendentry{\ouralgorithm{} w/ NeuroCuts};
\addlegendimage{fill=tuplemerge, area legend, postaction={pattern=grid, pattern color=tuplemerge_dark}};
\addlegendentry{\ouralgorithm{} w/ TupleMerge};

\addplot[draw=none, opacity=0.4, fill=gray] table {
	25.5 0
	25.5 8
	27 8
	27 0
};
\addplot[draw=none, opacity=0.4, fill=gray] table {
	12.5 0
	12.5 8
	13.5 8
	13.5 0
};
\addplot[color=gray, line width=1] table {
    13.5 0
    13.5 8
};

\addplot[ybar, bar width=\barwidth, fill=cutsplit, draw=none, area legend, postaction={pattern=crosshatch dots, pattern color=cutsplit_dark},bar shift=-\barshift] table {
1        3.14429
2        1.47147
3        2.29733
4        1.39096
5        0.779027
6        1.80357
7        1.38224
8        2.2139
9        1.66892
10       2.18285
11       1.6057
12       2.06473
13       1.73954
14       3.3478
15       1.81951
16       2.81436
17       2.7795
18       2.84786
19       2.04976
20       1.84277
21       2.49375
22       2.10013
23       2.33366
24       2.09365
25       2.28113
26       2.36045
}; 

\addplot[ybar, bar width=\barwidth, fill=neurocuts, draw=none, area legend, postaction={pattern=horizontal lines, pattern color=neurocuts_dark}] table {
1	1.104
2	2.5847
3	1.7393
4	0.90082
5	0.86531
6	4.4905
7	3.4574
8	0         
9	3.3032
10	2.8621
11	3.8218
12	0         
13	2.1560
14	3.7774
15	1.5599
16	3.116
17	2.9038
18	1.7099
19	4.4915
20	0        
21	1.995
22	0        
23	2.3626
24	0        
25	2.6895
26	2.5871
}; 

\addplot[ybar, bar width=\barwidth, fill=tuplemerge, draw=none, area legend, postaction={pattern=grid, pattern color=tuplemerge_dark},bar shift=+\barshift] table {
1	2.04674
2	1.67205
3	2.23608
4	1.64319
5	0.988932
6	1.7161
7	1.75989
8	1.37075
9	1.85272
10	1.52618
11	2.00597
12	1.84262
13	1.68876
14	1.79935
15	1.60044
16	1.87693
17	1.7744
18	1.7541
19	1.57098
20	1.57044
21	1.28873
22	1.61057
23	1.41857
24	1.98996
25	1.56897
26	1.64108
}; 

\node [anchor=north west, font={\bf \small}, opacity=0.8] at (axis cs:1.5,4.6) {100K Classifiers};
\node [anchor=north east, font={\bf \small}, opacity=0.8] at (axis cs:25.5,4.6) {500K Classifiers};

\end{axis}
\end{tikzpicture}%
\ifdefined\RELEASE
\else
\end{document}
\fi
    \caption{\changes{ClassBench: \ouralgorithm{} vs. CutSplit, NeuroCuts, and TupleMerge, using a single CPU core.}}
    \label{fig:performance_early_stopping}
\end{figure*}
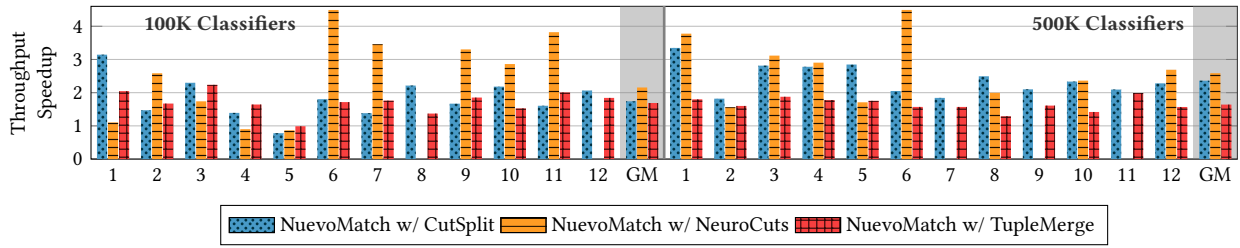

\begin{figure}[t]
	\centering
\ifdefined\RELEASE
\else
\documentclass[conference,compsoc]{IEEEtran}

\usepackage{graphicx}
\usepackage{subfig}

\usepackage[T1]{fontenc}
\usepackage[utf8]{inputenc}
\usepackage{pgfplots}
\usepackage{grffile}
\pgfplotsset{compat=newest}
\usetikzlibrary{plotmarks}
\usetikzlibrary{arrows.meta}
\usepgfplotslibrary{patchplots}
\usepackage{amsmath}

\newcommand{\ouralgorithm}[0]{NuevoMatch}
\newcommand{\iset}[0]{iSet}
\newcommand{\isets}[0]{iSets}
\begin{document}
\fi

\definecolor{tm}{HTML}{70b29d}
\definecolor{tm_dark}{HTML}{427b69}

\definecolor{nm}{HTML}{c8aa65}
\definecolor{nm_dark}{HTML}{6f5720}

\definecolor{text}{HTML}{24322e}

\definecolor{graybg}{HTML}{e0e0e0}
	
\pgfmathsetmacro{\barwidth}{0.3}
\pgfmathsetmacro{\barshift}{0.17}
\usetikzlibrary{patterns}

\newcommand{\scale}{1}

\begin{tikzpicture}[scale=\scale, every node/.style={scale=\scale}]

\pgfplotsset{
width=1.1in,
height=0.613in,
scale only axis,
xmin=0.5, xmax=4.5,
xticklabel style={font={\small}},
xtick style={draw=none},
xtick={1,2,3,4},
yticklabel style={font={\small}},
yminorticks=true,
ymajorgrids, yminorgrids,
ylabel style={font=\small, align=center},
xlabel style={font=\small, align=center},
legend style={
at={(1.07,-0.35)}, anchor=north, legend cell align=left, legend columns=-1, font={\small}
},
}

\begin{axis}[
at={(0in,0in)},
ymin=1e5, ymax=3e6,
ylabel={Throughput\\(pps)},
]

\addlegendimage{fill=tm, area legend, postaction={pattern=crosshatch dots, pattern color=tm_dark}};
\addlegendentry{TupleMerge};
\addlegendimage{fill=nm, area legend, postaction={pattern=horizontal lines, pattern color=nm_dark}};
\addlegendentry{\ouralgorithm{} w/ TupleMerge};

\addplot[ybar, bar width=\barwidth, fill=tm, draw=none, area legend, postaction={pattern=crosshatch dots, pattern color=tm_dark},bar shift=-\barshift] table {
1 687224
2 696345
3 691946
4 690193
}; 

\addplot[ybar, bar width=\barwidth, fill=nm, draw=none, area legend, postaction={pattern=horizontal lines, pattern color=nm_dark},bar shift=+\barshift] table {
1 2.41196e+06
2 2.42954e+06
3 2.34962e+06
4 2.45942e+06
};

\node [anchor=north, font={\scriptsize \bf}, align=center, color=text]
at (axis cs:1, 3.1e6) {3.51$\times$};
\node [anchor=north, font={\scriptsize \bf}, align=center, color=text]
at (axis cs:2, 3.1e6) {3.49$\times$};
\node [anchor=north, font={\scriptsize \bf}, align=center, color=text]
at (axis cs:3, 3.1e6) {3.40$\times$};
\node [anchor=north, font={\scriptsize \bf}, align=center, color=text]
at (axis cs:4, 3.1e6) {3.56$\times$};

\end{axis}

\begin{axis}[
at={(1.2in,0in)},
ymin=0, ymax=530,
yticklabel pos=right,
ylabel={Latency\\($\mu s$)},
]

\addplot[ybar, bar width=\barwidth, fill=tm, draw=none, area legend, postaction={pattern=crosshatch dots, pattern color=tm_dark},bar shift=-\barshift] table {
1 400.851
2 408.648
3 407.949
4 390.785
}; 

\addplot[ybar, bar width=\barwidth, fill=nm, draw=none, area legend, postaction={pattern=horizontal lines, pattern color=nm_dark}, bar shift=+\barshift] table {
1 53.3845
2 52.1131
3 53.7771
4 52.3093
};

\node [anchor=north, font={\scriptsize \bf}, align=center, color=text]
at (axis cs:1, 550) {7.51$\times$};
\node [anchor=north, font={\scriptsize \bf}, align=center, color=text]
at (axis cs:2, 550) {7.84$\times$};
\node [anchor=north, font={\scriptsize \bf}, align=center, color=text]
at (axis cs:3, 550) {7.59$\times$};
\node [anchor=north, font={\scriptsize \bf}, align=center, color=text]
at (axis cs:4, 550) {7.47$\times$};

\end{axis}

\end{tikzpicture}%
\ifdefined\RELEASE
\else
\end{document}
\fi
	\caption{End-to-end performance on real Stanford backbone data sets.}
	\label{fig:stanford}
\end{figure}
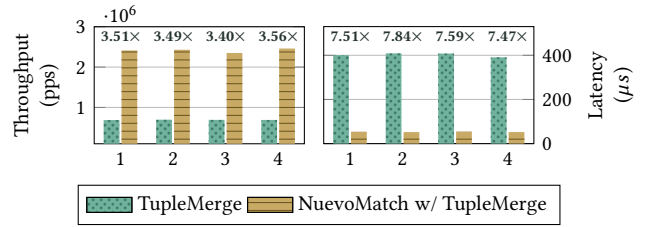

\begin{figure}[t]
	\centering
\ifdefined\RELEASE
\else
\documentclass[conference,compsoc]{IEEEtran}

\usepackage{graphicx}
\usepackage{subfig}
\usepackage[utf8]{inputenc}
\usepackage{xspace}
\usepackage{setspace}

\usepackage[T1]{fontenc}
\usepackage[utf8]{inputenc}
\usepackage{pgfplots}
\usepackage{grffile}
\pgfplotsset{compat=newest}
\usetikzlibrary{plotmarks}
\usetikzlibrary{arrows.meta}
\usepgfplotslibrary{patchplots}
\usepackage{amsmath}

\newcommand{\ouralgorithm}[0]{NuevoMatch}
\newcommand{\iset}[0]{iSet\xspace}
\newcommand{\isets}[0]{iSets\xspace}
\newcommand{\capital}[0]{\expandafter\MakeUppercase}
\begin{document}
	\fi

\definecolor{tm}{HTML}{42b357}
\definecolor{tm_dark}{HTML}{184320}

\definecolor{nm}{HTML}{184243}

\begin{tikzpicture}

\pgfmathsetmacro{\lineWidth}{1.0}

\pgfplotsset{
	width=2.5in,
	height=0.8in,
	scale only axis,
	ylabel style={font=\small, align=center},
	xlabel style={font=\small},
	xlabel={Number of Rules},
	ticklabel style={font=\small},
	legend style={legend cell align=left, align=left, draw=white!15!black, at={(0.5, -0.5)}, anchor=north, font={\small}, legend columns=-1},
	xmajorgrids, ymajorgrids,
}

\begin{semilogxaxis}[%
at={(0in,0in)},
ymin=1.5e6, ymax=6.4e6,
xmin=5e2, xmax=1e6,
ytick={2e6,3e6,4e6,5e6,6e6,7e6},
ylabel={Throughput (pps)},
]


\addplot [mark=*, color=tm, mark size=1.3, line width=\lineWidth] table{
943 5.36288e+06
9731 4.33902e+06
99619 3.35871e+06
497756 2.65675e+06
};\addlegendentry{TupleMerge};

\addplot [mark=triangle, color=nm, mark size=1.3, line width=\lineWidth] table{
943 4.24809e+06
9731 3.14136e+06
99619 5.49049e+06
497756 3.5868e+06
};\addlegendentry{\ouralgorithm{} w/ TupleMerge};

\addplot [dashed, mark=none, line width=1.2, opacity=0.8] table {
45000 1e6
45000 6.4e6
} node [anchor=north west, font={\bf \footnotesize \setstretch{.7}}, opacity=1] {L2 Size (1MB)};

\addplot [dashed, mark=none, line width=1.2, opacity=0.8] table {
1600 1e6
1600 6.4e6
} node [anchor=north west, font={\bf \footnotesize \setstretch{.7}}, opacity=1]  {L1 Size (32KB)};

\node [anchor=north, font={\scriptsize \bf \setstretch{.7}}, align=center, color=nm] at (axis cs:0.9e3, 3988307) {14.85:15.6\\25\%};
\node [anchor=north, font={\scriptsize \bf \setstretch{.7}}, align=center, color=nm] at (axis cs:9531, 3.1e6) {192.0:192.8\\6\%};
\node [anchor=south, font={\scriptsize \bf \setstretch{.7}}, align=center, color=nm] at (axis cs:99619, 3.8e6) {2.9:21.3\\99\%};
\node [anchor=south, font={\scriptsize \bf \setstretch{.7}}, align=center, color=nm] at (axis cs:497756, 3.6e6) {7.9:46.1\\99\%};

\node [anchor=south, font={\scriptsize \bf \setstretch{.7}}, align=center, color=tm_dark] at (axis cs:0.9e3, 5.3e6) {19.5~KB};
\node [anchor=south, font={\scriptsize \bf \setstretch{.7}}, align=center, color=tm_dark] at (axis cs:9531, 4.3e6) {205.0~KB};
\node [anchor=north, font={\scriptsize \bf \setstretch{.7}}, align=center, color=tm_dark] at (axis cs:99619, 3.3e6) {2~MB};
\node [anchor=north, font={\scriptsize \bf \setstretch{.7}}, align=center, color=tm_dark] at (axis cs:497756, 2.6e6) {10~MB};

\end{semilogxaxis}

\end{tikzpicture}
\ifdefined\RELEASE
\else
\end{document}
\fi
	\caption{Throughput vs. number of rules for TupleMerge and \ouralgorithm{}. Annotations are coverage (\%) and index memory size in KB (remainder : total). }
	\label{fig:thpt_tm_nm}
\end{figure}
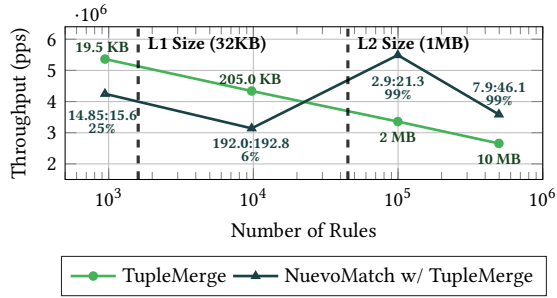

\changes{
    \paragraph{ClassBench: Skewed traffic}
    Figure \ref{fig:zipf} shows the evaluation of the early termination implementation on skewed packet traces. We report the throughput speedup of \NM compared to \CS and \TM; the results for \NC are similar to those of \CS.

    We perform 6000 experiments using 25 traces per rule-set: five traces per Zipf distribution plus five modified CAIDA traces. We evaluate over twelve 500K rule-sets and report the geometric mean.
    Additionally, we evaluate CAIDA traces in two settings.
    First, the classifier runs with access to the entire 16MB of the L3 cache (denoted as CAIDA).
    Second, the classifier use of L3 is restricted to 1.5\.MB via Intel's Cache Allocation Technology, emulating multi-tenant setting (denoted as CAIDA*).
    
    \ouralgorithm{} is significantly faster than \CS, but its benefits over \TM diminish for workloads with higher skews. Yet, the speedups are more pronounced under smaller L3 allocation. 
    
    Overall, we observe lower speedups for the skewed traffic than for the random trace. This is not surprising, as skewed traces induce a higher cache hit rate for all the methods, which in turn reduces the performance gains of \NM over both \CS and \TM, similar to the case of small rule-sets. Nevertheless, it is worth noting that classification algorithms are usually applied alongside caching mechanisms that catch the packets' temporal locality. For instance, Open vSwitch applies caching for most frequently used rules. It invokes Tuple Space Search upon cache misses \cite{Pfaff2015}. Therefore, if \ouralgorithm{} is applied at this stage, we expect it to yield the performance gains equivalent to those reported for unskewed workloads. Open vSwitch integration is the goal of our ongoing work. 
}

\begin{figure}[t]
	\centering
\ifdefined\RELEASE
\else
\documentclass[conference,compsoc]{IEEEtran}

\usepackage{graphicx}
\usepackage{subfig}

\usepackage[T1]{fontenc}
\usepackage[utf8]{inputenc}
\usepackage{pgfplots}
\usepackage{grffile}
\pgfplotsset{compat=newest}
\usetikzlibrary{plotmarks}
\usetikzlibrary{arrows.meta}
\usepgfplotslibrary{patchplots}
\usepackage{amsmath}

\newcommand{\ouralgorithm}[0]{NuevoMatch}
\newcommand{\isets}[0]{iSets}

\begin{document}
\fi

\definecolor{patternFg}{HTML}{292929}
\definecolor{text}{HTML}{292929}
\definecolor{linecolor}{HTML}{888888}

\definecolor{nmcs}{HTML}{d18710}
\definecolor{nmtm}{HTML}{de3b3b}
\definecolor{nmnc}{HTML}{007ec2}

\definecolor{barborder}{HTML}{2b2b2b}
\definecolor{graybg}{HTML}{bbbbbb}

\pgfmathsetmacro{\barwidth}{0.2}
\pgfmathsetmacro{\txtshift}{0.1}
\pgfmathsetmacro{\groupshift}{0.15}

\begin{tikzpicture}
\usetikzlibrary{patterns}

\pgfplotsset{
	at={(0in,0in)},
	ymode=log,
	width=2.6in,
	height=0.613in,
	xmin=0.5, xmax=6.5,
	ymin=0.5, ymax=2.7,
	scale only axis,
	legend style={
		legend cell align=left,
		align=center,
		at={(0.5, -0.6)},
		anchor=north,
		font={\small},
		legend columns=3
	}
}
	
\begin{axis}[
xmajorgrids, ymajorgrids, yminorgrids,
yminorticks=true,
ylabel style={font=\small, align=center},
xlabel style={font=\footnotesize, align=center},
ticklabel style={font=\footnotesize, align=center},
xtick={1,2,3,4,5,6,7,8},
ytick={0.5,1,1.5,2,2.5},
xticklabels={{Zipf 80\%\\($\alpha$=1.05)}, {Zipf 85\%\\($\alpha$=1.10)}, {Zipf 90\%\\($\alpha$=1.15)}, {Zipf 95\%\\($\alpha$=1.25)}, {CAIDA}, {CAIDA*}},
ylabel={Throughput\\Speedup}]

\addplot[draw=none, opacity=0.5, fill=graybg] table {
4.5 0.5
8.5 0.5
8.5 2.8
4.5 2.8
};

\addplot [color=linecolor, line width={0.8 pt}] table{
0.5 1
8.5 1
};

\node [anchor=south, font={\scriptsize \bf}, align=center, color=text]
at (axis cs:1-\groupshift+\txtshift, 1.96) {2.06$\times$};
\node [anchor=south, font={\scriptsize \bf}, align=center, color=text]
at (axis cs:1+\groupshift+\txtshift, 1.04) {1.14$\times$};

\node [anchor=south, font={\scriptsize \bf}, align=center, color=text]
at (axis cs:2-\groupshift+\txtshift, 1.85) {1.95$\times$};
\node [anchor=south, font={\scriptsize \bf}, align=center, color=text]
at (axis cs:2+\groupshift+\txtshift, 0.96) {1.06$\times$};

\node [anchor=south, font={\scriptsize \bf}, align=center, color=text]
at (axis cs:3-\groupshift+\txtshift, 1.74) {1.84$\times$};
\node [anchor=south, font={\scriptsize \bf}, align=center, color=text]
at (axis cs:3+\groupshift+\txtshift, 0.89) {0.99$\times$};

\node [anchor=south, font={\scriptsize \bf}, align=center, color=text]
at (axis cs:4-\groupshift+\txtshift, 1.52) {1.62$\times$};
\node [anchor=south, font={\scriptsize \bf}, align=center, color=text]
at (axis cs:4+\groupshift+\txtshift, 0.79) {0.89$\times$};

\node [anchor=south, font={\scriptsize \bf}, align=center, color=text]
at (axis cs:5-\groupshift+\txtshift, 1.69) {1.79$\times$};
\node [anchor=south, font={\scriptsize \bf}, align=center, color=text]
at (axis cs:5+\groupshift+\txtshift, 0.95) {1.05$\times$};

\node [anchor=south, font={\scriptsize \bf}, align=center, color=text]
at (axis cs:6-\groupshift+\txtshift, 2.16) {2.26$\times$};
\node [anchor=south, font={\scriptsize \bf}, align=center, color=text]
at (axis cs:6+\groupshift+\txtshift, 1.06) {1.16$\times$};

\end{axis}

\begin{axis}[
axis lines=none,
ticks=none
]


\addlegendimage{fill=nmcs, draw=barborder, area legend, postaction={pattern color=patternFg, pattern=grid}};
\addlegendentry{NuevoMatch w/ CutSplit};

\addlegendimage{fill=nmtm, draw=barborder, area legend, postaction={pattern color=patternFg, pattern=crosshatch dots}};
\addlegendentry{NuevoMatch w/ TupleMerge};

\addplot[ybar, bar width=\barwidth, bar shift=-\groupshift, fill=nmcs, draw=barborder, area legend, postaction={pattern color=patternFg, pattern=grid}] table{
1 2.06317
2 1.9513
3 1.84547
4 1.62894
5 1.79875
6 2.26189
};

\addplot[ybar, bar width=\barwidth, bar shift=+\groupshift, fill=nmtm, draw=barborder, area legend, postaction={pattern color=patternFg, pattern=crosshatch dots}] table{
1 1.14015
2 1.06487
3 0.999214
4 0.89831
5 1.05402
6 1.16732
};

\end{axis}

\end{tikzpicture}
	
\ifdefined\RELEASE
\else
\end{document}
\fi
	\caption{\changes{ClassBench: \ouralgorithm{} vs. CutSplit and TupleMerge with skewed traffic.}}
	\label{fig:zipf}
\end{figure}
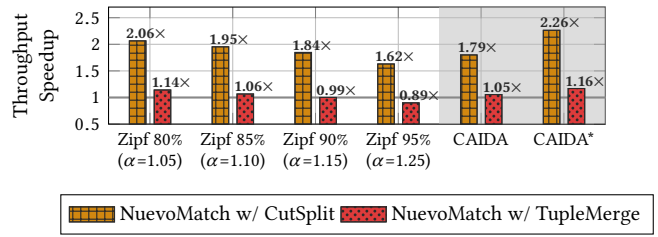

\subsubsection{Memory footprint comparison} \label{sec:eval_compression}

Figure \ref{fig:method_size_comparison} compares the memory footprint of the classifiers without and with \ouralgorithm{} (the two right-most bars in each bar cluster). We use the same number of \isets as in the end-to-end experiments. Note that a smaller footprint alone does not necessarily lead to higher performance if more \isets are used. Therefore, the results should be considered in conjunction with the end-to-end performance. 

The memory footprint includes only the index data structures but not the rules themselves. In particular, the memory footprint for \ouralgorithm{} includes both the \MSNN models and the remainder classifier. Each bar is the average of all the 12 application rule-sets of the same size. 

For \NM we show both the remainder index size (middle bar) and the total \MSNN size (right-most bar). Note that due to the logarithmic scale of the Y axis, the actual ratio bewteen the two is much higher than it might seem. For example, the remainder for 10K \TM is almost 100$\times$ the size of the \MSNNs. Note also that since we run \NM on two cores, both \MSNN and the remainder classifier use their own CPU caches.

Overall,  \ouralgorithm{} enables dramatic compression of the memory footprint, in particular for 500K rule-sets, with 4.9$\times$, 8$\times$, and 82$\times$ on average over \CS, \NC and \TM respectively.

The graph explains well the end-to-end performance results. For 1K rule-sets, the original classifiers fit into the L1 cache, so \NM is not effective. For 10K sets, even though the remainder index fits in L1, the ratio between L1 and L2 performance is insufficient to cover the \MSNN  overheads. For 100K, the situation is similar for \CS; however, for \NC, the remainder fits in L1, whereas the original \NC spills to L3. For \TM, the remainder is already in L2, yielding a lower overall speedup compared to \NC. Last, for 500K rule-sets, all the original classifiers spill to L3, whereas the remainder fits well in L2, yielding clear performance improvements.

\paragraph{Performance under L3 cache contention}
The small memory footprint of \NM plays an important role even when the rule-index fits in the L3 cache (16MB in our machine). L3 is shared among all the CPU cores; therefore, cache contention is not rare, in particular in data centers. \NM reduces the effects of L3 cache contention on  packet classification performance. In the experiment we use the 500\.K rule-set (1) and compare the performance of \CS and \NM (with \CS) while limiting the L3 to 1.5MB. \CS loses half of its performance, whereas \NM slows down by 30\%, increasing the original speedup. 

\begin{figure}[t]
	\centering
\ifdefined\RELEASE
\else
\documentclass[conference,compsoc]{IEEEtran}

\usepackage{graphicx}
\usepackage{subfig}

\usepackage[T1]{fontenc}
\usepackage[utf8]{inputenc}
\usepackage{pgfplots}
\usepackage{grffile}
\pgfplotsset{compat=newest}
\usetikzlibrary{plotmarks}
\usetikzlibrary{arrows.meta}
\usepgfplotslibrary{patchplots}
\usepackage{amsmath}

\newcommand{\ouralgorithm}[0]{NuevoMatch}
\newcommand{\isets}[0]{iSets}
\begin{document}
\fi
	
\definecolor{pattern_fg}{HTML}{292929}
\definecolor{linecolor}{HTML}{002561}

\definecolor{isets}{HTML}{de3b3b}
\definecolor{nuevomatch}{HTML}{d18710}
\definecolor{tuplemerge}{HTML}{00264A}
\definecolor{neurocuts}{HTML}{007ec2}
\definecolor{cutsplit}{HTML}{ABCDDE}

\definecolor{barborder}{HTML}{2b2b2b}

\pgfmathsetmacro{\barwidth}{0.074}
\pgfmathsetmacro{\groupshift}{0.32}
\pgfmathsetmacro{\nmshift}{0.076}

\begin{tikzpicture}
\usetikzlibrary{patterns}

\pgfplotsset{
width=2.6in,
height=0.8in,
xmin=0.6, xmax=4.53,
ymin=100, ymax=15000000,
scale only axis,
legend style={legend cell align=left, align=left, at={(0.45, -0.4)}, anchor=north, font={\small}, legend columns=3}
}

\begin{axis}[
ymode=log,
xmajorgrids, ymajorgrids, yminorgrids,
yminorticks=true,
ylabel style={font=\small, align=center},
xlabel style={font=\small},
ticklabel style={font=\small},
xtick={1,2,3,4},
ytick={1e2,1e3,1e4,1e5,1e6,1e7},
xticklabels={{1K},{10K},{100K},{500K}},
ylabel={Size (Bytes)},
xlabel={Number of rules}]

\addplot[draw=none, opacity=0.5, fill=gray] table {
1.55 1e2
2.57 1e2
2.57 1e8
1.55 1e8
};

\addplot[draw=none, opacity=0.5, fill=gray] table {
3.55 1e2
4.57 1e2
4.57 1e8
3.55 1e8
};

\addplot [color=linecolor, line width={0.8 pt}] table{
	4.6 32768
	0.6 32768
} node [font={\footnotesize}, anchor=south west, color=pattern_fg] {\bf 32KB L1 Cache Size};

\addplot [color=linecolor, line width={0.8 pt}] table{
	4.6 1048576
	0.6 1048576
} node [font={\footnotesize}, anchor=south west, color=pattern_fg] {\bf 1024KB L2 Cache Size};

\end{axis}

\begin{axis}[%
xticklabels=\empty,
yticklabels=\empty,
point meta min=1,
point meta max=5,
ymode=log,
]

\addlegendimage{fill=nuevomatch, draw=barborder, postaction={pattern=grid, pattern color=pattern_fg}, area legend};
\addlegendentry{\ouralgorithm{}: Remainder}

\addlegendimage{fill=cutsplit, fill=cutsplit, draw=barborder, postaction={pattern=north east lines, pattern color=pattern_fg}, area legend}; \addlegendentry{CutSplit}

\addlegendimage{fill=neurocuts, draw=barborder, postaction={pattern=horizontal lines, pattern color=pattern_fg}, area legend}; \addlegendentry{NeuroCuts}

\addlegendimage{fill=isets, draw=barborder, postaction={pattern=crosshatch dots, pattern color=pattern_fg}, area legend};
\addlegendentry{\ouralgorithm{}: \isets}

\addlegendimage{fill=tuplemerge, draw=barborder, postaction={pattern=grid, pattern color=pattern_fg}, area legend}; \addlegendentry{TupleMerge}

\addplot[ybar, bar width=\barwidth, fill=cutsplit, draw=barborder, postaction={pattern=north east lines, pattern color=pattern_fg}, area legend, bar shift=-\groupshift] table{
1 3317.67
2 27000
3 206520
4 1.23191e+06
};

\addplot[ybar, bar width=\barwidth, fill=nuevomatch, draw=barborder, postaction={pattern=grid, pattern color=pattern_fg}, area legend, bar shift=-\groupshift+\nmshift] table{
1 2404
2 16981.6
3 45533
4 215079
};

\addplot[ybar, bar width=\barwidth, fill=isets, draw=barborder, postaction={pattern=crosshatch dots, pattern color=pattern_fg}, area legend, bar shift=-\groupshift+2*\nmshift] table{
1 617.5
2 2279.67
3 19184
4 38179.7
};

\addplot[ybar, bar width=\barwidth, fill=neurocuts, draw=barborder, postaction={pattern=horizontal lines, pattern color=pattern_fg}, area legend] table{
1 10544.3
2 63002.3
3 1.02418e+06
4 3.06251e+06
};

\addplot[ybar, bar width=\barwidth, fill=nuevomatch, draw=barborder, postaction={pattern=grid, pattern color=pattern_fg}, area legend, bar shift=+\nmshift] table{
1 2964
2 154461
3 102500
4 343649
};

\addplot[ybar, bar width=\barwidth, fill=isets, draw=barborder, postaction={pattern=crosshatch dots, pattern color=pattern_fg}, area legend, bar shift=+2*\nmshift] table{
1 617.5
2 2279.67
3 19163
4 38117.4
};

\addplot[ybar, bar width=\barwidth, fill=tuplemerge, draw=barborder, postaction={pattern=grid, pattern color=pattern_fg}, area legend, bar shift=+\groupshift] table{
1 18625.3
2 193864
3 2033155
4 10185237
};

\addplot[ybar, bar width=\barwidth, fill=nuevomatch, draw=barborder, postaction={pattern=grid, pattern color=pattern_fg}, area legend, bar shift=+\groupshift+\nmshift] table{
1 11813.7
2 69725.7
3 40247
4 59601.2
};

\addplot[ybar, bar width=\barwidth, fill=isets, draw=barborder, postaction={pattern=crosshatch dots, pattern color=pattern_fg}, area legend, bar shift=+\groupshift+2*\nmshift] table{
1 1605.5
2 4929.83
3 35094.5
4 63989.8
};

\end{axis}

\end{tikzpicture}
\ifdefined\RELEASE
\else
\end{document}
\fi
	\caption{Memory size for CutSplit, NeuroCuts, TupleMerge vs.  \ouralgorithm{} with them indexing the remainder. Each bar is a geometric mean of 12 applications.}
	\label{fig:method_size_comparison}
\end{figure}
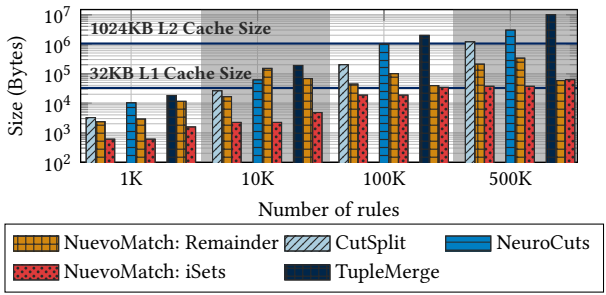

\subsection{Performance analysis}
\label{sec:evaluation:analys}

\subsubsection{\iset coverage} \label{sec:eval_rule_mixture}
 Table \ref{table:iset_analysis} shows the cumulative coverage achieved with up to 4 \isets  averaged over 12 rule-sets (ClassBench) of the same size.  The coverage of smaller rule-sets is worse on average, but improves with the size of the rule-set. 
 
 The last row shows a representative result for the Stanford backbone rule-set (the other three differ within 1\%). Two \isets are enough to achieve 90\% coverage and three are needed for 95\%. This data set differs from  ClassBench in that it contains only one field, providing fewer opportunities for \iset partitioning.

\begin{table}[t]
\small
	\caption{\iset coverage.}
	\label{table:iset_analysis}
	\begin{tabular}{c|cccc}    
		\toprule
		\midrule
		\multicolumn{1}{c}{} & 1 \iset & 2 \isets & 3 \isets & 4 \isets \\
		\midrule
        1K    & $20.2\pm18.6$ & $28.9\pm22.3$ & $34.6\pm25.6$ & $38.7\pm27.2$ \\
        10K   & $45.1\pm31.6$ & $59.6\pm38.9$ & $62.6\pm37.1$ & $65.1\pm35.7$ \\
        100K  & $80.0\pm14.5$ & $96.5\pm8.3$ & $98.1\pm4.8$ & $98.8\pm2.7$ \\
        500K  & $84.2\pm10.5$ & $98.8\pm1.5$ & $99.4\pm0.6$ & $99.7\pm0.2$ \\
		\midrule
		\midrule
        183,376 & 57.8 & 91.6 & 96.5 & 98.2 \\
		\bottomrule
	\end{tabular}
\end{table}

\subsubsection{Impact of the number of \isets } \label{sec:evaluation:analysis}
We seek to understand the tradeoff between the \iset coverage of the rule-set and the computational overheads of adding more \MSNNs.  All computations were performed on a single core to obtain the latency breakdown. We use \CS for indexing the remainder.
\begin{figure}[t]
	\centering
\ifdefined\RELEASE
\else
\documentclass[conference,compsoc]{IEEEtran}

\usepackage{graphicx}
\usepackage{subfig}
\usepackage[utf8]{inputenc}
\usepackage{xspace}

\usepackage[T1]{fontenc}
\usepackage[utf8]{inputenc}
\usepackage{pgfplots}
\usepackage{grffile}
\pgfplotsset{compat=newest}
\usetikzlibrary{plotmarks}
\usetikzlibrary{arrows.meta}
\usepgfplotslibrary{patchplots}
\usepackage{amsmath}

\newcommand{\ouralgorithm}[0]{Plasmus}
\newcommand{\iset}[0]{iSet\xspace}
\newcommand{\isets}[0]{iSets\xspace}
\newcommand{\capital}[0]{\expandafter\MakeUppercase}
\begin{document}
\fi

\definecolor{coverage}{HTML}{8f0c0a}

\definecolor{remainder}{HTML}{4ec0f9}
\definecolor{remainder_pattern}{HTML}{077cb6}

\definecolor{validation}{HTML}{d46a08}
\definecolor{validation_pattern}{HTML}{673404}

\definecolor{search}{HTML}{48913b}
\definecolor{search_pattern}{HTML}{20411b}

\definecolor{inference}{HTML}{c52b2b}
\definecolor{inference_pattern}{HTML}{641616}

\pgfmathsetmacro{\markWidth}{1.5}

\usetikzlibrary{patterns}

\begin{tikzpicture}

\pgfplotsset{
	width=2.6in,
	height=0.8in,
	scale only axis,
	xmin=0, xmax=6,
	xtick={0,1,2,3,4,5,6},
	legend style={at={(0.5,-0.4)}, anchor=north, legend cell align=left, align=left, draw=white!15!black, font={\small}, legend columns=3},
	ylabel style={font=\small},
	xlabel style={font=\small},
	ylabel style = {align=center},
	every tick label/.append style={font=\small},
}
	%


\begin{axis}[
at={(0in,1in)},
ylabel={Time (ns)},
xlabel={Number of \isets},
xmajorgrids, ymajorgrids, ymin=0, ymax=1000
]

\addplot[draw=remainder,fill=remainder,opacity=0.6,area legend, postaction={pattern=dots,pattern color=remainder_pattern}] table{
0 559.988
1 231.383
2 154.648
3 156.246
4 135.749
5 127.183
6 136.024
} \closedcycle;

\addplot[fill=search, draw=search,opacity=0.6,area legend, postaction={pattern=north east lines,pattern color=search_pattern}] table{
0 559.988
1 298.294
2 253.882
3 283.481
4 293.729
5 319.298
6 378.454
6 136.024
5 127.183
4 135.749
3 156.246
2 154.648
1 231.383
0 559.988
} \closedcycle;

\addplot[draw=validation,fill=validation,opacity=0.6,area legend,postaction={pattern=vertical lines,pattern color=validation_pattern}] table{
0 559.988
1 362.367
2 340.904
3 395.381
4 433.97
5 471.387
6 550.777
6 378.454
5 319.298
4 293.729
3 283.481
2 253.882
1 298.294
0 559.988
} \closedcycle;

\addplot[fill=inference, draw=inference,opacity=0.6,area legend,postaction={pattern=crosshatch dots,pattern color=inference_pattern}] table{
0 559.988
1 408.246
2 423.694
3 504.193
4 574.48
5 639.627
6 747.393
6 550.777
5 471.387
4 433.97
3 395.381
2 340.904
1 362.367
0 559.988
} \closedcycle;

\end{axis}

\begin{axis}[
at={(0in,1in)},
ticks=none,
axis y line*=right,
ylabel={Coverage (\%)}, ymin=0, ymax=100,
]

\addlegendimage{area legend, opacity=0.6, fill=remainder, postaction={pattern=dots,pattern color=remainder_pattern}};
\addlegendentry{Remainder};

\addlegendimage{area legend, opacity=0.6, fill=search, postaction={pattern=north east lines,pattern color=search_pattern}};
\addlegendentry{Secondary Search};

\addlegendimage{area legend, opacity=0.6, fill=validation,postaction={pattern=vertical lines,pattern color=validation_pattern}};
\addlegendentry{Validation};

\addlegendimage{area legend, opacity=0.6, fill=inference,postaction={pattern=crosshatch dots,pattern color=validation_pattern}};
\addlegendentry{Inference};

\addlegendimage{mark=x, color=coverage, mark size={\markWidth pt}};
\addlegendentry{Coverage};

\addplot[mark=x, mark options={}, mark size={\markWidth}, draw=coverage] table{
0 0
1 84.1842
2 98.64
3 99.282
4 99.613
5 99.743
6 99.74
};

\end{axis}

\end{tikzpicture}

\ifdefined\RELEASE
\else
\end{document}
\fi
	\caption{Coverage and execution time breakdown of \ouralgorithm{} vs. varying number of \isets.}
	\label{fig:iset_number_analysis}
\end{figure}
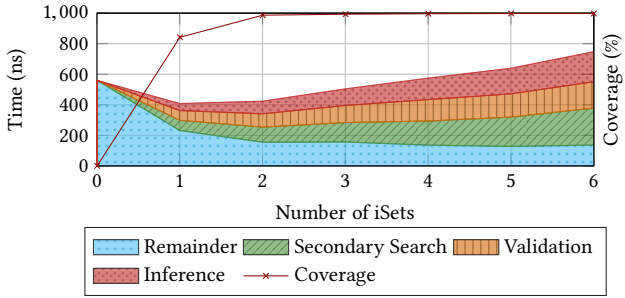

Figure~\ref{fig:iset_number_analysis} shows the geometric mean of the coverage  and the runtime breakdown over 12 rule-sets of 500K. The breakdown includes the runtime of the remainder classifier,  validation, secondary search, and \MSNN inference. Zero \isets implies that \CS was run alone. Adding more \isets shows diminishing returns because of their compute overhead, which is not compensated by the remainder runtime improvements because the coverage is already saturated to almost 100\%. Using one or two \isets shows the best trade-off. \NC shows similar results.

\TM behaved differently (not shown). \TM occupies much more memory than \CS; therefore, using more \isets to achieve higher coverage allowed us to further speed up the remainder by fitting it into an upper level cache. Thus, 4 \isets showed the best configuration.

We note that the runtime is split nearly equally between model inference and validation (which are compute-bound parts), and the secondary search and the remainder computations (which are memory-bound). We expect the compute performance of future processors to scale better than their cache capacity and memory access latency. Therefore, we believe \NM will provide better scaling than memory-bound state-of-the-art classifiers.

\subsubsection{Partitioning effectiveness} \label{sec:partitionining_effectiveness}

We seek to understand how low diversity rule-sets affect \ouralgorithm{}.
To analyze that, we synthetically generated a large rule-set as a Cartesian product of a small number of values per field (no ranges). We  blended them into a 500K ClassBench rule-set, replacing randomly selected rules with those from the Cartesian product,  while keeping the total number of rules the same.

Table \ref{table:cartesian_rules} shows the coverage and the speedup over \TM on the resulting mixed rule-sets for different fractions of low-diversity rules. 
%
The partitioning algorithm successfully segregates the low-diversity rules the best, achieving the coverage inversely proportional to their fraction in the rule-set. Note that  \ouralgorithm{} becomes effective when it offloads the processing of about 25\% of the rules.

\begin{table}[t]
\small
    \caption{Throughput and a single \iset coverage vs. the fraction of low-diversity rules in a 500K rule-set.}
    \label{table:cartesian_rules}
    \begin{tabular}{cccc}    
		\toprule
		\% Low diversity rules  & \% Coverage & Speedup (throughput)\\
		\midrule
		70\% & 25\% & 1.07$\times$ \\
		50\% & 50\% & 1.14$\times$ \\ 
		30\% & 70\% & 1.60$\times$ \\
		\bottomrule
	\end{tabular}
\end{table}

\subsubsection{Training time and secondary search range}
\label{sec:training_time}
\MSNN{}s are trained to minimize the prediction error bound to achieve a small secondary search distance. Recall that a secondary search involves a binary search within the error bound, where each rule is validated to match all the fields. 

The tradeoff between training time and secondary search performance is not trivial. A larger search distance enables faster training but slows down the secondary search. A smaller search distance results in a faster search but slows down the training. In extreme cases, the training does not converge, since a higher precision might require larger submodels. However, increasing the size of the submodels leads to a larger memory footprint and longer computations. 

Figure \ref{fig:error_training_time} shows the average \changes{end-to-end} training time in minutes of 500 models as a function of the secondary search distance and the rule-set size. \changes{The measurements include all training iterations as described in \S\ref{sec:full_msnn_training}.} As mentioned (\S\ref{sec:implmentation}), our training implementation can be dramatically accelerated, so the results here indicate the general trend. 

\begin{figure}
	\centering
\ifdefined\RELEASE
\else
\documentclass[conference,compsoc]{IEEEtran}

\usepackage{graphicx}
\usepackage{subfig}
\usepackage[utf8]{inputenc}
\usepackage{xspace}

\usepackage[T1]{fontenc}
\usepackage[utf8]{inputenc}
\usepackage{pgfplots}
\usepackage{grffile}
\pgfplotsset{compat=newest}
\usetikzlibrary{plotmarks}
\usetikzlibrary{arrows.meta}
\usepgfplotslibrary{patchplots}
\usepackage{amsmath}

\newcommand{\ouralgorithm}[0]{Plasmus}
\newcommand{\iset}[0]{iSet\xspace}
\newcommand{\isets}[0]{iSets\xspace}
\newcommand{\capital}[0]{\expandafter\MakeUppercase}
\begin{document}
	\fi

\definecolor{error64}{HTML}{0071bd}
\definecolor{traiaing100k}{HTML}{e68200}
\definecolor{traiaing500k}{HTML}{bd1000}
\definecolor{traiaing10k}{HTML}{4554de}

\begin{tikzpicture}

\pgfmathsetmacro{\lineWidth}{1.1}

\pgfplotsset{
	width=2.6in,
	height=0.6in,
	scale only axis,
	ylabel style={font=\small, align=center},
	xlabel style={font=\small},
	xlabel={Search Distance Bounds},
	ticklabel style={font=\small},
	xtick={0.5,1,2,4,8,16,32,64,128,256,512,1024},
	xticklabels={0,1,2,4,8,16,32,64,128,256,512,1024},
	legend style={legend cell align=left, align=left, draw=white!15!black, at={(1, 1)}, anchor=north east, font={\scriptsize}, legend columns=1},
	xmajorgrids, ymajorgrids,
}

\begin{semilogxaxis}[
log ticks with fixed point,
ylabel={Training\\Time\\(minutes)},
xmin=64, xmax=1024,
ymin=0,
ytick={0,10,20,30,40}
]

\addplot [mark=o, color=traiaing500k, line width={\lineWidth}, mark size=1.2] table{
1024 8.43333
512 8.48333
256 9.21667
128 10.3667
64 42.1667
};

\addplot [mark=triangle, color=traiaing100k, line width={\lineWidth}, mark size=2] table{
1024 4.26667
512 4.31667
256 4.45
128 4.58333
64 17.9167
};

\addplot [mark=square, color=traiaing10k, line width={\lineWidth}, mark size=1.2] table{
1024 0.683333
512 0.683333
256 0.7
128 0.716667
64 2.48333
};

\addlegendentry{500K rules}
\addlegendentry{100K rules}
\addlegendentry{10K rules}

\end{semilogxaxis}

\end{tikzpicture}%

\ifdefined\RELEASE
\else
\end{document}
\fi
	\caption{\MSNN training time in minutes vs. maximum search range bound.}
	\label{fig:error_training_time}
\end{figure}
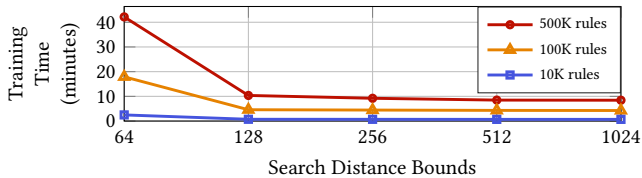

Training with the bound of 64 is expensive, but is it really necessary?  To answer,  we  evaluate the performance impact of the search distance on the secondary search time. We measure 40\.ns for retrieving a rule with a precise prediction (no search). 
For 64, 128 and 256 distances the search time varies between 75 to 80\.ns thanks to the binary search. Last, it turns out that the \emph{actual} search distance from the predicted index is often much smaller than the worst-case one enforced in training. Our analysis shows that in practice, training with a relatively large bound of 128 leads to 80\% of the lookups with a search distance of 64, and 60\% with 32.

We conclude that training with larger bounds is likely to have a minor effect on the end-to-end performance, but significantly accelerate training. This property is important to support more frequent retraining and faster updates~(\S\ref{sec:updates}).

\subsubsection{Performance with more fields} 

Adding fields to an existing classifier will not harm its coverage, so it will not affect the \MSNN performance. 
Nonetheless, more fields will increase validation time. 

Unfortunately, we did not find public rule-sets that have a large number of fields. Thus, we ran a microbenchmark by increasing the number of fields and measuring the validation stage performance. 
As expected, we observed almost linear growth in the validation time, from 25ns for one field to 180ns for 40 fields.

\section{Related Work}

\paragraph{Hardware-based classifiers}
Hardware-based solutions for classification such as TCAMs and FPGAs achieve a
very high throughput ~\cite{Spitznagel2003, Firestone2018}. Consequently, many
software algorithms take advantage of them, further improving classification
performance \cite{Taylor2005, Lakshminarayanan2005, Narodytska2019, Liu2008,
Katta2016, Rottenstreich2017, Ma2012}. Our work is complementary, but can be
used to improve scaling of these solutions. For example, if the original
classifier required large TCAMs, the remainder set would fit a much smaller
TCAM.

\paragraph{GPUs for classification}
Accelerating classification on GPUs was suggested by numerous works.
PacketShader~\cite{Han2010} uses GPU for packet forwarding and provides integration with Open vSwitch. 
However, packet forwarding is a single-dimensional problem, so it is easier than multi-field classification \cite{Gupta2001}. 
Varvello et al. \cite{Varvello2016} implemented various packet classification algorithms in GPUs, 
including linear search, Tuple Space Search, and bloom search. Nonetheless,
these techniques suffer from poor scalability for large classifiers with
wildcard rules, which \ouralgorithm aims to
alleviate.

\paragraph{ML techniques for networking}
Recent works suggest using ML techniques for solving networking problems, such as TCP congestion control \cite{Zarchy2018, Jay2018, Yasir2015}, 
resource management \cite{Mao2016}, quality of experience in video streaming \cite{Hongzi2017, Hyunho2018},
routing \cite{Valadarsky2017}, 
and decision tree optimization for packet classification \cite{Liang2019}.
\ouralgorithm{} is different in that it uses an ML technique for building 
space-efficient representations of the rules that fit in the CPU cache.

\section{Conclusions}
We have presented \ouralgorithm{}, the first packet classification technique that uses \emph{Range-Query RMI} machine learning model for \emph{accelerating} packet classification. We have shown an efficient way of training \MSNN models, making them learn the matching ranges of large rule-sets, via sampling and analytical error bound computations. We demonstrated the application of \MSNNs to multi-field packet classification using rule-set partitioning. We evaluated \ouralgorithm{} on synthetic and real-world rule-sets and confirmed its benefits for large rule-sets over state-of-the-art techniques.

\ouralgorithm introduces a new point in the design space of packet classification algorithms and opens up new ways to scale it on commodity processors. We believe that its compute-bound nature and the use of neural networks will enable further scaling with future CPU generations, which will feature powerful compute capabilities targeting faster execution of neural network-related computations.

\changes{
 \section{Acknowledgements} 
We thank the anonymous reviewers of SIGCOMM'20 and our shepherd Minlan Yu for their helpful comments and feedback.
We would also like to thank Isaac Keslassy and Leonid Ryzhyk for their feedback on the early draft of the paper.

This work was partially supported by the Technion Hiroshi Fujiwara Cyber Security Research Center and the Israel National Cyber Directorate, by the Alon fellowship and by the Taub Family Foundation. We gratefully acknowledge support from Israel Science Foundation (Grant 1027/18) and Israeli Innovation Authority.
}

\bibliographystyle{ACM-Reference-Format}
\bibliography{bibliography}


\begin{thebibliography}{46}


\ifx \showCODEN    \undefined \def \showCODEN     #1{\unskip}     \fi
\ifx \showDOI      \undefined \def \showDOI       #1{#1}\fi
\ifx \showISBNx    \undefined \def \showISBNx     #1{\unskip}     \fi
\ifx \showISBNxiii \undefined \def \showISBNxiii  #1{\unskip}     \fi
\ifx \showISSN     \undefined \def \showISSN      #1{\unskip}     \fi
\ifx \showLCCN     \undefined \def \showLCCN      #1{\unskip}     \fi
\ifx \shownote     \undefined \def \shownote      #1{#1}          \fi
\ifx \showarticletitle \undefined \def \showarticletitle #1{#1}   \fi
\ifx \showURL      \undefined \def \showURL       {\relax}        \fi
\providecommand\bibfield[2]{#2}
\providecommand\bibinfo[2]{#2}
\providecommand\natexlab[1]{#1}
\providecommand\showeprint[2][]{arXiv:#2}

\bibitem[\protect\citeauthoryear{Abadi, Barham, Chen, Chen, Davis, Dean, Devin,
  Ghemawat, Irving, Isard, Kudlur, Levenberg, Monga, Moore, Murray, Steiner,
  Tucker, Vasudevan, Warden, Wicke, Yu, and Zheng}{Abadi et~al\mbox{.}}{2016}]%
        {tensor_flow}
\bibfield{author}{\bibinfo{person}{Mart{\'\i}n Abadi}, \bibinfo{person}{Paul
  Barham}, \bibinfo{person}{Jianmin Chen}, \bibinfo{person}{Zhifeng Chen},
  \bibinfo{person}{Andy Davis}, \bibinfo{person}{Jeffrey Dean},
  \bibinfo{person}{Matthieu Devin}, \bibinfo{person}{Sanjay Ghemawat},
  \bibinfo{person}{Geoffrey Irving}, \bibinfo{person}{Michael Isard},
  \bibinfo{person}{Manjunath Kudlur}, \bibinfo{person}{Josh Levenberg},
  \bibinfo{person}{Rajat Monga}, \bibinfo{person}{Sherry Moore},
  \bibinfo{person}{Derek~G. Murray}, \bibinfo{person}{Benoit Steiner},
  \bibinfo{person}{Paul Tucker}, \bibinfo{person}{Vijay Vasudevan},
  \bibinfo{person}{Pete Warden}, \bibinfo{person}{Martin Wicke},
  \bibinfo{person}{Yuan Yu}, {and} \bibinfo{person}{Xiaoqiang Zheng}.}
  \bibinfo{year}{2016}\natexlab{}.
\newblock \showarticletitle{TensorFlow: A System for Large-Scale Machine
  Learning}. In \bibinfo{booktitle}{\emph{{USENIX} {OSDI}}}.
\newblock


\bibitem[\protect\citeauthoryear{CAIDA}{CAIDA}{[n.d.]}]%
        {CAIDA}
\bibfield{author}{\bibinfo{person}{CAIDA}.} \bibinfo{year}{[n.d.]}\natexlab{}.
\newblock \bibinfo{booktitle}{\emph{The {CAIDA} {UCSD} Anonymized Internet
  Traces 2019}}.
\newblock
\urldef\tempurl%
\url{http://www.caida.org/data/passive/passive_dataset.xml}
\showURL{%
Retrieved June 15, 2020 from \tempurl}


\bibitem[\protect\citeauthoryear{Daly, Bruschi, Linguaglossa, Pontarelli,
  Rossi, Tollet, Torng, and Yourtchenko}{Daly et~al\mbox{.}}{2019}]%
        {Daly2019}
\bibfield{author}{\bibinfo{person}{James Daly}, \bibinfo{person}{Valerio
  Bruschi}, \bibinfo{person}{Leonardo Linguaglossa}, \bibinfo{person}{Salvatore
  Pontarelli}, \bibinfo{person}{Dario Rossi}, \bibinfo{person}{Jerome Tollet},
  \bibinfo{person}{Eric Torng}, {and} \bibinfo{person}{Andrew Yourtchenko}.}
  \bibinfo{year}{2019}\natexlab{}.
\newblock \showarticletitle{TupleMerge: Fast Software Packet Processing for
  online Packet Classification}.
\newblock \bibinfo{journal}{\emph{IEEE/ACM Transactions on Networking ({TON})}}
  \bibinfo{volume}{27}, \bibinfo{number}{4} (\bibinfo{year}{2019}),
  \bibinfo{pages}{1417--1431}.
\newblock


\bibitem[\protect\citeauthoryear{Dong, Meng, Zarchy, Arslan, Gilad, Godfrey,
  and Schapira}{Dong et~al\mbox{.}}{2018}]%
        {Zarchy2018}
\bibfield{author}{\bibinfo{person}{Mo Dong}, \bibinfo{person}{Tong Meng},
  \bibinfo{person}{Doron Zarchy}, \bibinfo{person}{Engin Arslan},
  \bibinfo{person}{Yossi Gilad}, \bibinfo{person}{Brighten Godfrey}, {and}
  \bibinfo{person}{Michael Schapira}.} \bibinfo{year}{2018}\natexlab{}.
\newblock \showarticletitle{{PCC} Vivace: Online-Learning Congestion Control}.
  In \bibinfo{booktitle}{\emph{USENIX NSDI}}.
\newblock


\bibitem[\protect\citeauthoryear{Firestone}{Firestone}{2017}]%
        {Firestone2017}
\bibfield{author}{\bibinfo{person}{Daniel Firestone}.}
  \bibinfo{year}{2017}\natexlab{}.
\newblock \showarticletitle{{VFP}: A Virtual Switch Platform for Host {SDN} in
  the Public Cloud}. In \bibinfo{booktitle}{\emph{{USENIX} {NSDI}}}.
\newblock


\bibitem[\protect\citeauthoryear{Firestone, Putnam, Mundkur, Chiou, Dabagh,
  Andrewartha, Angepat, Bhanu, Caulfield, Chung, Chandrappa, Chaturmohta,
  Humphrey, Lavier, Lam, Liu, Ovtcharov, Padhye, Popuri, Raindel, Sapre, Shaw,
  Silva, Sivakumar, Srivastava, Verma, Zuhair, Bansal, Burger, Vaid, Maltz, and
  Greenberg}{Firestone et~al\mbox{.}}{2018}]%
        {Firestone2018}
\bibfield{author}{\bibinfo{person}{Daniel Firestone}, \bibinfo{person}{Andrew
  Putnam}, \bibinfo{person}{Sambrama Mundkur}, \bibinfo{person}{Derek Chiou},
  \bibinfo{person}{Alireza Dabagh}, \bibinfo{person}{Mike Andrewartha},
  \bibinfo{person}{Hari Angepat}, \bibinfo{person}{Vivek Bhanu},
  \bibinfo{person}{Adrian~M. Caulfield}, \bibinfo{person}{Eric~S. Chung},
  \bibinfo{person}{Harish~Kumar Chandrappa}, \bibinfo{person}{Somesh
  Chaturmohta}, \bibinfo{person}{Matt Humphrey}, \bibinfo{person}{Jack Lavier},
  \bibinfo{person}{Norman Lam}, \bibinfo{person}{Fengfen Liu},
  \bibinfo{person}{Kalin Ovtcharov}, \bibinfo{person}{Jitu Padhye},
  \bibinfo{person}{Gautham Popuri}, \bibinfo{person}{Shachar Raindel},
  \bibinfo{person}{Tejas Sapre}, \bibinfo{person}{Mark Shaw},
  \bibinfo{person}{Gabriel Silva}, \bibinfo{person}{Madhan Sivakumar},
  \bibinfo{person}{Nisheeth Srivastava}, \bibinfo{person}{Anshuman Verma},
  \bibinfo{person}{Qasim Zuhair}, \bibinfo{person}{Deepak Bansal},
  \bibinfo{person}{Doug Burger}, \bibinfo{person}{Kushagra Vaid},
  \bibinfo{person}{David~A. Maltz}, {and} \bibinfo{person}{Albert~G.
  Greenberg}.} \bibinfo{year}{2018}\natexlab{}.
\newblock \showarticletitle{Azure Accelerated Networking: SmartNICs in the
  Public Cloud}. In \bibinfo{booktitle}{\emph{{USENIX} {NSDI}}}.
\newblock


\bibitem[\protect\citeauthoryear{Gupta and McKeown}{Gupta and McKeown}{1999}]%
        {Gupta1999}
\bibfield{author}{\bibinfo{person}{Pankaj Gupta} {and} \bibinfo{person}{Nick
  McKeown}.} \bibinfo{year}{1999}\natexlab{}.
\newblock \showarticletitle{Packet Classification on Multiple Fields}. In
  \bibinfo{booktitle}{\emph{{ACM} {SIGCOMM}}}.
\newblock


\bibitem[\protect\citeauthoryear{Gupta and McKeown}{Gupta and McKeown}{2000}]%
        {Gupta2000}
\bibfield{author}{\bibinfo{person}{Pankaj Gupta} {and} \bibinfo{person}{Nick
  McKeown}.} \bibinfo{year}{2000}\natexlab{}.
\newblock \showarticletitle{Classifying Packets with Hierarchical Intelligent
  Cuttings}.
\newblock \bibinfo{journal}{\emph{IEEE Micro}} \bibinfo{volume}{20},
  \bibinfo{number}{1} (\bibinfo{year}{2000}), \bibinfo{pages}{34--41}.
\newblock


\bibitem[\protect\citeauthoryear{Gupta and McKeown}{Gupta and McKeown}{2001}]%
        {Gupta2001}
\bibfield{author}{\bibinfo{person}{Pankaj Gupta} {and} \bibinfo{person}{Nick
  McKeown}.} \bibinfo{year}{2001}\natexlab{}.
\newblock \showarticletitle{Algorithms for Packet Classification}.
\newblock \bibinfo{journal}{\emph{IEEE Network}} \bibinfo{volume}{15},
  \bibinfo{number}{2} (\bibinfo{year}{2001}), \bibinfo{pages}{24--32}.
\newblock


\bibitem[\protect\citeauthoryear{Han, Jang, Park, and Moon}{Han
  et~al\mbox{.}}{2010}]%
        {Han2010}
\bibfield{author}{\bibinfo{person}{Sangjin Han}, \bibinfo{person}{Keon Jang},
  \bibinfo{person}{KyoungSoo Park}, {and} \bibinfo{person}{Sue Moon}.}
  \bibinfo{year}{2010}\natexlab{}.
\newblock \showarticletitle{PacketShader: {A} {GPU}-accelerated software
  router}. In \bibinfo{booktitle}{\emph{ACM SIGCOMM}}.
\newblock


\bibitem[\protect\citeauthoryear{Intel}{Intel}{2019}]%
        {IntelNervana}
\bibfield{author}{\bibinfo{person}{Intel}.} \bibinfo{year}{2019}\natexlab{}.
\newblock \bibinfo{booktitle}{\emph{Intel Nervana Neural Network Processors}}.
\newblock
\urldef\tempurl%
\url{https://www.intel.ai/nervana-nnp/}
\showURL{%
Retrieved September 25, 2019 from \tempurl}


\bibitem[\protect\citeauthoryear{Jay, Rotman, Godfrey, Schapira, and Tamar}{Jay
  et~al\mbox{.}}{2018}]%
        {Jay2018}
\bibfield{author}{\bibinfo{person}{Nathan Jay}, \bibinfo{person}{Noga~H.
  Rotman}, \bibinfo{person}{Philip~Brighten Godfrey}, \bibinfo{person}{Michael
  Schapira}, {and} \bibinfo{person}{Aviv Tamar}.}
  \bibinfo{year}{2018}\natexlab{}.
\newblock \showarticletitle{Internet Congestion Control via Deep Reinforcement
  Learning}.
\newblock \bibinfo{journal}{\emph{arXiv preprint arXiv:1810.03259}}
  (\bibinfo{year}{2018}).
\newblock


\bibitem[\protect\citeauthoryear{Katta, Alipourfard, Rexford, and Walker}{Katta
  et~al\mbox{.}}{2016}]%
        {Katta2016}
\bibfield{author}{\bibinfo{person}{Naga~Praveen Katta}, \bibinfo{person}{Omid
  Alipourfard}, \bibinfo{person}{Jennifer Rexford}, {and}
  \bibinfo{person}{David Walker}.} \bibinfo{year}{2016}\natexlab{}.
\newblock \showarticletitle{CacheFlow: Dependency-Aware Rule-Caching for
  Software-Defined Networks}. In \bibinfo{booktitle}{\emph{{ACM SOSR}}}.
\newblock


\bibitem[\protect\citeauthoryear{Kingma and Ba}{Kingma and Ba}{2014}]%
        {Kingma2014}
\bibfield{author}{\bibinfo{person}{Diederik~P Kingma} {and}
  \bibinfo{person}{Jimmy Ba}.} \bibinfo{year}{2014}\natexlab{}.
\newblock \showarticletitle{Adam: A Method for Stochastic Optimization}.
\newblock \bibinfo{journal}{\emph{arXiv:1412.6980}} (\bibinfo{year}{2014}).
\newblock


\bibitem[\protect\citeauthoryear{Kleinberg and Tardos}{Kleinberg and
  Tardos}{2006}]%
        {Kleinberg2006}
\bibfield{author}{\bibinfo{person}{Jon~M. Kleinberg} {and}
  \bibinfo{person}{{\'{E}}va Tardos}.} \bibinfo{year}{2006}\natexlab{}.
\newblock \bibinfo{booktitle}{\emph{Algorithm Design}}.
\newblock \bibinfo{publisher}{Addison-Wesley}, \bibinfo{pages}{116--125}.
\newblock
\showISBNx{978-0-321-37291-8}


\bibitem[\protect\citeauthoryear{Kogan, Nikolenko, Rottenstreich, Culhane, and
  Eugster}{Kogan et~al\mbox{.}}{2014}]%
        {KoganKirill2014}
\bibfield{author}{\bibinfo{person}{Kirill Kogan}, \bibinfo{person}{Sergey
  Nikolenko}, \bibinfo{person}{Ori Rottenstreich}, \bibinfo{person}{William
  Culhane}, {and} \bibinfo{person}{Patrick Eugster}.}
  \bibinfo{year}{2014}\natexlab{}.
\newblock \showarticletitle{{SAX-PAC} (Scalable and expressive packet
  classification)}. In \bibinfo{booktitle}{\emph{ACM SIGCOMM}}.
\newblock


\bibitem[\protect\citeauthoryear{Kraska, Alizadeh, Beutel, Chi, Ding, Kristo,
  Leclerc, Madden, Mao, and Nathan}{Kraska et~al\mbox{.}}{2019}]%
        {Kraska2019}
\bibfield{author}{\bibinfo{person}{Tim Kraska}, \bibinfo{person}{Mohammad
  Alizadeh}, \bibinfo{person}{Alex Beutel}, \bibinfo{person}{Ed~H. Chi},
  \bibinfo{person}{Jialin Ding}, \bibinfo{person}{Ani Kristo},
  \bibinfo{person}{Guillaume Leclerc}, \bibinfo{person}{Samuel Madden},
  \bibinfo{person}{Hongzi Mao}, {and} \bibinfo{person}{Vikram Nathan}.}
  \bibinfo{year}{2019}\natexlab{}.
\newblock \showarticletitle{Sage{DB}: A Learned Database System}.
\newblock


\bibitem[\protect\citeauthoryear{Kraska, Beutel, Chi, Dean, and
  Polyzotis}{Kraska et~al\mbox{.}}{2018}]%
        {Kraska2018}
\bibfield{author}{\bibinfo{person}{Tim Kraska}, \bibinfo{person}{Alex Beutel},
  \bibinfo{person}{Ed~H. Chi}, \bibinfo{person}{Jeffrey Dean}, {and}
  \bibinfo{person}{Neoklis Polyzotis}.} \bibinfo{year}{2018}\natexlab{}.
\newblock \showarticletitle{The Case for Learned Index Structures}. In
  \bibinfo{booktitle}{\emph{{ACM} {SIGMOD}}}.
\newblock


\bibitem[\protect\citeauthoryear{Labs}{Labs}{2019}]%
        {Habana}
\bibfield{author}{\bibinfo{person}{Habana Labs}.}
  \bibinfo{year}{2019}\natexlab{}.
\newblock \bibinfo{booktitle}{\emph{Habana {AI} Processors}}.
\newblock
\urldef\tempurl%
\url{https://habana.ai/product}
\showURL{%
Retrieved September 25, 2019 from \tempurl}


\bibitem[\protect\citeauthoryear{Lakshminarayanan, Rangarajan, and
  Venkatachary}{Lakshminarayanan et~al\mbox{.}}{2005}]%
        {Lakshminarayanan2005}
\bibfield{author}{\bibinfo{person}{Karthik Lakshminarayanan},
  \bibinfo{person}{Anand Rangarajan}, {and} \bibinfo{person}{Srinivasan
  Venkatachary}.} \bibinfo{year}{2005}\natexlab{}.
\newblock \showarticletitle{Algorithms for Advanced Packet Classification with
  Ternary {CAM}s}. In \bibinfo{booktitle}{\emph{ACM SIGCOMM}}.
\newblock


\bibitem[\protect\citeauthoryear{Li, Li, Li, and Xie}{Li et~al\mbox{.}}{2018}]%
        {Wenjun2018}
\bibfield{author}{\bibinfo{person}{Wenjun Li}, \bibinfo{person}{Xianfeng Li},
  \bibinfo{person}{Hui Li}, {and} \bibinfo{person}{Gaogang Xie}.}
  \bibinfo{year}{2018}\natexlab{}.
\newblock \showarticletitle{CutSplit: A Decision-Tree Combining Cutting and
  Splitting for Scalable Packet Classification}. In
  \bibinfo{booktitle}{\emph{IEEE INFOCOM}}.
\newblock


\bibitem[\protect\citeauthoryear{Liang, Zhu, Jin, and Stoica}{Liang
  et~al\mbox{.}}{2019}]%
        {Liang2019}
\bibfield{author}{\bibinfo{person}{Eric Liang}, \bibinfo{person}{Hang Zhu},
  \bibinfo{person}{Xin Jin}, {and} \bibinfo{person}{Ion Stoica}.}
  \bibinfo{year}{2019}\natexlab{}.
\newblock \showarticletitle{Neural Packet Classification}. In
  \bibinfo{booktitle}{\emph{ACM SIGCOMM}}.
\newblock


\bibitem[\protect\citeauthoryear{Liu, Meiners, and Zhou}{Liu
  et~al\mbox{.}}{2008}]%
        {Liu2008}
\bibfield{author}{\bibinfo{person}{Alex~X Liu}, \bibinfo{person}{Chad~R
  Meiners}, {and} \bibinfo{person}{Yun Zhou}.} \bibinfo{year}{2008}\natexlab{}.
\newblock \showarticletitle{All-Match Based Complete Redundancy Removal for
  Packet Classifiers in {TCAM}s}. In \bibinfo{booktitle}{\emph{IEEE INFOCOM}}.
\newblock


\bibitem[\protect\citeauthoryear{Ma and Banerjee}{Ma and Banerjee}{2012}]%
        {Ma2012}
\bibfield{author}{\bibinfo{person}{Yadi Ma} {and} \bibinfo{person}{Suman
  Banerjee}.} \bibinfo{year}{2012}\natexlab{}.
\newblock \showarticletitle{A Smart Pre-classifier to Reduce Power Consumption
  of TCAMs for Multi-dimensional Packet Classification}. In
  \bibinfo{booktitle}{\emph{{ACM} {SIGCOMM}}}.
\newblock


\bibitem[\protect\citeauthoryear{Mao, Alizadeh, Menache, and Kandula}{Mao
  et~al\mbox{.}}{2016}]%
        {Mao2016}
\bibfield{author}{\bibinfo{person}{Hongzi Mao}, \bibinfo{person}{Mohammad
  Alizadeh}, \bibinfo{person}{Ishai Menache}, {and} \bibinfo{person}{Srikanth
  Kandula}.} \bibinfo{year}{2016}\natexlab{}.
\newblock \showarticletitle{Resource Management with Deep Reinforcement
  Learning}. In \bibinfo{booktitle}{\emph{{ACM} {SIGCOMM} HotNets Workshop}}.
\newblock


\bibitem[\protect\citeauthoryear{Mao, Netravali, and Alizadeh}{Mao
  et~al\mbox{.}}{2017}]%
        {Hongzi2017}
\bibfield{author}{\bibinfo{person}{Hongzi Mao}, \bibinfo{person}{Ravi
  Netravali}, {and} \bibinfo{person}{Mohammad Alizadeh}.}
  \bibinfo{year}{2017}\natexlab{}.
\newblock \showarticletitle{Neural Adaptive Video Streaming with Pensieve}. In
  \bibinfo{booktitle}{\emph{{ACM} {SIGCOMM}}}.
\newblock


\bibitem[\protect\citeauthoryear{McKeown, Anderson, Balakrishnan, Parulkar,
  Peterson, Rexford, Shenker, and Turner}{McKeown et~al\mbox{.}}{2008}]%
        {McKeown2008}
\bibfield{author}{\bibinfo{person}{Nick McKeown}, \bibinfo{person}{Tom
  Anderson}, \bibinfo{person}{Hari Balakrishnan}, \bibinfo{person}{Guru
  Parulkar}, \bibinfo{person}{Larry Peterson}, \bibinfo{person}{Jennifer
  Rexford}, \bibinfo{person}{Scott Shenker}, {and} \bibinfo{person}{Jonathan
  Turner}.} \bibinfo{year}{2008}\natexlab{}.
\newblock \showarticletitle{OpenFlow: {E}nabling Innovation in Campus
  Networks}.
\newblock \bibinfo{journal}{\emph{ACM SIGCOMM CCR}} \bibinfo{volume}{38},
  \bibinfo{number}{2} (\bibinfo{year}{2008}), \bibinfo{pages}{69--74}.
\newblock


\bibitem[\protect\citeauthoryear{Narodytska, Ryzhyk, Ganichev, and
  Sevinc}{Narodytska et~al\mbox{.}}{2019}]%
        {Narodytska2019}
\bibfield{author}{\bibinfo{person}{Nina Narodytska}, \bibinfo{person}{Leonid
  Ryzhyk}, \bibinfo{person}{Igor Ganichev}, {and} \bibinfo{person}{Soner
  Sevinc}.} \bibinfo{year}{2019}\natexlab{}.
\newblock \showarticletitle{{BDD}-Based Algorithms for Packet Classification}.
  In \bibinfo{booktitle}{\emph{Formal Methods in Computer Aided Design
  {FMCAD}}}.
\newblock


\bibitem[\protect\citeauthoryear{Nvidia}{Nvidia}{2019}]%
        {NvidiaDLInference}
\bibfield{author}{\bibinfo{person}{Nvidia}.} \bibinfo{year}{2019}\natexlab{}.
\newblock \bibinfo{booktitle}{\emph{Nvidia Deep Learning Inference Platform}}.
\newblock
\urldef\tempurl%
\url{https://www.nvidia.com/en-us/deep-learning-ai/solutions/inference-platform/}
\showURL{%
Retrieved September 25, 2019 from \tempurl}


\bibitem[\protect\citeauthoryear{Pfaff, Pettit, Koponen, Jackson, Zhou,
  Rajahalme, Gross, Wang, Stringer, Shelar, Amidon, and Casado}{Pfaff
  et~al\mbox{.}}{2015}]%
        {Pfaff2015}
\bibfield{author}{\bibinfo{person}{Ben Pfaff}, \bibinfo{person}{Justin Pettit},
  \bibinfo{person}{Teemu Koponen}, \bibinfo{person}{Ethan Jackson},
  \bibinfo{person}{Andy Zhou}, \bibinfo{person}{Jarno Rajahalme},
  \bibinfo{person}{Jesse Gross}, \bibinfo{person}{Alex Wang},
  \bibinfo{person}{Joe Stringer}, \bibinfo{person}{Pravin Shelar},
  \bibinfo{person}{Keith Amidon}, {and} \bibinfo{person}{Martin Casado}.}
  \bibinfo{year}{2015}\natexlab{}.
\newblock \showarticletitle{The Design and Implementation of {O}pen v{S}witch}.
  In \bibinfo{booktitle}{\emph{{USENIX} {NSDI}}}.
\newblock


\bibitem[\protect\citeauthoryear{Rashelbach}{Rashelbach}{2020}]%
        {sourcecode}
\bibfield{author}{\bibinfo{person}{Alon Rashelbach}.}
  \bibinfo{year}{2020}\natexlab{}.
\newblock \bibinfo{booktitle}{\emph{NeuvoMatch source code}}.
\newblock
\urldef\tempurl%
\url{https://github.com/acsl-technion/nuevomatch}
\showURL{%
Retrieved June 21, 2020 from \tempurl}


\bibitem[\protect\citeauthoryear{Rottenstreich and Tapolcai}{Rottenstreich and
  Tapolcai}{2015}]%
        {Rottenstreich2017}
\bibfield{author}{\bibinfo{person}{Ori Rottenstreich} {and}
  \bibinfo{person}{J{\'{a}}nos Tapolcai}.} \bibinfo{year}{2015}\natexlab{}.
\newblock \showarticletitle{Lossy Compression of Packet Classifiers}. In
  \bibinfo{booktitle}{\emph{ACM/IEEE ANCS}}.
\newblock


\bibitem[\protect\citeauthoryear{Sarrar, Uhlig, Feldmann, Sherwood, and
  Huang}{Sarrar et~al\mbox{.}}{2012}]%
        {Sarrar2012}
\bibfield{author}{\bibinfo{person}{Nadi Sarrar}, \bibinfo{person}{Steve Uhlig},
  \bibinfo{person}{Anja Feldmann}, \bibinfo{person}{Rob Sherwood}, {and}
  \bibinfo{person}{Xin Huang}.} \bibinfo{year}{2012}\natexlab{}.
\newblock \showarticletitle{Leveraging Zipf's law for traffic offloading}.
\newblock \bibinfo{journal}{\emph{Computer Communication Review}}
  \bibinfo{volume}{42}, \bibinfo{number}{1} (\bibinfo{year}{2012}),
  \bibinfo{pages}{16--22}.
\newblock


\bibitem[\protect\citeauthoryear{Singh, Baboescu, Varghese, and Wang}{Singh
  et~al\mbox{.}}{2003}]%
        {Singh2003}
\bibfield{author}{\bibinfo{person}{Sumeet Singh}, \bibinfo{person}{Florin
  Baboescu}, \bibinfo{person}{George Varghese}, {and} \bibinfo{person}{Jia
  Wang}.} \bibinfo{year}{2003}\natexlab{}.
\newblock \showarticletitle{Packet Classification Using Multidimensional
  Cutting}. In \bibinfo{booktitle}{\emph{ACM SIGCOMM}}.
\newblock


\bibitem[\protect\citeauthoryear{Spitznagel, Taylor, and Turner}{Spitznagel
  et~al\mbox{.}}{2003}]%
        {Spitznagel2003}
\bibfield{author}{\bibinfo{person}{Ed Spitznagel}, \bibinfo{person}{David~E
  Taylor}, {and} \bibinfo{person}{Jonathan~S Turner}.}
  \bibinfo{year}{2003}\natexlab{}.
\newblock \showarticletitle{Packet classification using extended {TCAM}s}. In
  \bibinfo{booktitle}{\emph{IEEE ICNP}}.
\newblock


\bibitem[\protect\citeauthoryear{Srinivasan, Suri, and Varghese}{Srinivasan
  et~al\mbox{.}}{1999}]%
        {Srinivasan1999}
\bibfield{author}{\bibinfo{person}{Venkatachary Srinivasan},
  \bibinfo{person}{Subhash Suri}, {and} \bibinfo{person}{George Varghese}.}
  \bibinfo{year}{1999}\natexlab{}.
\newblock \showarticletitle{Packet Classification Using Tuple Space Search}. In
  \bibinfo{booktitle}{\emph{ACM SIGCOMM}}.
\newblock


\bibitem[\protect\citeauthoryear{Taylor}{Taylor}{2005}]%
        {Taylor2005}
\bibfield{author}{\bibinfo{person}{David~E Taylor}.}
  \bibinfo{year}{2005}\natexlab{}.
\newblock \showarticletitle{Survey and Taxonomy of Packet Classification
  Techniques}.
\newblock \bibinfo{journal}{\emph{ACM Computing Surveys (CSUR)}}
  \bibinfo{volume}{37}, \bibinfo{number}{3} (\bibinfo{year}{2005}),
  \bibinfo{pages}{238--275}.
\newblock


\bibitem[\protect\citeauthoryear{Taylor and Turner}{Taylor and Turner}{2005}]%
        {DCFLTaylor2005}
\bibfield{author}{\bibinfo{person}{David~E. Taylor} {and}
  \bibinfo{person}{Jonathan~S. Turner}.} \bibinfo{year}{2005}\natexlab{}.
\newblock \showarticletitle{Scalable packet classification using distributed
  crossproducing of field labels}. In \bibinfo{booktitle}{\emph{IEEE INFOCOM}}.
\newblock


\bibitem[\protect\citeauthoryear{Taylor and Turner}{Taylor and Turner}{2007}]%
        {TaylorTurner2007}
\bibfield{author}{\bibinfo{person}{David~E Taylor} {and}
  \bibinfo{person}{Jonathan~S Turner}.} \bibinfo{year}{2007}\natexlab{}.
\newblock \showarticletitle{Classbench: A Packet Classification Benchmark}.
\newblock \bibinfo{journal}{\emph{IEEE/ACM Transactions on Networking (TON)}}
  \bibinfo{volume}{15}, \bibinfo{number}{3} (\bibinfo{year}{2007}),
  \bibinfo{pages}{499--511}.
\newblock


\bibitem[\protect\citeauthoryear{Valadarsky, Schapira, Shahaf, and
  Tamar}{Valadarsky et~al\mbox{.}}{2017}]%
        {Valadarsky2017}
\bibfield{author}{\bibinfo{person}{Asaf Valadarsky}, \bibinfo{person}{Michael
  Schapira}, \bibinfo{person}{Dafna Shahaf}, {and} \bibinfo{person}{Aviv
  Tamar}.} \bibinfo{year}{2017}\natexlab{}.
\newblock \showarticletitle{Learning to Route with Deep {RL}}. In
  \bibinfo{booktitle}{\emph{NIPS Deep Reinforcement Learning Symposium}}.
\newblock


\bibitem[\protect\citeauthoryear{Vamanan, Voskuilen, and Vijaykumar}{Vamanan
  et~al\mbox{.}}{2010}]%
        {VamananBalajee2011}
\bibfield{author}{\bibinfo{person}{Balajee Vamanan}, \bibinfo{person}{Gwendolyn
  Voskuilen}, {and} \bibinfo{person}{T.~N. Vijaykumar}.}
  \bibinfo{year}{2010}\natexlab{}.
\newblock \showarticletitle{Effi{C}uts: {O}ptimizing Packet Classification for
  Memory and Throughput}. In \bibinfo{booktitle}{\emph{ACM SIGCOMM}}.
\newblock


\bibitem[\protect\citeauthoryear{Varvello, Laufer, Zhang, and
  Lakshman}{Varvello et~al\mbox{.}}{2016}]%
        {Varvello2016}
\bibfield{author}{\bibinfo{person}{Matteo Varvello}, \bibinfo{person}{Rafael
  Laufer}, \bibinfo{person}{Feixiong Zhang}, {and} \bibinfo{person}{T.~V.
  Lakshman}.} \bibinfo{year}{2016}\natexlab{}.
\newblock \showarticletitle{Multilayer Packet Classification with Graphics
  Processing Units}.
\newblock \bibinfo{journal}{\emph{IEEE/ACM Transactions on Networking ({TON})}}
  \bibinfo{volume}{24}, \bibinfo{number}{5} (\bibinfo{year}{2016}),
  \bibinfo{pages}{2728--2741}.
\newblock


\bibitem[\protect\citeauthoryear{Yeo, Jung, Kim, Shin, and Han}{Yeo
  et~al\mbox{.}}{2018}]%
        {Hyunho2018}
\bibfield{author}{\bibinfo{person}{Hyunho Yeo}, \bibinfo{person}{Youngmok
  Jung}, \bibinfo{person}{Jaehong Kim}, \bibinfo{person}{Jinwoo Shin}, {and}
  \bibinfo{person}{Dongsu Han}.} \bibinfo{year}{2018}\natexlab{}.
\newblock \showarticletitle{Neural Adaptive Content-aware Internet Video
  Delivery}. In \bibinfo{booktitle}{\emph{{USENIX} {OSDI}}}.
\newblock


\bibitem[\protect\citeauthoryear{Yingchareonthawornchai, Daly, Liu, and
  Torng}{Yingchareonthawornchai et~al\mbox{.}}{2018}]%
        {YingSoDa2018}
\bibfield{author}{\bibinfo{person}{Sorrachai Yingchareonthawornchai},
  \bibinfo{person}{James Daly}, \bibinfo{person}{Alex~X. Liu}, {and}
  \bibinfo{person}{Eric Torng}.} \bibinfo{year}{2018}\natexlab{}.
\newblock \showarticletitle{A Sorted-Partitioning Approach to Fast and Scalable
  Dynamic Packet Classification}.
\newblock \bibinfo{journal}{\emph{IEEE/ACM Transactions on Networking ({TON})}}
  \bibinfo{volume}{26}, \bibinfo{number}{4} (\bibinfo{year}{2018}),
  \bibinfo{pages}{1907--1920}.
\newblock


\bibitem[\protect\citeauthoryear{Zaki, P{\"{o}}tsch, Chen, Subramanian, and
  G{\"{o}}rg}{Zaki et~al\mbox{.}}{2015}]%
        {Yasir2015}
\bibfield{author}{\bibinfo{person}{Yasir Zaki}, \bibinfo{person}{Thomas
  P{\"{o}}tsch}, \bibinfo{person}{Jay Chen}, \bibinfo{person}{Lakshminarayanan
  Subramanian}, {and} \bibinfo{person}{Carmelita G{\"{o}}rg}.}
  \bibinfo{year}{2015}\natexlab{}.
\newblock \showarticletitle{Adaptive Congestion Control for Unpredictable
  Cellular Networks}. In \bibinfo{booktitle}{\emph{{ACM} {SIGCOMM}}}.
\newblock


\bibitem[\protect\citeauthoryear{Zeng, Kazemian, Varghese, and McKeown}{Zeng
  et~al\mbox{.}}{2012}]%
        {Zeng2012}
\bibfield{author}{\bibinfo{person}{Hongyi Zeng}, \bibinfo{person}{Peyman
  Kazemian}, \bibinfo{person}{George Varghese}, {and} \bibinfo{person}{Nick
  McKeown}.} \bibinfo{year}{2012}\natexlab{}.
\newblock \showarticletitle{Automatic Test Packet Generation}. In
  \bibinfo{booktitle}{\emph{ACM CoNEXT}}.
\newblock


\end{thebibliography}

\vspace{3pt}
\noindent \changes{Appendices are supporting material that has not been peer-reviewed.}

\appendix

\section{\MSNN Correctness} \label{appendix:omitted_proofs}

\subsection{Responsibility of a submodel}

Denote the input domain of an \MSNN model as $D \subset \mathbb{R}$ and its number of stages as $n$.

\begin{theorem} [Responsibility Theorem] \label{theorem:responsibility}
Let $s_i$ be a trained stage such that $i<n-1$.
The responsibilities of submodels in $s_{i+1}$ can be calculated by evaluating a finite set of inputs over the stage $s_i$.
\end{theorem}

The intuition behind Theorem \ref{theorem:responsibility} is based on Corollary \ref{corollary:piecewise_linear}, namely that submodels output piecewise linear functions. Proving it requires some additional definitions.

\begin{definition} [Stage Output] \label{definition:stage_output}
    The output of stage $s_i$ is defined for $x \in D$ as
    $S_i(x)=M_{i,f_i(x)}(x)$
    where $f_i(x)$ is the index of the submodel in $s_i$ that is responsible for input $x$, and defined as
    \begin{equation*}
        f_i(x)=
        \begin{cases}
            0 & i=0 \\
            \big\lfloor S_{i-1}(x) \cdot W_{i} \big\rfloor & i=\{1,2,...,n-1\}
        \end{cases}
    \end{equation*}
\end{definition}

\begin{definition} [Submodel Responsibility] \label{definition:responsibility}
    The responsibility of a submodel $m_{i,j}$ is defined as
    \begin{equation*}
        R_{i,j}=
        \begin{cases} 
        D & i=0 \\
        \big\{ x \: \big\rvert \: f_i(x)=j \big\} & i=\{1,2,...,n-1\}
        \end{cases}
    \end{equation*}
    \noindent
    Note that the responsibilities of every two submodels in the same stage are disjoint.
\end{definition}

\begin{definition} [Left and Right Slopes]
For a range $R$, if points $\min_{x \in R} x$ or $\max_{x \in R} {x}$ are defined, we refer to them as the boundaries of the range.
For all other points, we refer to as internal points of the range.
For a piecewise linear function defined over some range $R$, for every internal point $x \in R$, there exists $\delta > 0$ such that the function is linear in each of $(x-\delta,x), (x, x+\delta)$.
Accordingly, we can refer to the left slope and the right slope of a point, defined as those of the two linear functions.
\end{definition}


\begin{definition} [Trigger Inputs]  \label{definition:trigger_inputs}
We say that an input $g \in D$ is a \emph{trigger input} of a submodel $m_{i,j}$ if one of the following holds: \emph{(i)} $g$ is a boundary point of $D$ (namely, $g = \min_{y \in D} y$ or  $g = \max_{y \in D} {y}$).
\emph{(ii)} $g$ is an internal point of $D$ and the left and right slopes of $M_{i,j}(g)$ differ.
\end{definition}

\begin{definition} [Transition Inputs] \label{definition:transition_inputs}
We say that an input $t \in D$ is a \emph{transition input} of a submodel $m_{i,j}$ if it changes submodel selection in the following stage. Formally, there exists $\epsilon>0$ such that for all $0<\delta<\epsilon$:
\begin{equation*}
    \big\lfloor M_{i,j}(t-\delta) \cdot W_{i+1} \big\rfloor
    \neq
    \big\lfloor M_{i,j}(t+\delta) \cdot W_{i+1} \big\rfloor
\end{equation*}
\end{definition}

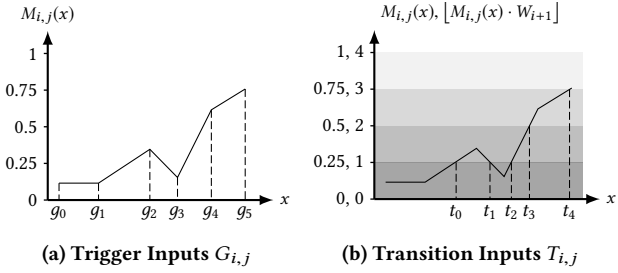
\begin{figure}
    \centering
    \begin{subfigure}[b]{0.23\textwidth}
        \centering
\ifdefined\RELEASE
\else
\documentclass[conference,compsoc]{IEEEtran}

\usepackage{graphicx}
\usepackage{subfig}

\usepackage[T1]{fontenc}
\usepackage[utf8]{inputenc}
\usepackage{pgfplots}
\usepackage{grffile}
\pgfplotsset{compat=newest}
\usetikzlibrary{plotmarks}
\usetikzlibrary{arrows.meta}
\usepgfplotslibrary{patchplots}
\usetikzlibrary{decorations.markings}
\usepackage{amsmath}

\newcommand{\ouralgorithm}[0]{Plasmus}
\begin{document}
\fi

\newcommand{\scale}{0.75}

\begin{tikzpicture}[scale=\scale, every node/.style={scale=\scale}]

	\tikzstyle{densely dashed}= [dash pattern=on 3pt off 1.3pt]
	
		
	\pgfmathsetmacro{\base}{0};
	\pgfmathsetmacro{\width}{3.7};
	
	\draw[-latex, line width=0.8] (0,0) -- (4,0) node [anchor=west, align=left] {$x$};
	\draw[-latex, line width=0.8] (0,0) -- (0,3) node [anchor=south, align=left] {$M_{i,j}(x)$};
	
	\draw[-, line width=0.5] (0,2.6) -- (-0.07, 2.6) node [anchor=east, font={\large}] {1};
	\draw[-, line width=0.5] (0,1.95) -- (-0.07, 1.95) node [anchor=east, font={\large}] {0.75};
	\draw[-, line width=0.5] (0,1.30) -- (-0.07, 1.30) node [anchor=east, font={\large}] {0.5};
	\draw[-, line width=0.5] (0,0.65) -- (-0.07, 0.65) node [anchor=east, font={\large}] {0.25};
	\draw[-, line width=0.5] (0,0.0) -- (-0.07, 0.0) node [anchor=east, font={\large}] {0};
	
	\pgfmathsetmacro{\a}{\base+0.2};
	\pgfmathsetmacro{\b}{\base+0.9};
	\pgfmathsetmacro{\c}{\base+1.81};
	\pgfmathsetmacro{\d}{\base+2.3};
	\pgfmathsetmacro{\e}{\base+2.9};
	\pgfmathsetmacro{\f}{\base+3.5};
	
	\coordinate (A) at (\a, 0.3);
	\coordinate (B) at (\b, 0.3);
	\coordinate (C) at (\c, 0.9);
	\coordinate (D) at (\d, 0.4);
	\coordinate (E) at (\e, 1.6);
	\coordinate (F) at (\f, 1.97);

	\draw[-] (A) -- (B) -- (C) -- (D) -- (E) -- (F);
	
	\draw[densely dashed] (A) -- (\a,0) node [anchor=north, font={\large}] {$g_0$};
	\draw[densely dashed] (B) -- (\b,0) node [anchor=north, font={\large}] {$g_1$};
	\draw[densely dashed] (C) -- (\c,0) node [anchor=north, font={\large}] {$g_2$};
	\draw[densely dashed] (D) -- (\d,0) node [anchor=north, font={\large}] {$g_3$};
	\draw[densely dashed] (E) -- (\e,0) node [anchor=north, font={\large}] {$g_4$};
	\draw[densely dashed] (F) -- (\f,0) node [anchor=north, font={\large}] {$g_5$};

	
\end{tikzpicture}

\ifdefined\RELEASE
\else
\end{document}
\fi
        \caption{Trigger Inputs $G_{i,j}$}
    \end{subfigure}
    \begin{subfigure}[b]{0.23\textwidth}
        \centering
\ifdefined\RELEASE
\else
\documentclass[conference,compsoc]{IEEEtran}

\usepackage{graphicx}
\usepackage{subfig}

\usepackage[T1]{fontenc}
\usepackage[utf8]{inputenc}
\usepackage{pgfplots}
\usepackage{grffile}
\pgfplotsset{compat=newest}
\usetikzlibrary{plotmarks}
\usetikzlibrary{arrows.meta}
\usepgfplotslibrary{patchplots}
\usetikzlibrary{decorations.markings}
\usepackage{amsmath}

\newcommand{\ouralgorithm}[0]{Plasmus}
\begin{document}
\fi

\newcommand{\scale}{0.75}

\begin{tikzpicture}[scale=\scale, every node/.style={scale=\scale}]

	\tikzstyle{densely dashed}= [dash pattern=on 3pt off 1.3pt]
	
	
	\pgfmathsetmacro{\base}{6.5};
	\pgfmathsetmacro{\width}{3.7};
	
	\draw[draw=none, fill=gray, opacity=0.7](\base,0.0) rectangle (\base+\width, 0.65);
	\draw[draw=none, fill=gray, opacity=0.5](\base,0.65) rectangle (\base+\width, 1.3);
	\draw[draw=none, fill=gray, opacity=0.3](\base,1.3) rectangle (\base+\width, 1.95);
	\draw[draw=none, fill=gray, opacity=0.1](\base,1.95) rectangle (\base+\width, 2.6);
	
	\draw[-latex, line width=0.8] (\base,0) -- (\base+4,0) node [anchor=west, align=left] {$x$};
	\draw[-latex, line width=0.8] (\base,0) -- (\base,3) node [anchor=south west, align=left] {$M_{i,j}(x), \left\lfloor  M_{i,j}(x)  \cdot W_{i+1} \right\rfloor$};
	
	\draw[-, line width=0.5] (\base,2.6) -- (\base-0.07, 2.6) node [anchor=east, font={\large}] {1, 4};
	\draw[-, line width=0.5] (\base,1.95) -- (\base-0.07, 1.95) node [anchor=east, font={\large}] {0.75, 3};
	\draw[-, line width=0.5] (\base,1.30) -- (\base-0.07, 1.30) node [anchor=east, font={\large}] {0.5, 2};
	\draw[-, line width=0.5] (\base,0.65) -- (\base-0.07, 0.65) node [anchor=east, font={\large}] {0.25, 1};
	\draw[-, line width=0.5] (\base,0.0) -- (\base-0.07, 0.0) node [anchor=east, font={\large}] {0, 0};

	\pgfmathsetmacro{\a}{\base+0.2};
	\pgfmathsetmacro{\b}{\base+0.9};
	\pgfmathsetmacro{\c}{\base+1.81};
	\pgfmathsetmacro{\d}{\base+2.3};
	\pgfmathsetmacro{\e}{\base+2.9};
	\pgfmathsetmacro{\f}{\base+3.5};
	
	\coordinate (A) at (\a, 0.3);
	\coordinate (B) at (\b, 0.3);
	\coordinate (C) at (\c, 0.9);
	\coordinate (D) at (\d, 0.4);
	\coordinate (E) at (\e, 1.6);
	\coordinate (F) at (\f, 1.97);
		
	\draw[-] (A) -- (B) -- (C) -- (D) -- (E) -- (F);
	
	\pgfmathsetmacro{\a}{\base+1.45};
	\pgfmathsetmacro{\b}{\base+2.05};
	\pgfmathsetmacro{\c}{\base+2.43};
	\pgfmathsetmacro{\d}{\base+2.75};
	\pgfmathsetmacro{\e}{\base+3.46};
	
	\draw[densely dashed] (\a,0.65) -- (\a,0) node [anchor=north, font={\large}] {$t_0$};
	\draw[densely dashed] (\b,0.65) -- (\b,0) node [anchor=north, font={\large}] {$t_1$};
	\draw[densely dashed] (\c,0.65) -- (\c,0) node [anchor=north, font={\large}] {$t_2$};
	\draw[densely dashed] (\d,1.30) -- (\d,0) node [anchor=north, font={\large}] {$t_3$};
	\draw[densely dashed] (\e,1.95) -- (\e,0) node [anchor=north, font={\large}] {$t_4$};
	
	
\end{tikzpicture}

\ifdefined\RELEASE
\else
\end{document}
\fi
        \caption{Transition Inputs $T_{i,j}$}
    \end{subfigure}
    \caption{Illustration of the trigger inputs ($g_0,...,g_5$) and transition inputs ($t_0,...,t_4$) for graph $M_{i,j}(x)$ of submodel $m_{i,j}$. Note that $W_{i+1}$, namely the number of submodels in stage $i+1$, affects the transition inputs of $m_{i,j}$ and equals $4$.}
    \label{fig:trigger_transition_inputs}
\end{figure}

\begin{definition} [The function $B_i(x)$]
    We define the function $B_i$ for $i \in \{0,1,...,n-1\}$. $B_i$ is a staircase function of values $[0, W_{i+1}-1]$, and defined as
    $B_i(x)= \lfloor x \cdot W_{i+1} \rfloor$ for $x \in [0,1)$.
\end{definition}

For a submodel $m_{i,j}$, we term the set of its trigger inputs as $G_{i,j}$ and the set of its transition inputs as $T_{i,j}$. See Figure \ref{fig:trigger_transition_inputs} for illustration.
From submodel definition and Corollary \ref{corollary:piecewise_linear}, we can tell that a submodel's ReLU operations determine its trigger inputs. Consequently, any set of trigger inputs is finite and can be calculated using a few linear equations. Nonetheless, calculating the transition inputs of a submodel is not straightforward. We show a fast and efficient way for doing so in the following lemma:

\begin{lemma} \label{lemma:submodel_transition_inputs}
Let $m_{i,j}$ be an \MSNN submodel, and $a<b \in G_{i,j}$ two adjacent trigger inputs of $m_{i,j}$. Then the set $S = [a,b] \cap T_{i,j}$ is finite and can be calculated using the inputs $a$ and $b$ alone.
\end{lemma}

\begin{proof}
        We divide the construction of $S$ to two subsets $S = S_0\cup S_1$. First we handle $S_0$. For each $x \in \{a,b\}$, $x \in S_0$ if and only if there exists $\epsilon>0$ such that for all $0<\delta<\epsilon$:
        \begin{displaymath}
        B_i\big(M_{i,j}(x-\delta)\big) \neq B_i\big(M_{i,j}(x+\delta)\big)
        \end{displaymath}
        
        \noindent
        Now to $S_1$. Without loss of generality, $M_{i,j}(a) \leq M_{i,j}(b)$.
        From Corollary \ref{corollary:piecewise_linear} and Definition \ref{definition:trigger_inputs}, $M_{i,j}$ is linear in $[a,b]$. If $B_i(M_{i,j}(a)) = B_i(M_{i,j}(b))$, then $S_1=\emptyset$.
        Otherwise, $M_{i,j}(a) \neq M_{i,j}(b)$. $B_i(x)$ outputs discrete values between $B_i(M_{i,j}(a))$ and $B_i(M_{i,j}(b))$ for all $x \in (a,b)$. Denote this finite set of discrete values as $M$. For any $y \in M$ there exists a value $d \in (a,b]$ such that $M_{i,j}(d) \cdot W_{i+1} = y$. By the linearity of $M_{i,j}$ in $[a,b]$:
        \begin{equation} \nonumber
        d=\Big(\frac{y}{W_{i+1}} - M_{i,j}\left(a\right)\Big)\cdot\frac{ b - a }{M_{i,j}(b) - M_{i,j}(a)}+a
        \end{equation}
        
        \noindent
        We construct $S_1$ as follows:
        \begin{equation} \nonumber
        S_1=\bigg\{ \Big(\frac{y}{W_{i+1}} - M_{i,j}(a)\Big)\cdot\frac{ b - a }{M_{i,j}(b) - M_{i,j}(a)}+a \;\Big\rvert\; \forall y \in M \bigg\}
        \end{equation}
\end{proof}

\begin{corollary} \label{corollary:transition_input_construction}
The set of transition inputs $T_{i,j}$ can be calculated using $G_{i,j}$ and its size is bounded such that $|T_{i,j}| \leq W_{i+1} \cdot |G_{i,j}|$.
\end{corollary}

Not all transition inputs of all submodels are reachable, as some exist outside of their corresponding submodel's responsibility. Therefore, we define the set of reachable transition inputs of a stage $s_i$ as the \emph{transition set} of a stage:

\begin{definition} [Transition Set] \label{definition:transition_set}
    The \emph{transition set} $U_i$ of a stage $s_i$ is an ordered set, defined as:
    \begin{equation*} 
    U_i = \{ \min(D) \} \cup \{ \bigcup_{j=0}^{W_i-1} T_{i,j} \cap R_{i,j} \} \cup \{ \max(D) \}
    \end{equation*}
\end{definition}

The proof of Theorem \ref{theorem:responsibility} directly follows from the next two lemmas:

\begin{lemma} \label{lemma:one_submodel_between_transitions}
Let $s_i, s_{i+1}$ be two adjacent stages. For any two adjacent values $u_0 < u_1 \in U_i$ there exists a submodel $m_{i+1,j}$ such that $S_{i+1}(x)$ is piecewise linear and equal to $M_{i+1,j}(x)$ for all $x \in (u_0,u_1)$.
\end{lemma}

\begin{proof}
        We show that there exists a submodel $m_{i+1,j}$ such that any $x \in (u_0, u_1)$ satisfies $x \in R_{i+1,j}$, which implies $f_{i+1}(x)=j$ and so $S_{i+1}(x)=M_{i+1,j}(x)$. By Corollary \ref{corollary:piecewise_linear}, $S_{i+1}$ is piecewise linear for all $x \in (u_0,u_1)$.
        
        Let $x<y \in (u_0, u_1)$.
        Assume by contradiction there exist two submodels $m_{i+1,j_0}$ and $m_{i+1,j_1}$ such that $x \in R_{i+1,j_0}$ and $y \in R_{i+1,j_1}$.
        From Definition \ref{definition:responsibility}, $f_{i+1}(x) \neq f_{i+1}(y)$ implies $B_i(S_i(x)) \neq B_i(S_i(y))$.
        Thus, there exists an input $z \in (x,y]$ and $\epsilon>0$ such that for all $0<\delta<\epsilon$:
        \begin{equation} \nonumber
        B_{i}\big( S_{i}(z - \delta) \big)
        \neq
        B_{i}\big( S_{i}(z + \delta) \big)
        \end{equation}
        
        \noindent
        Since $S_i$ consists of the outputs of submodels in $s_i$, there exists a submodel $m_{i,k}$ such that $S_i(z)=M_{i,k}(z)$.
        Therefore, $z \in T_{i,k}$ and $z \in R_{i,k}$, which means $z \in U_i$, in contradiction to definition of $u_0$ and $u_1$.
\end{proof}

\begin{lemma} \label{lemma:fi_calculated_using_ui_over_si}
Let $s_i$ be an \MSNN stage such that $i \in \{0,1,...,n-2\}$. The function $f_{i+1}$ defined over the space $D$ can be calculated using the inputs $U_i$ over $S_i$.
\end{lemma}

\begin{proof}
    	Let $u_0<u_1 \in U_i$ be two adjacent values.
    	By Lemma \ref{lemma:one_submodel_between_transitions} there exists a submodel $m_{i+1,j}$ such that $S_{i+1}(x)=M_{i,j}(x)$ for all $x \in (u_0, u_1)$.
    	From Definition \ref{definition:stage_output}, $f_{i+1}(x)=j$ for all $x \in (u_0,u_1)$. By calculating $B_i(S_i(u_0))$ and $B_i(S_i(u_1))$, $f_{i+1}(x)$ is known for all $x \in [u_0,u_1]$.
    	Since $\min\{D\} \in U_i$ and $max\{D\} \in U_i$, $f_{i+1}(x)$ is known for all $x \in D$.
\end{proof}

\subsection{Submodel prediction error}

\begin{theorem} [Submodel Prediction Error] \label{theorem:prediction_error}
Let $s_{n-1}$ be the last stage of an \MSNN model. The maximum prediction error of any submodel in $s_{n-1}$ can be calculated using a finite set of inputs over the stage $s_{n-1}$.
\end{theorem}

The intuition behind Theorem \ref{theorem:prediction_error} is to address the set of range-value pairs as an additional, virtual, stage in the model.

\begin{definition} [Range-Value Pair]
\label{definition:range_value_pairs}
A range-value pair $\langle r, v \rangle$ is defined such that $r$ is an interval in $D$ and $v \in \{0,1,2,...\}$ is unique to that pair.
\end{definition}

We term $W_n$ the number of range-value pairs an \MSNN model should index. Similar to the definitions for submodels, we extend $f_i$ such that $f_n(x)=\lfloor S_{n-1}(x) \cdot W_n \rfloor$, and say that the responsibility $R_p$ of a pair $p=\langle r,v \rangle$ is the set of inputs $\{x \rvert f_n(x)=v \}$. Consequently, we make the following two observations. First, all inputs in the range $r \setminus R_p$ should have reached $p$ but did not. Second, all inputs in the range $R_p \setminus r$ did reach $p$ but should not.

\begin{definition} [Misclassified Pair Set]
Let $m$ be a submodel in $s_{n-1}$ with a responsibility $R_m$. Denote $P_m$ as the set of all pairs such that a pair $p=\langle r,v \rangle \in P_m$ satisfies $(r \setminus R_p) \cup (R_p \setminus r) \cap R_m \neq \emptyset$. In other words, $P_m$ holds all pairs that were misclassified by $m$, and termed the \emph{misclassified pair set} of $m$.
\end{definition}


\begin{definition} [Maximum Prediction Error]
Let $m$ be a submodel in $s_{n-1}$ with a responsibility $R_m$ and a misclassified pair set $P_m$.
The maximum prediction error of $m$ is defined as:

\begin{equation*}
\max\big\{
|f_{n}(x)-v|
\,\big\rvert\,
\langle r,v \rangle \in P_m, x \in R_m
\big\}
\end{equation*}
\end{definition}


\begin{lemma} \label{lemma:misprediction_pairs_of_submodel}
The misclassified pair sets of all submodels in $s_{n-1}$ can be calculated using $U_{n-1}$ over $S_{n-1}$.
\end{lemma}

\begin{proof}
        Let $q_0<q_1$ be two adjacent values in $U_{n-1}$.
        From Lemma \ref{lemma:one_submodel_between_transitions} there exists a single submodel $m_{n-1,j}$, $j \in W_{n-1}$ s.t $S_{n-1}(x)=M_{n-1,j}(x)$ for all $x \in (q_0, q_1)$.
        Hence, using Corollary \ref{corollary:piecewise_linear}, $S_{n-1}$ is linear in $(q_0, q_1)$.
        Therefore, the values of $S_{n-1}$ in $[q_0, q_1]$ can be calculated using $q_0$ and $q_1$ alone.
        Consequently, according to the definitions of $f_n$ and the responsibility of a pair, the set of pairs $P_j$ with responsibilities in $[q_0, q_1]$ can also be calculated using $q_0$ and $q_1$.
        Calculating the responsibilities of all pairs is performed by repeating the process for any two adjacent points in $U_{n-1}$.
        
        At this point, as we know $R_p$ for all $p=\langle r,v \rangle$, calculating the set $(r \setminus R_p)\cup(R_p \setminus r)$ is trivial.
        Acquiring the responsibility of any submodel in $s_{n-1}$ using Theorem \ref{theorem:responsibility} enables us to calculate its misclassified pair set immediately.
\end{proof}
        
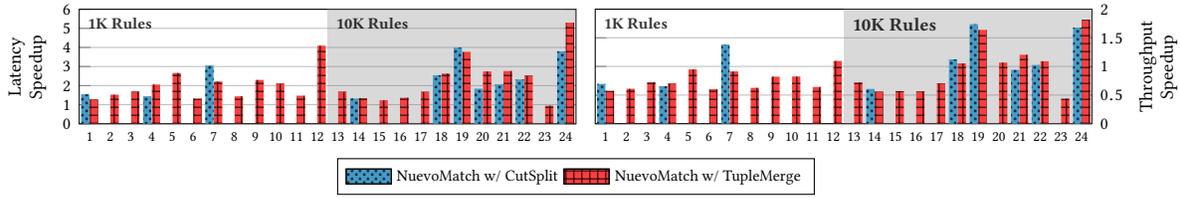
\begin{figure*}
	\centering
\ifdefined\RELEASE
\else
\documentclass[conference,compsoc]{IEEEtran}

\usepackage{graphicx}
\usepackage{subfig}

\usepackage[T1]{fontenc}
\usepackage[utf8]{inputenc}
\usepackage{pgfplots}
\usepackage{grffile}
\pgfplotsset{compat=newest}
\usetikzlibrary{plotmarks}
\usetikzlibrary{arrows.meta}
\usepgfplotslibrary{patchplots}
\usepackage{amsmath}

\newcommand{\ouralgorithm}[0]{NuevoMatch}
\newcommand{\iset}[0]{iSet}
\newcommand{\isets}[0]{iSets}
\begin{document}
\fi

\definecolor{cutsplit}{HTML}{489ac7}
\definecolor{cutsplit_dark}{HTML}{003242}

\definecolor{neurocuts}{HTML}{ff8b0f}
\definecolor{neurocuts_dark}{HTML}{1a0d00}

\definecolor{tuplemerge}{HTML}{f83f3f}
\definecolor{tuplemerge_dark}{HTML}{280606}

\definecolor{graybg}{HTML}{e0e0e0}
	
\pgfmathsetmacro{\barwidth}{0.4}
\pgfmathsetmacro{\barshift}{0.42}
\usetikzlibrary{patterns}

\newcommand{\scale}{1}

\begin{tikzpicture}[scale=\scale, every node/.style={scale=\scale}]

\pgfplotsset{
width=2.6in,
height=0.6in,
scale only axis,
xmin=0.5, xmax=24.5,
xticklabel style={font={\tiny}},
xtick style={draw=none},
yticklabel style={font={\scriptsize}},
yminorticks=true,
ymajorgrids, yminorgrids,
xtick={1,2,3,4,5,6,7,8,9,10,11,12,13,14,15,16,17,18,19,20,21,22,23,24},
ylabel style={font=\footnotesize, align=center},
legend style={
at={(1,-0.3)}, anchor=north, legend cell align=left, legend columns=-1, font={\scriptsize}
},
}

\begin{axis}[
at={(0in,0in)},
ymin=0, ymax=6,
ytick={0,1,2,3,4,5,6},
ylabel={Latency\\Speedup},
]

\addlegendimage{fill=cutsplit, area legend, postaction={pattern=crosshatch dots, pattern color=cutsplit_dark}};
\addlegendentry{\ouralgorithm{} w/ CutSplit};
\addlegendimage{fill=tuplemerge, area legend, postaction={pattern=grid, pattern color=tuplemerge_dark}};
\addlegendentry{\ouralgorithm{} w/ TupleMerge};

\addplot[draw=none, opacity=0.3, fill=gray] table {
	12.5 0
	12.5 6
	25 6
	25 0
};
\addplot[ybar, bar width=\barwidth, fill=cutsplit, draw=none, area legend, postaction={pattern=crosshatch dots, pattern color=cutsplit_dark},bar shift=-\barshift/2] table {
1 1.54564
4 1.44111
7 3.05463

14 1.32865
18 2.54386
19 3.99317
20 1.8397
21 2.06563
22 2.32044
24 3.79467

};

\addplot[ybar, bar width=\barwidth, fill=tuplemerge, draw=none, area legend, postaction={pattern=grid, pattern color=tuplemerge_dark},bar shift=+\barshift/2] table {
1 1.27037
2 1.52563
3 1.68518
4 2.05608
5 2.66187
6 1.32207
7 2.21475
8 1.43763
9 2.29661
10 2.10195
11 1.47252
12 4.09127

13 1.67661
14 1.34433
15 1.23339
16 1.36405
17 1.67495
18 2.63269
19 3.75229
20 2.74264
21 2.76341
22 2.51834
23 0.947012
24 5.29901
};

\node [anchor=north west, font={\scriptsize \bf}, opacity=0.8] at (axis cs:0.5,6) {1K Rules};
\node [anchor=north west, font={\scriptsize \bf}, opacity=0.8] at (axis cs:12.5,6) {10K Rules};

\end{axis}


\begin{axis}[
at={(2.7in,0in)},
ymin=0, ymax=2,
ylabel={Throughput\\Speedup},
yticklabel pos=right,
]
\addplot[draw=none, opacity=0.3, fill=gray] table {
	12.5 0
	12.5 6
	25 6
	25 0
};

\addplot[ybar, bar width=\barwidth, fill=cutsplit, draw=none, area legend, postaction={pattern=crosshatch dots, pattern color=cutsplit_dark},bar shift=-\barshift/2] table {
1 0.694826
4 0.656062
7 1.38289

14 0.607448
18 1.12367
19 1.73684
29 0.842083
21 0.940678
22 1.02564
24 1.67881
};

\addplot[ybar, bar width=\barwidth, fill=tuplemerge, draw=none, area legend, postaction={pattern=grid, pattern color=tuplemerge_dark},bar shift=+\barshift/2] table {
1 0.574082
2 0.609313
3 0.723505
4 0.69927
5 0.949231
6 0.603424
7 0.912674
8 0.626273
9 0.821339
10 0.824301
11 0.641441
12 1.09581

13 0.719974
14 0.559879
15 0.569011
16 0.562887
17 0.701145
18 1.05389
19 1.63811
20 1.06728
21 1.20382
22 1.0906
23 0.438439
24 1.81565
};

\node [anchor=north west, font={\scriptsize \bf}, opacity=0.8] at (axis cs:0.5,2) {1K Rules};
\node [anchor=north west, font={\footnotesize\bf}, opacity=0.8] at (axis cs:12.5,2) {10K Rules};

\end{axis}

\end{tikzpicture}%
\ifdefined\RELEASE
\else
\end{document}
\fi
	\caption{A detailed version of end-to-end performance for \emph{small} rule-sets. Speedup in throughput and latency of \ouralgorithm{} against stand-alone versions of CutSplit and TupleMerge. Classifiers with no valid \isets are not displayed.}
	\label{fig:performance_speedup_small}
\end{figure*}
        
\noindent\textbf{Proof of Theorem \ref{theorem:prediction_error}}
\begin{proof}
        Let $m$ be a submodel in $s_{n-1}$ with a responsibility $R_m$. For simplicity, we address the case where $R_m$ is a continuous range. Extension to the general case is possible by repeating the proof for any continuous range in $R_m$.
        
        Denote the submodel's finite set of trigger inputs as $G_m$. Define the set $Q$ as follows:
        \begin{equation*}
            Q=\min{R_m} \cup (G_m \cap R_m) \cup \max{R_m}
        \end{equation*}
        
        Let $q_0<q_1$ be two adjacent values in $Q$. From the definition of trigger inputs, $m$ outputs a linear function in $[q_0, q_1]$.
        Hence, the set of values $S_0=\{ f_n(x) \rvert x \in [q_0, q_1]\}$ can be calculated using only $q_0$ and $q_1$ over $S_{n-1}$.
        From Lemma \ref{lemma:misprediction_pairs_of_submodel}, the misclassified pair set $P_m$ can be calculated using the finite set $U_{n-1}$. Denote the set
        \begin{equation*}
        \hat{P}_0=\{
        \langle r,v \rangle
        \,\rvert\,
        \langle r,v \rangle \in P_m ,\,
        r \cap [q_0,q_1] \neq \emptyset
        \}
        \end{equation*}
        
        \noindent
        Calculating $\max\{s-v \rvert s \in S_0 ,\, \langle r,v \rangle \in \hat{P}_0\}$ yields the maximum error of $m$ in $[q_0, q_1]$.
        Repeating the process for any two adjacent points in $Q$ yields the maximum error of $m$ for all $R_m$.
\end{proof}

\begin{table}[b]
\small
    \caption{\MSNN configurations for different input rule-set sizes.}
    \label{table:msnn_structure}
    \begin{tabular}{ccc}    
		\toprule
		\#Rules & \#Stages & Stage Widths \\
		\midrule
		Less than $10^3$ & 2 & [1, 4] \\
		$10^3$ to $10^4$ & 3 & [1, 4, 16] \\
		$10^4$ to $10^5$ & 3 & [1, 4, 128] \\
		More than $10^5$ & 3 & [1, 8, 256] or [1, 8, 512] \\
		\bottomrule
	\end{tabular}
\end{table}

\noindent\textbf{Rule-set names} in Figures \ref{fig:performance_speedup_large} and \ref{fig:performance_speedup_small}, by order: ACL1, ACL2, ACL3, ACL4, ACL5, FW1, FW2, FW3, FW4, FW5, IPC1, IPC2.

\end{document}